\documentclass[a4paper,11pt]{article} 

\pdfoutput=1

\usepackage[x11names, svgnames, rgb]{xcolor}

\usepackage[T1]{fontenc}
\usepackage[english]{babel}
\usepackage{tikz}
\usepackage[compat=1.1.0]{tikz-feynman}
\usepackage{slashed}
\usepackage{tabularx}
\usepackage{wrapfig}
\usepackage{subfig}
\usepackage{subcaption}
\usepackage{afterpage}
\usepackage{physics}
\usepackage{comment}
\usepackage{emptypage}
\usepackage{listings}
\usepackage{pdfpages}
\usepackage{enumerate}
\usepackage{appendix}
\usepackage{bbold}
\usepackage{bm}
\usepackage{mathrsfs}
\usepackage{setspace}
\usepackage{pifont}
\usepackage{xfrac}
\usepackage{cancel}
\usepackage{mathtools}
\usepackage{multirow}
\usepackage{booktabs}
\usepackage{makecell}
\usepackage[export]{adjustbox}
\usepackage{lmodern}

\usetikzlibrary{snakes,arrows,shapes}
\usepackage{jheppub} 

\newcommand{\agl}[2]{\langle#1 #2 \rangle}
\newcommand{\sqr}[2]{\lbrack #1 #2 \rbrack}

\newcommand{\tildee}[1]{\widetilde{#1}}

\DeclareMathOperator*{\dime}{\text{dim}}
\makeatletter\g@addto@macro\bfseries{\boldmath}\makeatother

\definecolor{nicered}{rgb}{0.7,0.1,0.1}
\definecolor{nicegreen}{rgb}{0.1,0.5,0.1}
\definecolor{violet}{rgb}{0.7,0.3,0.3}
\hypersetup{colorlinks,citecolor= nicegreen,linkcolor= nicered}

\preprint{CERN-TH-2025-032}

\title{Anomalous Dimension of a General Effective Gauge Theory I: Bosonic Sector}

\author[a]{Jason Aebischer,}
\author[b,c]{Luigi C.~Bresciani,}
\author[c]{Nud\v zeim Selimovi\'c}
\affiliation[a]{Theoretical Physics Department, CERN, 1211 Geneva 23, Switzerland}
\affiliation[b]{Dipartimento di Fisica e Astronomia ``G.~Galilei'', Università degli Studi di Padova,\\Via F.~Marzolo 8,
I-35131 Padova, Italy}
\affiliation[c]{Istituto Nazionale di Fisica Nucleare, Sezione di Padova,\\Via F.~Marzolo 8,
I-35131 Padova, Italy}

\emailAdd{jason.aebischer@cern.ch}
\emailAdd{luigicarlo.bresciani@phd.unipd.it}
\emailAdd{nudzeim.selimovic@pd.infn.it}

\abstract{
We classify the physical operators of the most general bosonic effective gauge theory up to dimension six using on-shell methods. Based on this classification, we compute the complete one-loop anomalous dimension employing both on-shell unitarity-based and geometric techniques. Our analysis fully accounts for the mixing of operators with different dimensions. The results broadly apply to any Effective Field Theory with arbitrary gauge symmetry and bosonic degrees of freedom. To illustrate their utility, we perform a complete cross-check of results on the renormalization of the Standard Model Effective Field Theory (SMEFT), $O(n)$ scalar theory, and the SMEFT extended with an axion-like particle. Additionally, we present new results for axion-like particles with CP-violating interactions.
}
\allowdisplaybreaks

\begin{document}
\maketitle
\pagestyle{myplain}

\section{Introduction}

Anomalous Dimension Matrices (ADMs) play a key role in Effective Field Theories (EFTs). Due to the renormalization procedure, the coupling constants and Wilson coefficients of a given EFT depend on the renormalization scale. This scale dependence is governed by the Renormalization Group Equations (RGEs) of the EFT. This set of differential equations describes the dependence of the Wilson coefficients and couplings on the unphysical renormalization scale $\mu$, which is introduced in dimensional regularization. The RGEs for a given theory can be derived using different computational methods. The traditional one is the diagrammatic approach, in which the divergent part of Feynman diagrams contributing to a given process is extracted. The RGEs are proportional to the coefficient of the $1/\epsilon$-poles of the corresponding Feynman diagrams, regularized in $d=4-2\epsilon$ space-time dimensions. This approach was adopted for instance to derive the dimension-six RGEs of the Standard Model Effective Field Theory (SMEFT) \cite{Buchmuller:1985jz,Grzadkowski:2010es,Brivio:2017vri,Isidori:2023pyp} at the one-loop level \cite{Jenkins:2013zja,Jenkins:2013wua,Alonso:2013hga,Alonso:2014zka}, and similar for the Weak Effective Theory (WET) \cite{Jenkins:2017dyc}.\footnote{Many contributions to the RGEs of operators with dimensionality different from six have been computed in the literature \cite{Chankowski:1993tx,Babu:1993qv,Antusch:2001ck,Davidson:2018zuo,Liao:2016hru,Liao:2019tep,Zhang:2023kvw,Zhang:2023ndw,Chala:2021pll,DasBakshi:2022mwk,DasBakshi:2023htx,Chala:2023xjy,Jiang:2020mhe,Helset:2022pde,Assi:2023zid,Boughezal:2024zqa} and also some two-loop results are known \cite{Aebischer:2022anv,Ibarra:2024tpt,Aebischer:2025hsx,Naterop:2024ydo,Jenkins:2023bls,Jenkins:2023rtg,Born:2024mgz}.} Very recently this approach was also used to derive the RGEs for a general bosonic EFT up to dimension six in \cite{Fonseca:2025zjb,Misiak:2025xzq}.

RGEs can however also be computed using functional methods \cite{Fuentes-Martin:2024agf}. This approach was used to compute the one-loop RGEs of the bosonic sector of the SMEFT \cite{Buchalla:2019wsc}, and since recently even the complete set of two-loop RGEs for the bosonic SMEFT operators is known \cite{Born:2024mgz}. Another method to compute RGEs is based on geometric arguments. Interpreting the scalar and gauge fields as coordinates of a manifold, the RGEs are related to the curvature in this field space \cite{Helset:2022pde,Assi:2023zid,Jenkins:2023rtg,Jenkins:2023bls}. Finally, a fourth and very powerful method to compute RGEs is to employ the spinor-helicity formalism, which has been formulated in general terms in \cite{Caron-Huot:2016cwu}.\footnote{A pedagogical introduction to this framework can be found for instance in \cite{Baratella:2020lzz}.} This method was used for instance to compute several results in the SMEFT: The two-loop RGEs for electroweak dipole operators were computed in \cite{Panico:2018hal}, dimension-eight results were presented in~\cite{AccettulliHuber:2021uoa,Jiang:2020mhe}, and parts of the one-loop SMEFT RGEs as well as various vanishing entries were derived in \cite{Bern:2020ikv,Bern:2019wie, Jiang:2020mhe,Cheung:2015aba}. Finally, in \cite{EliasMiro:2020tdv} several two-loop mixing contributions were computed. Furthermore, a generalization of the approach in \cite{Caron-Huot:2016cwu} accounting for the mixing of operators with different dimensions and leading mass effects was presented in~\cite{Bresciani:2023jsu}, and the method was applied to study the RGEs of EFTs with axion-like particles~\cite{Bresciani:2024shu}. 

In this article, we do not concentrate on a particular EFT, but instead adopt a general approach. We consider a general EFT involving scalar and vector fields, containing all conceivable interactions up to mass-dimension six. For this general EFT, we compute the complete set of one-loop RGEs for all couplings up to $\mathcal{O}(1/\Lambda^2)$ in the power counting, where $\Lambda$ denotes the new physics scale. Hence, double insertions of dimension-six operators as well as products of dimension-five and dimension-six Wilson coefficients will be neglected in the running. To obtain these general RGEs we mainly adopt the spinor-helicity formalism, although parts of the RGEs were obtained using the geometric or diagrammatic approach. For the renormalizable part of such general theories, several results exist in the literature. Some early work on this topic at the two-loop level can be found in \cite{Machacek:1983tz,Machacek:1983fi,Machacek:1984zw,Jack:1984vj}, which was later corrected in \cite{Luo:2002ti,Schienbein:2018fsw}. Two-loop renormalization of vacuum expectation values in gauge theories was computed in \cite{Sperling:2013eva,Sperling:2013xqa}. The three-loop beta function for Yukawa and gauge interactions were derived in \cite{Davies:2021mnc,Poole:2019txl} and \cite{Pickering:2001aq,Poole:2019kcm,Mihaila:2012pz}, respectively, and the generic four-loop gauge beta function is given in \cite{Bednyakov:2021qxa}. The three-loop RGEs for gaugeless models containing scalars and fermions were derived in \cite{Jack:2023zjt,Steudtner:2021fzs}. Finally, the quartic beta-functions were computed up to three loops in \cite{Steudtner:2024teg} and for scalar theories, the renormalizable part is even known up to six loops \cite{Bednyakov:2021ojn,Bednyakov:2025sri}. 

In this article, we compute the full RGEs for the rest of the couplings of a general EFT up to dimension six, where we include one-loop mixing effects of operators with different dimensionalities. Our results can be used to derive the RGEs for specific models at the one-loop level, which we explicitly demonstrate for the case of the SMEFT, $O(n)$ scalar theory, and the SMEFT extended with a CP-violating axion-like particle. Besides a valuable consistency check, this exercise serves as a guide to use our results for specific theories.

The rest of the article is structured as follows: In Sec.~\ref{sec:EFT} we describe the general EFT up to mass dimension six, which is used for the calculation of the RGEs. In Sec.~\ref{sec:method} we review the on-shell approach used to compute the RGEs. The results of our calculation are presented in Sec.~\ref{sec:results} and are applied to specific examples in Sec.~\ref{sec:comp}. We conclude in Sec.~\ref{sec:concl}. Conventions and useful relations are collected in the appendix~\ref{app:conventions}.

\section{General effective gauge theory}
\label{sec:EFT}

In this section, we set the stage for the general EFT used to compute the RGEs. We introduce the renormalizable
interactions and the higher-dimensional operators, together with the corresponding notation that is used in subsequent sections.
The complete Lagrangian takes the form
\begin{equation}\label{eq:fullEFT}
    \mathcal{L}_{\text{EFT}} = \mathcal{L}^{(4)}+\mathcal{L}^{(5)}+\mathcal{L}^{(6)}\,, 
\end{equation}
where the individual parts are defined in the following subsections in Eqs.~\eqref{eq:L4}, \eqref{eq:L5}, and \eqref{eq:L6}.

\subsection{Renormalizable interactions}

The most general renormalisable Lagrangian involving an arbitrary number of real scalar fields $\phi_a$ and gauge fields with field strengths $F_{\mu\nu}^{A_\alpha}$ is given by
\begin{align}\label{eq:L4}
    \mathcal{L}^{(4)} = &-\frac{1}{4}\sum_{\alpha=1}^{N_G}\sum_{\beta=1}^{N_G}\xi^{A_\alpha B_\beta} F^{A_\alpha}_{\mu\nu}F^{{B_\beta}\,\mu\nu}+ \sum_{\alpha=1}^{N_G}\sum_{\beta=1}^{N_G} \tilde\xi^{A_\alpha B_\beta} F^{A_\alpha}_{\mu\nu}\tildee F^{{B_\beta}\,\mu\nu}  + \frac{1}{2} \eta_{ab} (D_\mu\phi)_{a}(D^\mu\phi)_b\nonumber\\
    &-\Lambda-t_a \phi_a-\frac{m^2_{ab}}{2!}\phi_{a}\phi_{b}-\frac{h_{abc}}{3!}\phi_a \phi_b\phi_c
    -\frac{\lambda_{abcd}}{4!}\phi_{a}\phi_{b}\phi_{c}\phi_{d}\,. 
\end{align}
For completeness, we have also included the vacuum energy and tadpole contributions.
Here, we allow for an arbitrary number of gauge groups, $N_G$, and label each gauge group by an index $\alpha\in \{1,\dotsc,N_G\}$. Hence, the direct product of all considered gauge groups determines the full gauge group $G$ of the theory
\begin{equation}
    G = \prod_{\alpha=1}^{N_G} G_\alpha\,.
    \label{eq:full_gauge_group}
\end{equation}
On the other hand, $A_\alpha,B_\alpha,\dotsc \in \{1,\dotsc,\dime G_\alpha\}$ are indices of the adjoint representation of the gauge group $G_\alpha$, while $a_{\alpha},b_{\alpha},\dotsc \in \{1,\dotsc,\dime R_{\alpha}\}$ span the dimension of the $G_\alpha$-representation $R_{\alpha}$ of the real scalar field, which can be reducible. With $a,b,\dotsc$ we denote the collection of $N_G$ indices $\{a_{\alpha}\},\{b_{\alpha}\},\dotsc$ and we keep the summation over these indices implicit. Furthermore, the covariant derivative and the field strength tensors in Eq.~\eqref{eq:L4} are defined as\,\footnote{We use the convention $\epsilon^{0123} =-\epsilon_{0123}= 1$.}
\begin{gather}
    (D_\mu\phi)_{a} = \partial_\mu\phi_{a} -i\sum_{\alpha=1}^{N_G} g_\alpha A^{A_\alpha}_\mu \theta^{A_\alpha}_{ab}\phi_{b}\,,\label{eq:cov_der_phi} \\ F^{A_\alpha}_{\mu\nu}=\partial_\mu A^{A_\alpha}_\nu - \partial_\nu A^{A_\alpha}_\mu + g_\alpha f^{A_\alpha B_\alpha C_\alpha}A^{B_\alpha}_\mu A^{C_\alpha}_\nu\,,\label{eq:field_strength}\\
    \tildee F^{A_\alpha}_{\mu\nu}= \frac{1}{2}\epsilon_{\mu\nu\rho\sigma}  F^{A_\alpha\,\rho\sigma}\,,
\end{gather}
with the gauge couplings $g_\alpha$ and structure constants $f^{A_\alpha B_\alpha C_\alpha}$ and where we have
\begin{equation}
    \theta^{A_\alpha}_{ab} = \delta_{a_{1}b_{1}}\dotsc \delta_{a_{\alpha-1}b_{\alpha-1}}\theta^{A_\alpha}_{a_{\alpha}b_{\alpha}} \delta_{a_{\alpha+1}b_{\alpha+1}}\dotsc \delta_{a_{N_G}b_{N_G}}\,.
\end{equation}
Here, $\theta^{A_\alpha}_{a_{\alpha}b_{\alpha}}$ denote the generators of the representation $R_{\alpha}$, which we assume to be imaginary and antisymmetric. They satisfy 
\begin{equation}
    [\theta^{A_\alpha},\theta^{B_\alpha}] = i f^{A_\alpha B_\alpha C_\alpha}\theta^{C_\alpha} \,.
\end{equation}
For convenience, the following definitions are introduced
\begin{align}
    T^\alpha_{abcd} &= \theta^{A_\alpha}_{ab}\theta^{A_\alpha}_{cd}\,,\\
    Y^{A_\alpha B_\beta}_{ab} &= \theta^{A_\alpha}_{ac}\theta^{B_\beta}_{cb}\,,\\
    F^{A_\alpha B_\alpha C_\alpha D_\alpha} &= f^{A_\alpha B_\alpha E_\alpha} f^{C_\alpha D_\alpha E_\alpha}\,,
\end{align}
which occur in several of the RGEs. Furthermore, identities involving generators, as well as their explicit form in the case of the Standard Model (SM), will be discussed in more detail in Sec.~\ref{subsec:SMEFT_RGEs}.

We also consider the possibility of incorporating off-diagonal kinetic terms for gauge and scalar fields as required by wavefunction renormalization.
In the case of the gauge kinetic term and the topological density, gauge invariance imposes the condition
\begin{equation}
    \xi^{A_\alpha B_\beta} = \delta_{\alpha\beta}\delta^{A_\alpha B_\beta}\,,\quad \tilde \xi^{A_\alpha B_\beta} = \frac{\vartheta_\alpha g_\alpha^2}{32\pi^2} \delta_{\alpha\beta}\delta^{A_\alpha B_\beta}  
\end{equation}
where $A_\alpha$ and $B_\beta$ are adjoint indices of non-Abelian gauge factors within the gauge group $G$ in Eq.~\eqref{eq:full_gauge_group}, and we normalised the topological angle $\vartheta_\alpha$ in a canonical way using $g^2/32\pi^2\int \text{d}^4x\, F\tildee F \in \mathbb{Z}$. However, for multiple $U(1)$ factors, gauge invariance permits the presence of off-diagonal entries~\cite{Holdom:1985ag,delAguila:1988jz}. In our approach,  this feature is realized by promoting the gauge couplings $g_{\alpha}$ to a matrix-valued coupling $g^{A_\alpha B_\beta}$. Consequently, in scenarios involving $U(1)$ mixing, the results presented in Sec.~\ref{sec:results} can be generalized via the following substitution
\begin{equation}
    g_\alpha \theta^{A_\alpha} \to \sum_{\beta=1}^{N_G} g^{A_\alpha B_\beta} \theta^{B_\beta}\,.
\end{equation} 

\subsection{Operator classification}

A basis of physical operators at a given mass dimension can be conveniently constructed by analyzing the independent kinematic structures associated with the contact amplitudes of the EFT~\cite{Shadmi:2018xan,Cheung:2016drk,Falkowski:2019zdo,Durieux:2019siw,Li:2022tec,DeAngelis:2022qco,Durieux:2019eor}.
In the following, we perform such a construction for operators up to mass-dimension six.

We work with the massless spinor-helicity formalism, where a lightlike momentum $p$ is decomposed in terms of the helicity spinors $\lambda$ and $\tildee \lambda$ as
\begin{equation}
    p_{\alpha\dot\alpha} = p_\mu \tau^\mu_{\alpha\dot\alpha} = \lambda_\alpha \tildee \lambda_{\dot\alpha} \,.
\end{equation}
Here, $\tau^\mu = (\mathbb 1,\vec \tau)$ with the Pauli matrices $\vec \tau$, and the spinors $\lambda_\alpha$ and $\tildee \lambda^{\dot\alpha}$ transform in the $(1/2,0)$ and $(0,1/2)$ representations of $SL(2,\mathbb C)$, respectively.
In this framework the Lorentz-invariant inner products are given by\,\footnote{We use the convention $\epsilon^{12}=\epsilon^{\dot 1 \dot 2} = -\epsilon_{12}=-\epsilon_{\dot 1 \dot 2} = 1$.}
\begin{gather}
    \agl{i}{j} = \lambda_i{}^\alpha \lambda_j{}_\alpha = \epsilon_{\alpha\beta} \lambda_i{}^\alpha \lambda_j{}^\beta = -\agl{j}{i}\,,\\
    \sqr{i}{j} = \tildee \lambda_i{}_{\dot\alpha} \tildee \lambda_j{}^{\dot \alpha} = -\epsilon_{\dot\alpha\dot\beta} \tildee \lambda_i{}^{\dot\alpha} \tildee \lambda_j{}^{\dot \beta} = -\sqr{j}{i}\,.
\end{gather}
The Mandelstam invariants can then be expressed as 
\begin{equation}
    s_{ij}=2p_i\cdot p_j = \agl{i}{j}\sqr{j}{i}\,.
\end{equation}
Additionally, for real momenta, square and angle products are related via 
\begin{equation}
    \agl{i}{j} = \sqr{j}{i}^*\,.
\end{equation}
Such angle and square brackets are very useful when decomposing scattering amplitudes. In four spacetime dimensions, $n$-particle amplitudes have mass-dimension~$[\mathcal M_n]=4-n$. Furthermore, contact amplitudes $\mathcal{M}$ are polynomials in the inner products and can be schematically written as
\begin{equation}
    \mathcal M = C_i \times \mathcal K_i\,,
\end{equation}
where $C_i$ denotes the Wilson coefficient (WC) associated to the operator $\mathcal O_i$, and $\mathcal K_i$ is a kinematic polynomial, $\mathcal K_i = \mathcal K_i(\{\agl{i}{j},\sqr{i}{j}\})$. The mass dimension of $\mathcal K_i$ is fixed by the properties of the operator $\mathcal O_i$:
\begin{equation}
    [\mathcal K_i] = [\mathcal O_i] - \ell(\mathcal O_i) \ge 0\,.
\end{equation}
Here, $\ell(\mathcal O_i)$ denotes the length of the operator, namely the number of fundamental fields it is composed of.
Different kinematic structures can be related via momentum conservation, $\sum_{i=1}^n p_i = 0$, or Schouten identities
\begin{align}
    \agl{i}{j}\agl{k}{\ell} + \agl{k}{i}\agl{j}{\ell} + \agl{j}{k}\agl{i}{\ell} &= 0\,,\\
    \sqr{i}{j}\sqr{k}{\ell} + \sqr{k}{i}\sqr{j}{\ell} + \sqr{j}{k}\sqr{i}{\ell} &= 0\,.
\end{align}
Moreover, on-shell three-particle amplitudes vanish due to momentum conservation.
Nevertheless, they can be constructed by assuming complex momenta~\cite{Witten:2003nn}, and, once the particle helicities $\{h_1,h_2,h_3\}$ are fixed, they are uniquely determined by Poincaré invariance, locality, and dimensional analysis~\cite{Benincasa:2007xk}:
\begin{equation}
    \mathcal M(1^{h_1},2^{h_2},3^{h_3}) = 
    \begin{cases}
        C_{\text{H}}\, \agl{1}{2}^{a_3} \agl{2}{3}^{a_1} \agl{3}{1}^{a_2} & \text{if }h_1+h_2+h_3<0\,,\\  C_{\text{A}}\, \sqr{1}{2}^{-a_3} \sqr{2}{3}^{-a_1} \sqr{3}{1}^{-a_2} & \text{if }h_1+h_2+h_3>0\,,
    \end{cases}
\end{equation}
with
\begin{equation}
    a_1 = h_1-h_2-h_3\,,\qquad
    a_2 = h_2-h_3-h_1\,,\qquad
    a_3 = h_3-h_1-h_2\,,
\end{equation}
and $C_{\text{H}}$ and $C_{\text{A}}$ are arbitrary coefficients with mass-dimension given by
\begin{equation}
    [C_{\text{H}}] = 1+h_1+h_2+h_3\,,\qquad 
    [C_{\text{A}}] = 1-h_1-h_2-h_3\,.
\end{equation}
Therefore, we can classify all \emph{on-shell} independent kinematic polynomials and the corresponding \emph{physical} operators as follows:
\begin{itemize}
    \item For dimension-5 operators, where we have $[\mathcal K_i] = 5 - \ell(\mathcal O_i)$:
    \begin{align}
        \ell_i &= 5
        &&
        \implies 
        &
        \mathcal K_i &=
        \begin{cases}
            1
        \end{cases} 
        &&
        \implies 
        &
        \mathcal O_i &=
        \begin{cases}
            \phi_1\, \phi_2\, \phi_3\, \phi_4\, \phi_5\,,
        \end{cases}
        \\
        \ell_i &= 4 
        &&
        \implies
        &
        \mathcal K_i &= 
        \begin{cases}
             \agl{1}{2}
        \end{cases}
        &&
        \implies 
        &
        \mathcal O_i &=
        \begin{cases}
            \psi_1{}_L\,  \psi_2{}_L\,  \phi_3\,  \phi_4  \,,
        \end{cases}
        \\
        \ell_i &= 3 
        &&
        \implies
        &
        \mathcal K_i &=
        \begin{cases}
            \agl{1}{2}^2 \\
            \agl{1}{2}\agl{1}{3}
        \end{cases}
        &&
        \implies 
        &
        \mathcal O_i &= 
        \begin{cases}
            F_1{}_L\,  F_2{}_L\,  \phi_3\,, \\
            F_1{}_L\,  \psi_2{}_L\,  \psi_3{}_L \,.
        \end{cases}
    \end{align}
    
    \item For dimension-6 operators $[\mathcal K_i] = 6 -\ell(\mathcal O_i)$:
    \begin{align}
        \ell_i &= 6
        &&
        \implies 
        &
        \mathcal K_i &=
        \begin{cases}
            1
        \end{cases} 
        &&
        \implies 
        &
        \mathcal O_i &=
        \begin{cases}
            \phi_1\, \phi_2\, \phi_3\, \phi_4\, \phi_5\, \phi_6\,,
        \end{cases}
         \\
        \ell_i &= 5
        &&
        \implies 
        &
        \mathcal K_i &=
        \begin{cases}
            \agl{1}{2}
        \end{cases}
        &&
        \implies 
        &
        \mathcal O_i &=
        \begin{cases}
            \psi_1{}_L\,  \psi_2{}_L\,  \phi_3\,  \phi_4\,  \phi_5 \,,
        \end{cases}
        \\
        \ell_i &= 4
        &&
        \implies 
        &
        \mathcal K_i &= 
        \begin{cases}
            \agl{1}{2}^2 \\
            \agl{1}{2}\agl{1}{3}\\
            \agl{1}{2}\agl{3}{4}\\
            \agl{1}{2}\sqr{3}{4}\\
            \agl{1}{2}\sqr{2}{3} \\
            \agl{1}{2}\sqr{1}{2}
        \end{cases}
        &&
        \implies 
        &
        \mathcal O_i &= 
        \begin{cases}
            F_1{}_L\,  F_2{}_L\,  \phi_3\,  \phi_4\,, \\
            F_1{}_L\,  \psi_2{}_L\,  \psi_3{}_L\,  \phi_4 \,,\\
            \psi_1{}_L\,  \psi_2{}_L\, \psi_3{}_L\,  \psi_4{}_L\,,\\
            \psi_1{}_L\,  \psi_2{}_L\, \psi_3{}_R\,  \psi_4{}_R\,,\\
            \psi_1{}_L\,  D\phi_2\,  \psi_3{}_R\,  \phi_4\,,\\
            D\phi_1\,  D\phi_2\,  \phi_3\,  \phi_4\,.
        \end{cases}
        \\
        \ell_i &= 3
        &&
        \implies 
        &
        \mathcal K_i &=
        \begin{cases}
            \agl{1}{2}\agl{2}{3}\agl{1}{3}
        \end{cases}
        &&
        \implies 
        &
        \mathcal O_i &=
        \begin{cases}
            F_1{}_L\,  F_2{}_L\,  F_3{}_L\,.
        \end{cases}
    \end{align}
\end{itemize}
The scalar and fermion fields are denoted with $\phi$ and $\psi$, respectively, while we use $F$ for the gauge field strengths. The kinematic polynomials with angle brackets substituted by square ones correspond to operators with right-handed fields. Finally, the symmetries of the WCs are inherited by those of the kinematic structures.
This implies for bosonic operators that if a kinematic polynomial is (anti-)symmetric under the exchange of two spinors, then the WC must also be (anti-)symmetric in the associated pair of gauge indices.

In the following, we report the basis of bosonic operators used in this work.
We choose to work with Hermitian operators so that all the components of the WCs are real.

\subsubsection{Dimension-5 operators}

The dimension-five Lagrangian is given by:
\begin{align}\label{eq:L5}
    \mathcal{L}^{(5)}&=\sum_{\alpha=1}^{N_G}\sum_{\beta=1}^\alpha \left[\left[C_{\phi F^2}\right]^{A_\alpha B_\beta}_{a}\phi_{a}F^{A_\alpha}_{\mu\nu}F^{B_\beta \, \mu\nu}+\left[C_{\phi \tildee F^2}\right]^{A_\alpha B_\beta}_{a}\phi_{a}F^{A_\alpha}_{\mu\nu}\tildee F^{B_\beta \, \mu\nu}\right]
    \nonumber\\&\quad 
    + \left[C_{\phi^5}\right]_{abcde}\phi_{a}\phi_{b}\phi_{c}\phi_{d}\phi_{e}\,,
\end{align}
where the operators, their symmetry properties as well as the corresponding form factors --- precisely defined in Eq.~\eqref{eq:FF} --- are collected in Tab.~\ref{tab:dim5}. 
\begin{table}[tbp]
\renewcommand{\arraystretch}{2}
\begin{align*}
\resizebox{\textwidth}{!}{
\begin{array}[t]{|c|c|c|c|}
\toprule
\text{Name} & \text{Operator} &  \text{Symmetry} & \text{Form factor} \\
\midrule\midrule
\mathcal{O}_{\phi^5} & \phi_a\phi_b\phi_c\phi_d\phi_e & [C_{\phi^5}]_{abcde}=[C_{\phi^5}]_{(abcde)} & F_{\phi^5}(1_{a},2_{b},3_{c},4_{d},5_{e}) = 5! \left[C_{\phi^5}\right]_{abcde}\\
\mathcal{O}_{\phi F^2} & \phi_{a}F^{A_\alpha}_{\mu\nu}F^{B_\beta \, \mu\nu} & \left[C_{\phi F^2}\right]^{A_\alpha B_\beta}_{a}=\left[C_{\phi F^2}\right]^{(A_\alpha B_\beta)}_{a} &
F_{\phi F^2_{\alpha\beta}}(1_{a},2^-_{A_\alpha},3^-_{B_\beta}) = - \mathcal S_{\alpha\beta} \left[C_{\phi F^2}\right]^{A_\alpha B_\beta}_{a}\agl{2}{3}^2\\
\mathcal{O}_{\phi \tildee F^2} & \phi_{a}F^{A_\alpha}_{\mu\nu}\tildee F^{B_\beta \, \mu\nu} & \left[C_{\phi \tildee F^2}\right]^{A_\alpha B_\beta}_{a}=\left[C_{\phi \tildee F^2}\right]^{(A_\alpha B_\beta)}_{a} 
 & F_{\phi \tildee F^2_{\alpha\beta}}(1_{a},2^-_{A_\alpha},3^-_{B_\beta})  = - i\mathcal S_{\alpha\beta} \left[C_{\phi \tildee F^2}\right]^{A_\alpha B_\beta}_{a}\agl{2}{3}^2\\
\bottomrule
\end{array}
}
\end{align*}
\caption{Dimension-five operators of the general EFT and their form factors considering negative helicities for the gauge bosons. The symmetry factors $\mathcal{S}_{\alpha\beta}$ are defined in App.~\ref{sec:symm_factors}.}
\label{tab:dim5}
\end{table}

\subsubsection{Dimension-6 operators}\label{subsec:dim6_general}
The dimension-six operators are collected in the following Lagrangian
\begin{align}\label{eq:L6}
    \mathcal{L}^{(6)}&=\sum_{\alpha=1}^{N_G}\sum_{\beta=1}^\alpha \left[\left[C_{\phi^2 F^2}\right]^{A_\alpha B_\beta}_{ab}\phi_{a}\phi_{b}F^{A_\alpha}_{\mu\nu}F^{B_\beta \, \mu\nu}+\left[C_{\phi^2 \tildee F^2}\right]^{A_\alpha B_\beta}_{ab}\phi_{a}\phi_{b}F^{A_\alpha}_{\mu\nu}\tildee F^{B_\beta \, \mu\nu}\right]
    \nonumber\\&\quad  \nonumber
    + \left[C_{\phi^6}\right]_{abcdef}\phi_{a}\phi_{b}\phi_{c}\phi_{d}\phi_{e}\phi_{f}+\left[C_{D^2\phi^4}\right]_{abcd}(D_\mu \phi)_{a}(D^\mu \phi)_{b}\phi_{c}\phi_{d}\\&\quad
    +\sum_{\alpha=1}^{N_G}\left[\left[C_{F^3}\right]^{A_\alpha B_\alpha C_\alpha}F^{A_\alpha\,\nu}_{\mu}F^{B_\alpha\,\rho}_{ \nu} F^{C_\alpha\,\mu}_{\rho}+\left[C_{\widetilde F^3}\right]^{A_\alpha B_\alpha C_\alpha}F^{A_\alpha\,\nu}_{\mu}F^{B_\alpha\,\rho}_{ \nu} \widetilde F^{C_\alpha\,\mu}_{\rho}\right]\,.
\end{align}
The symmetries of the Wilson coefficients are given in Tab.~\ref{tab:dim6} and the form factors of the operators are collected in Tab.~\ref{tab:dim6FFs}.
\begin{table}[tbp]
\renewcommand{\arraystretch}{2.3}
\small
\begin{align*}
\begin{array}[t]{|c|c|c|c|}
\toprule
\text{Name} & \text{Operator} &  \text{Symmetry} \\
\midrule\midrule
\mathcal O_{\phi^6} & \phi_a\phi_b\phi_c\phi_d\phi_e\phi_f & [C_{\phi^6}]_{abcdef}=[C_{\phi^6}]_{(abcdef)} \\
\mathcal O_{D^2\phi^4}  & (D_\mu \phi)_{a}(D^\mu \phi)_{b}\phi_{c}\phi_{d} & [C_{D^2\phi^2}]_{abcd} = [C_{D^2\phi^2}]_{(ab)cd} = [C_{D^2\phi^2}]_{ab(cd)}
\\
\mathcal{O}_{\phi^2 F^2} & \phi_{a}\phi_{b}F^{A_\alpha}_{\mu\nu}F^{B_\beta \, \mu\nu} & \left[C_{\phi^2F^2}\right]^{A_\alpha B_\beta}_{ab}=\left[C_{\phi^2F^2}\right]^{(A_\alpha B_\beta)}_{ab}=\left[C_{\phi^2F^2}\right]^{A_\alpha B_\beta}_{(ab)} \\
\mathcal{O}_{\phi^2\tildee F^2} & \phi_{a}\phi_{b}F^{A_\alpha}_{\mu\nu}\tildee F^{B_\beta \, \mu\nu} & \left[C_{\phi^2 \tildee F^2}\right]^{A_\alpha B_\beta}_{ab}=\left[C_{\phi^2\tildee F^2}\right]^{(A_\alpha B_\beta)}_{ab}=\left[C_{\phi^2\tildee F^2}\right]^{A_\alpha B_\beta}_{(ab)}\\
 \mathcal{O}_{F^3} & F^{A_\alpha\,\nu}_{\mu}F^{B_\alpha\,\rho}_{ \nu}F^{C_\alpha\,\mu}_{\rho} & \left[C_{F^3}\right]^{A_\alpha B_\alpha C_\alpha}=\left[C_{F^3}\right]^{[A_\alpha B_\alpha C_\alpha]}\\
  \mathcal{O}_{\tildee F^3} & F^{A_\alpha\,\nu}_{\mu}F^{B_\alpha\,\rho}_{ \nu}{\tildee F}^{C_\alpha\,\mu}_{\rho} & \left[C_{\tildee F^3}\right]^{A_\alpha B_\alpha C_\alpha}=\left[C_{\tildee F^3}\right]^{[A_\alpha B_\alpha C_\alpha]}\\
\bottomrule
\end{array}
\end{align*}
\caption{Dimension-six operators of the general EFT together with the corresponding Wilson coefficients.}
\label{tab:dim6}
\end{table}
\begin{table}[tbp]
\renewcommand{\arraystretch}{2.5}
\small
\begin{align*}
\begin{array}[t]{|c|c|c|c|}
\toprule
\text{Name}  & \text{Form factor} \\
\midrule\midrule
\mathcal O_{\phi^6} & F_{\phi^5}(1_{a},2_{b},3_{c},4_{d},5_{e},6_{f}) = 6! \left[C_{\phi^6}\right]_{abcdef}\\
\mathcal O_{D^2\phi^4}  & F_{D^2\phi^4}(1_a,2_b,3_c,4_d) =-2\left(\left[\widehat C_{D^2\phi^4}\right]_{abcd} s_{12}  + \left[\widehat C_{D^2\phi^4}\right]_{acbd} s_{13}\right)\\
\mathcal{O}_{\phi^2 F^2} &
F_{\phi^2 F^2_{\alpha\beta}}(1_a,2_b,3^-_{A_\alpha},4^-_{B_\beta}) = -2 \mathcal S_{\alpha\beta} \left[C_{\phi^2 F^2}\right]^{A_\alpha B_\beta}_{a b}\agl{3}{4}^2\\
\mathcal{O}_{\phi^2\tildee F^2} & F_{\phi^2 \tildee F^2_{\alpha\beta}}(1_a,2_b,3^-_{A_\alpha},4^-_{B_\beta}) = -2i \mathcal S_{\alpha\beta} \left[C_{\phi^2 \tildee F^2}\right]^{A_\alpha B_\beta}_{a b}\agl{3}{4}^2\\
 \mathcal{O}_{F^3} & F_{F^3_\alpha}(1^-_{A_\alpha},2^-_{B_\alpha},3^-_{C_\alpha}) =-3i\sqrt{2} \left[C_{F^3}\right]^{A_\alpha B_\alpha C_\alpha} \agl{1}{2}\agl{2}{3}\agl{3}{1}\\
  \mathcal{O}_{\tildee F^3} & F_{\tildee F^3_\alpha}(1^-_{A_\alpha},2^-_{B_\alpha},3^-_{C_\alpha}) =3\sqrt{2} \left[C_{\tildee F^3}\right]^{A_\alpha B_\alpha C_\alpha} \agl{1}{2}\agl{2}{3}\agl{3}{1}\\
\bottomrule
\end{array}
\end{align*}
\caption{Dimension-six operators of the general EFT and their form factors considering negative helicities for the gauge bosons. The symmetry factors $\mathcal{S}_{\alpha\beta}$ are defined in App.~\ref{sec:symm_factors}.}
\label{tab:dim6FFs}
\end{table}

Regarding the operator $[\mathcal O_{D^2\phi^4}]_{abcd} = (D_\mu \phi)_a (D^\mu \phi)_b \phi_c \phi_d$, some comments are in order. 
First, we define the following combination of WCs which is present in the corresponding form factor
\begin{equation}\label{eq:hatCD2phi4}
\left[\widehat C_{D^2\phi^4}\right]_{abcd}=\left[C_{D^2\phi^4}\right]_{abcd}+\left[C_{D^2\phi^4}\right]_{cdab}-\left[C_{D^2\phi^4}\right]_{adbc}-\left[C_{D^2\phi^4}\right]_{bcad}\,.
\end{equation}
This structure is proportional to the Riemann tensor derived from a generalized scalar metric defined on the scalar manifold,\footnote{More precisely, for a metric given by $g_{ab}(\phi) = \delta_{ab} + 2[C_{D^2\phi^4}]_{abcd}\phi^c \phi^d$, the Riemann tensor $R_{abcd}$ is proportional to $[\widehat C_{D^2\phi^4}]_{acbd} = [C_{D^2\phi^4}]_{acbd}+[C_{D^2\phi^4}]_{bdac}-[C_{D^2\phi^4}]_{adbc}-[C_{D^2\phi^4}]_{bcad}$.} and hence has the following symmetry properties
\begin{gather}
    \left[\widehat C_{D^2\phi^4}\right]_{abcd} = \left[\widehat C_{D^2\phi^4}\right]_{badc} = -\left[\widehat C_{D^2\phi^4}\right]_{adcb} = -\left[\widehat C_{D^2\phi^4}\right]_{cbad}\,,\\
    \left[\widehat C_{D^2\phi^4}\right]_{abcd} + \left[\widehat C_{D^2\phi^4}\right]_{acdb} + \left[\widehat C_{D^2\phi^4}\right]_{adbc} = 0\,.
\end{gather}
Second, using integration by parts one finds
\begin{equation}
    \left[\mathcal O_{D^2\phi^4}\right]_{abcd} + \left[\mathcal O_{D^2\phi^4}\right]_{adbc} +
    \left[\mathcal O_{D^2\phi^4}\right]_{acdb}  = -\left[\mathcal R_{D^2\phi^4}\right]_{abcd}\,,\label{eq:D2phi4symmetric}
\end{equation}
and
\begin{align}
    \left[\mathcal O_{D^2\phi^4}\right]_{abcd} - \left[\mathcal O_{D^2\phi^4}\right]_{cdab} &= -\frac{1}{2}\left(
    \left[\mathcal R_{D^2\phi^4}\right]_{abcd} + \left[\mathcal R_{D^2\phi^4}\right]_{bacd}\right.\nonumber\\
    &\left.\qquad\,\,-\left[\mathcal R_{D^2\phi^4}\right]_{cabd}-\left[\mathcal R_{D^2\phi^4}\right]_{dabc}\,
    \right)\,,
\end{align}
where we defined
\begin{equation}
    \left[\mathcal R_{D^2\phi^4}\right]_{abcd} = (\Box \phi)_a \phi_b \phi_c \phi_d \,.
\end{equation}
In particular, Eq.~\eqref{eq:D2phi4symmetric} implies
\begin{equation}
      \left[\mathcal O_{D^2\phi^4}\right]_{(abcd)} = -\frac{1}{3}\left[\mathcal R_{D^2\phi^4}\right]_{(abcd)}\,.
\end{equation}
From the equation of motion
\begin{equation}
    (\Box \phi)_a = -t_a -m^2_{ab}\,\phi_b - \frac{1}{2!}h_{abc}\,\phi_b\phi_c - \frac{1}{3!}\lambda_{abcd}\,\phi_b\phi_c\phi_d\,,
\end{equation}
it follows that $\mathcal R_{D^2\phi^4}$ is a redundant operator that can be written as a linear combination of $\mathcal O_{\phi^6}$, $\mathcal O_{\phi^5}$, $\mathcal O_{\phi^4}$, and $\mathcal O_{\phi^3}$.
Hence, if $\mathcal R_{D^2\phi^4}$ is generated at the loop level, it affects the running of $C_{\phi^6}$, $C_{\phi^5}$, $\lambda$, and $h$.
In particular, if we decompose $[C_{D^2\phi^4}]_{abcd}$ as
\begin{equation}
    \left[C_{D^2\phi^4}\right]_{abcd} = \frac{1}{6}\left(
    2\left[\widehat C_{D^2\phi^4}\right]_{abcd}-\left[\widehat C_{D^2\phi^4}\right]_{acbd}
    \right)+\left[\tildee C_{D^2\phi^4}\right]_{abcd}+\left[\overline C_{D^2\phi^4}\right]_{abcd}\,,
    \label{eq:decomposition}
\end{equation}
with
\begin{align}
    \left[\tildee C_{D^2\phi^4}\right]_{abcd} &= \left[ C_{D^2\phi^4}\right]_{(abcd)}\,,\\
    \left[\overline C_{D^2\phi^4}\right]_{abcd} &= \frac{1}{2}\left(
    \left[C_{D^2\phi^4}\right]_{abcd} - \left[C_{D^2\phi^4}\right]_{cdab}
    \right)\,,
\end{align}
one finds
\begin{align}
    \left[\dot C_{\phi^6}\right]_{abcdef} &= \dotsc + \frac{1}{6!}\frac{1}{3!}\sum_{\sigma(\{a,b,c,d,e,f\})}\lambda_{abcg}\left(
    \frac{1}{3}\left[\dot{\tildee C}_{D^2\phi^4}\right]_{gdef}+\left[\dot{\overline C}_{D^2\phi^4}\right]_{gdef}
    \right)\,,\label{eq:dotCphi6<-Ctildeandbar}\\
    \left[\dot C_{\phi^5}\right]_{abcde} &= \dotsc + \frac{1}{5!}\frac{1}{2!}\sum_{\sigma(\{a,b,c,d,e\})}h_{abf}\left(
    \frac{1}{3}\left[\dot{\tildee C}_{D^2\phi^4}\right]_{fcde}+\left[\dot{\overline C}_{D^2\phi^4}\right]_{fcde}
    \right)\,,\label{eq:dotCphi5<-Ctildeandbar}\\
    \dot \lambda_{abcd} &= \dotsc - \sum_{\sigma(\{a,b,c,d\})}m^2_{ae}\left(
    \frac{1}{3}\left[\dot{\tildee C}_{D^2\phi^4}\right]_{ebcd}+\left[\dot{\overline C}_{D^2\phi^4}\right]_{ebcd}
    \right)\,,\label{eq:dotlambda<-Ctildeandbar}\\
    \dot h_{abc} &= \dotsc -  \sum_{\sigma(\{a,b,c\})} t_d \left(
    \frac{1}{3}\left[\dot{\tildee C}_{D^2\phi^4}\right]_{dabc}+\left[\dot{\overline C}_{D^2\phi^4}\right]_{dabc}
    \right)\,.
    \label{eq:dotCphi3<-Ctildeandbar}
\end{align}
The reason for presenting the above equations is that we have computed the running of $\mathcal{O}_{D^2\phi^4}$ into operators consisting solely of scalar fields, denoted as $\mathcal{O}_{\phi^n}$ with $n=1,\dotsc, 6$, using the geometric approach. In principle, if one were to employ on-shell techniques, as described in the subsequent section, it would not be necessary to account for redundant operators or utilize equations of motion. However, the on-shell approach becomes exceedingly cumbersome when renormalizing operators with multiple field insertions. This is particularly the case for purely scalar operators, where helicity selection rules that impose additional constraints are not available. Conversely, the geometric approach interprets scalars as coordinates on a field-space manifold, allowing one-loop divergences to be expressed in terms of geometric invariants. This significantly simplifies the computation of RGEs for operators involving only scalar fields, and for this reason, we adopt this approach when computing the ${\phi^n}\leftarrow {D^2\phi^4}$ mixing. Finally, we have validated the results obtained via the geometric approach through an independent calculation using the diagrammatic method, finding full agreement. 

In the subsequent section, we provide a review of the on-shell approach to computing RGEs, as all other results in this work were derived within this framework. 

\section{Review of the method}
\label{sec:method}

The on-shell method to compute RGEs is rooted in the non-perturbative relation~\cite{Caron-Huot:2016cwu}
\begin{equation}
    e^{-i\pi D} F_i^* = S F_i^*\,,
    \label{eq:D_and_S}
\end{equation}
which involves the $S$-matrix $S=\mathbb 1 + i \mathcal M$, the dilatation operator $D=\sum_k p_k \cdot \partial/\partial p_k$ as well as form factors associated to local and gauge-invariant operators
\begin{equation}
    F_i(\vec n;q) = \mel{\vec n}{\mathcal O_i(q)}{0}\,,
    \label{eq:FF}
\end{equation}
where $\bra{\vec n}$ is an outgoing on-shell state.
Eq.~\eqref{eq:D_and_S} is a direct consequence of unitarity, the CPT theorem as well as analyticity of the form factors~\cite{EliasMiro:2020tdv}.

Moreover, form factors in dimensional regularization satisfy the Callan-Symanzik equation
\begin{equation}
    \left(\delta_{ij}\mu\frac{\partial }{\partial \mu} + \frac{\partial \beta_i}{\partial C_j} - \delta_{ij}\gamma_{i,\text{IR}}+\delta_{ij}\beta_g \frac{\partial}{\partial g}\right) F_i = 0\,,
    \label{eq:CSeq}
\end{equation}
where $g$ are the couplings of the renormalizable Lagrangian, and 
\begin{equation}
    \beta_i(\{C_k\}) =\frac{1}{16\pi^2} \dot C_i= \mu \dv[]{C_i}{\mu} = \sum_{n>0}\frac{1}{n!}\gamma_{i\leftarrow j_1,\dotsc,j_n} C_{j_1}\dotsc C_{j_n}\,,
\end{equation}
denotes the $\beta$-functions of the WCs of the EFT Lagrangian $\mathcal L_{\text{EFT}} = \sum_i C_i \mathcal O_i/\Lambda^{[\mathcal O_i]-4}$.
The infrared anomalous dimension $\gamma_{i,\text{IR}}$ results from soft and collinear particle emissions~\cite{Sterman:2002qn,Becher:2009cu,Chiu:2009mg}.
While the soft divergences are directly taken into account by Stokes integration, which is presented below, the collinear anomalous dimensions, $\gamma_{c}$, depend only on the external fields associated with the operator $\mathcal{O}_i$.
Their expressions for scalar and vector fields are reported in App.~\ref{sec:coll_ADM}.

In the massless limit, Eqs.~\eqref{eq:D_and_S} and \eqref{eq:CSeq} can be combined and expanded at the one-loop order to give the master formulae~\cite{Bresciani:2023jsu}
\begin{align}
    \left(\gamma_{i\leftarrow j}-\delta_{ij}\gamma_{i,\text{IR}}\right) F_i|_* = -\frac{1}{\pi} (\mathcal M F_j)|_*\,,
    \qquad
    \gamma_{i\leftarrow j,k} F_i|_* = -\frac{1}{\pi}\left.\frac{\partial }{\partial c_k} \right|_* (\mathcal M F_j)\,,
    \label{eq:masterformulae}
\end{align}
for a single and double operator insertion, respectively.
The symbol ``$*$'' denotes the Gaussian fixed point, where all WCs are vanishing.
The RHS of these formulae involve the convolution between tree-level amplitudes and form factors, which correspond to two-particle unitarity cuts
\begin{equation}
    (\mathcal M F_j)(1,\dots,n) = 
   \sum_{k=2}^n \sum_{\{x,y\}}\int \text{dLIPS}_2
   \sum_{h_1,h_2} F_j(x^{h_1},y^{h_2},k + 1,\dots,n) \mathcal M(1,\dots,k;x^{h_1},y^{h_2})\, ,
\end{equation}
where $\mathcal M(\vec n;\vec m) = \mel{\vec n}{\mathcal M}{\vec m}$, and $\text{dLIPS}_2$ is the two-body Lorentz invariant phase-space measure. Although the above master formulae are valid only in the massless limit, leading mass effects can still be taken into account by following the procedure outlined in \cite{Bresciani:2023jsu}.

The computation of the phase-space integrals can be performed using Stokes parametrization \cite{Mastrolia:2009dr,Bresciani:2024shu,Jiang:2020mhe}, which has proven to be computationally efficient, especially when dealing with infrared divergences. The virtual spinors are parameterized in terms of the external ones as
\begin{equation}
    \begin{pmatrix}
 \lambda_x \\
 \lambda_y
    \end{pmatrix}
     =  
   \frac{1}{\sqrt{1+ z \bar{z}}}\left(
\begin{array}{cc}
 1 & \bar{z} \\
 -z & 1 \\
\end{array}
\right)
\begin{pmatrix}
 \lambda_{a} \\
 \lambda_{b}
    \end{pmatrix}
    \,, \qquad 
    \begin{pmatrix}
 \tildee\lambda_x \\
 \tildee\lambda_y
    \end{pmatrix}
     =  
   \frac{1}{\sqrt{1+ z \bar{z}}}\left(
\begin{array}{cc}
 1 & z \\
 -\bar z & 1 \\
\end{array}
\right)
\begin{pmatrix}
 \tildee\lambda_{a} \\
 \tildee\lambda_{b}
    \end{pmatrix}\,,\label{eq:stokesparam}
\end{equation}
such that $p_x+p_y = p_a+p_b$, and, accordingly,
\begin{equation}
    \int \text{dLIPS}_2 = -\frac{1}{8 \pi} \oint \frac{\text{d}z}{2\pi i} \int  \frac{\text{d}\bar{z}}{(1 + z \bar{z})^2} \, .
\end{equation}

In order to handle helicity spinors and perform these integrals, we used the \texttt{S@M} package~\cite{Maitre:2007jq}.
Furthermore, we exploited the application based on a machine learning algorithm presented in~\cite{Cheung:2024svk} to simplify spinor-helicity expressions for five- and six-point amplitudes.

As an illustration of how to apply the master formulae in Eq.~\eqref{eq:masterformulae} and Stokes integration, we discuss in detail the self-mixing of the dimension-six operator $\mathcal{O}_{\phi^2F^2}$ defined in Tab.~\ref{tab:dim6}.

\subsection*{Example: \texorpdfstring{$\phi^2 F^2 \leftarrow \phi^2 F^2$}{phi2tophi2example}}

Following Eq.~\eqref{eq:masterformulae}, the master formula to obtain the anomalous dimension can be written in a diagrammatic way as follows
\begin{align}
&-\pi \left(  \gamma_{\phi^2F^2\leftarrow \phi^2F^2}-\gamma_{\phi^2F^2,\mathrm{IR}}  \right) 
\begin{tikzpicture}[baseline = (a)]
            \begin{feynman}[small]
            \node [blue, crossed dot, thick] (a);
            \vertex [below right=1.3cm of a] (b) {\(4_{B_\beta}^-\)};
            \vertex [above right=1.3cm  of a] (c) {\(3_{A_\alpha}^-\)} ;
            \vertex [below left=1.3cm of a] (d) {\(2_b\)} ;
            \vertex [above left=1.3cm of a] (e) {\(1_a\)};
            \diagram*{
                (a) -- [photon] (b),
                (a) -- [photon] (c),
                (a) -- [scalar] (d),
                (a) -- [scalar] (e),
            };
            \end{feynman}
            \end{tikzpicture}
             =\nonumber \\
		 &  \delta_{\alpha\beta}\sum_{h_x,h_y=\pm}
            \begin{tikzpicture}[baseline = (a)]
            \begin{feynman}[small]
            \node [blue, crossed dot, thick] (a);
            \vertex [below right=1.3cm of a] (b) {\(y_{D_\alpha}^{h_y}\)};
            \vertex [above right=1.3cm  of a] (c) {\(x_{C_\alpha}^{h_x}\)} ;
            \vertex [below left=1.3cm of a] (d) {\(2_b\)} ;
            \vertex [above left=1.3cm of a] (e) {\(1_a\)};
            \diagram*{
                (a) -- [photon] (b),
                (a) -- [photon] (c),
                (a) -- [scalar] (d),
                (a) -- [scalar] (e),
            };
            \end{feynman}
            \end{tikzpicture}
            \includegraphics[valign=c,height=30pt]{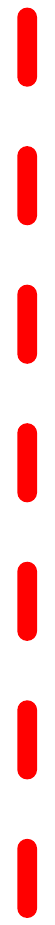}
            \begin{tikzpicture}[baseline = (a)]
                \begin{feynman}[small]
                \node [blob] (a);
                \vertex [below right=1.3cm of a] (b) {\(4_{B_\alpha}^{-}\)};
                \vertex [above right=1.3cm of a] (c) {\(3_{A_\alpha}^{-}\)} ;
                \vertex [below left=1.3cm of a] (d) {\(-y_{D_\alpha}^{-h_y}\)} ;
                \vertex [above left=1.3cm of a] (e) {\(-x_{C_\alpha}^{-h_x}\)};
                \diagram*{
                    (a) -- [photon] (b),
                    (a) -- [photon] (c),
                    (a) -- [photon] (d),
                    (a) -- [photon] (e),
                };
                \end{feynman}
                \end{tikzpicture}
                +
		\begin{tikzpicture}[baseline = (a)]
                \begin{feynman}[small]
                \node [blue, crossed dot, thick] (a);
                \vertex [below right=1.3cm of a] (b) {\(y_d\)};
                \vertex [above right=1.3cm  of a] (c) {\(x_c\)} ;
                \vertex [below left=1.3cm of a] (d) {\(4_{B_\beta}^-\)} ;
                \vertex [above left=1.3cm of a] (e) {\(3_{A_\alpha}^-\)};
                \diagram*{
                    (a) -- [scalar] (b),
                    (a) -- [scalar] (c),
                    (a) -- [photon] (d),
                    (a) -- [photon] (e),
                };
                \end{feynman}
                \end{tikzpicture}
                \includegraphics[valign=c,height=30pt]{images/dashed-line.pdf}
                                \begin{tikzpicture}[baseline = (a)]
                    \begin{feynman}[small]
                    \node [blob] (a);
                    \vertex [below right=1.3cm of a] (b) {\(2_b\)};
                    \vertex [above right=1.3cm of a] (c) {\(1_a\)} ;
                    \vertex [below left=1.3cm of a] (d) {\(-y_d\)} ;
                    \vertex [above left=1.3cm of a] (e) {\(-x_c\)};
                    \diagram*{
                        (a) -- [scalar] (b),
                        (a) -- [scalar] (c),
                        (a) -- [scalar] (d),
                        (a) -- [scalar] (e),
                    };
                    \end{feynman}
                    \end{tikzpicture}\nonumber \\
		 &+
                    \sum_{\sigma(\{a,b\}\times \{A_\alpha,B_\beta\})}\sum_{\gamma=1}^{N_G}
                    \sum_{h_y=\pm}
                \begin{tikzpicture}[baseline = (a)]
                    \begin{feynman}[small]
                    \node [blue, crossed dot, thick] (a);
                    \vertex [below right=1.3cm of a] (b) {\(y_{C_\gamma}^{h_y}\)};
                    \vertex [above right=1.3cm of a] (c) {\(x_c\)} ;
                    \vertex [below left=1.3cm of a] (d) {\(3_{A_\alpha}^-\)} ;
                    \vertex [above left=1.3cm of a] (e) {\(1_a\)};
                    \diagram*{
                        (a) -- [photon] (b),
                        (a) -- [scalar] (c),
                        (a) -- [photon] (d),
                        (a) -- [scalar] (e),
                    };
                    \end{feynman}
                    \end{tikzpicture}
                    \includegraphics[valign=c,height=30pt]{images/dashed-line.pdf}
                    \begin{tikzpicture}[baseline = (a)]
                        \begin{feynman}[small]
                        \node [blob] (a);
                        \vertex [below right=1.3cm of a] (b) {\(4_{B_\beta}^-\)};
                        \vertex [above right=1.3cm of a] (c) {\(2_b\)} ;
                        \vertex [below left=1.3cm of a] (d) {\(-y_{C_\gamma}^{-h_y}\)} ;
                        \vertex [above left=1.3cm of a] (e) {\(-x_c\)};
                        \diagram*{
                            (a) -- [photon] (b),
                            (a) -- [scalar] (c),
                            (a) -- [photon] (d),
                            (a) -- [scalar] (e),
                        };
                        \end{feynman}
                        \end{tikzpicture} 
                        \label{eq:diag-master-formula}
\end{align}
where blue crossed dots and blobs denote insertions of $\mathcal{O}_{\phi^2F^2}$ and amplitudes constructed from marginal interactions, respectively.
The $\delta_{\alpha\beta}$ in the first term ensures that the four-vector amplitude is nonvanishing only if the simple gauge group associated with each gauge boson is the same, while the sum $\sum_{\sigma(\{a,b\}\times \{A_\alpha,B_\beta\})} = \sum_{\sigma(\{a,b\})} \sum_{\sigma( \{A_\alpha,B_\beta\})} $ permutes the external scalars and gauge bosons.

\noindent The first term of Eq.~\eqref{eq:diag-master-formula} is given by
\begin{align}
    \mathcal{A}_1 = \sum_{h_x,h_y=\pm}
            \begin{tikzpicture}[baseline = (a)]
            \begin{feynman}[small]
            \node [blue, crossed dot, thick] (a);
            \vertex [below right=1.3cm of a] (b) {\(y_{D_\alpha}^{h_y}\)};
            \vertex [above right=1.3cm  of a] (c) {\(x_{C_\alpha}^{h_x}\)} ;
            \vertex [below left=1.3cm of a] (d) {\(2_b\)} ;
            \vertex [above left=1.3cm of a] (e) {\(1_a\)};
            \diagram*{
                (a) -- [photon] (b),
                (a) -- [photon] (c),
                (a) -- [scalar] (d),
                (a) -- [scalar] (e),
            };
            \end{feynman}
            \end{tikzpicture}
            \times 
            \begin{tikzpicture}[baseline = (a)]
                \begin{feynman}[small]
                \node [blob] (a);
                \vertex [below right=1.3cm of a] (b) {\(4_{B_\alpha}^{-}\)};
                \vertex [above right=1.3cm of a] (c) {\(3_{A_\alpha}^{-}\)} ;
                \vertex [below left=1.3cm of a] (d) {\(-y_{D_\alpha}^{-h_y}\)} ;
                \vertex [above left=1.3cm of a] (e) {\(-x_{C_\alpha}^{-h_x}\)};
                \diagram*{
                    (a) -- [photon] (b),
                    (a) -- [photon] (c),
                    (a) -- [photon] (d),
                    (a) -- [photon] (e),
                };
                \end{feynman}
                \end{tikzpicture}
                =
                \begin{tikzpicture}[baseline = (a)]
            \begin{feynman}[small]
            \node [blue, crossed dot, thick] (a);
            \vertex [below right=1.3cm of a] (b) {\(y_{D_\alpha}^{-}\)};
            \vertex [above right=1.3cm  of a] (c) {\(x_{C_\alpha}^{-}\)} ;
            \vertex [below left=1.3cm of a] (d) {\(2_b\)} ;
            \vertex [above left=1.3cm of a] (e) {\(1_a\)};
            \diagram*{
                (a) -- [photon] (b),
                (a) -- [photon] (c),
                (a) -- [scalar] (d),
                (a) -- [scalar] (e),
            };
            \end{feynman}
            \end{tikzpicture}
            \times 
            \begin{tikzpicture}[baseline = (a)]
                \begin{feynman}[small]
                \node [blob] (a);
                \vertex [below right=1.3cm of a] (b) {\(4_{B_\alpha}^{-}\)};
                \vertex [above right=1.3cm of a] (c) {\(3_{A_\alpha}^{-}\)} ;
                \vertex [below left=1.3cm of a] (d) {\(-y_{D_\alpha}^{+}\)} ;
                \vertex [above left=1.3cm of a] (e) {\(-x_{C_\alpha}^{+}\)};
                \diagram*{
                    (a) -- [photon] (b),
                    (a) -- [photon] (c),
                    (a) -- [photon] (d),
                    (a) -- [photon] (e),
                };
                \end{feynman}
                \end{tikzpicture}
\end{align}
where only one helicity configuration contributes, namely~$(h_x,h_y)=(-,-)$. For all other combinations the amplitude vanishes.
The form of the amplitude is given by the Parke-Taylor formula~\cite{Parke:1986gb}.\footnote{In order to relate amplitudes with incoming and outgoing states, we exploited crossing symmetry, with the following convention for spinors associated with flipped momenta: $\lambda_{-p} = i \lambda_p$ and $\tildee \lambda_{-p} = i \tildee \lambda_p$.} Together with the form factor in Tab.~\ref{tab:dim6FFs} the final expression reads
\begin{equation}
    \mathcal{A}_1 = 8 g_\alpha^2 \left[C_{\phi^2 F^2}\right]^{C_\alpha D_\alpha}_{ab}\frac{\agl{3}{4}^3 \agl{x}{y}}{\agl{3}{x}\agl{3}{y}\agl{4}{x}\agl{4}{y}}\left(F^{A_\alpha C_\alpha B_\alpha D_\alpha}\agl{3}{4}\agl{x}{y} - F^{A_\alpha B_\alpha C_\alpha D_\alpha}\agl{3}{x}\agl{4}{y}\right)\,,
\end{equation}
with $F^{A_\alpha B_\alpha C_\alpha D_\alpha} = f^{A_\alpha B_\alpha E_\alpha}f^{C_\alpha D_\alpha E_\alpha}$.
By parametrizing the internal spinors $(x,y)$ as linear combinations of the spinors $(3,4)$ as in Eq.~\eqref{eq:stokesparam} we find 
\begin{equation}
    \mathcal{A}_1(z,\bar z) = 
    8\frac{1+z\bar z}{z\bar z}{g^2_\alpha }{\left[C_{\phi^2F^2}\right]_{ab}^{C_\alpha D_\alpha}}\agl{3}{4}^2\left[z\bar z F^{A_\alpha B_\alpha C_\alpha D_\alpha }-(1+z\bar z)F^{A_\alpha C_\alpha B_\alpha D_\alpha }\right]\,.
\end{equation}

\noindent The second term of Eq.~\eqref{eq:diag-master-formula} is
\begin{align}
    \mathcal{A}_2 &= \begin{tikzpicture}[baseline = (a)]
                \begin{feynman}[small]
                \node [blue, crossed dot, thick] (a);
                \vertex [below right=1.3cm of a] (b) {\(y_d\)};
                \vertex [above right=1.3cm  of a] (c) {\(x_c\)} ;
                \vertex [below left=1.3cm of a] (d) {\(4_{B_\beta}^-\)} ;
                \vertex [above left=1.3cm of a] (e) {\(3_{A_\alpha}^-\)};
                \diagram*{
                    (a) -- [scalar] (b),
                    (a) -- [scalar] (c),
                    (a) -- [photon] (d),
                    (a) -- [photon] (e),
                };
                \end{feynman}
                \end{tikzpicture}
                \times
                                \begin{tikzpicture}[baseline = (a)]
                    \begin{feynman}[small]
                    \node [blob] (a);
                    \vertex [below right=1.3cm of a] (b) {\(2_b\)};
                    \vertex [above right=1.3cm of a] (c) {\(1_a\)} ;
                    \vertex [below left=1.3cm of a] (d) {\(-y_d\)} ;
                    \vertex [above left=1.3cm of a] (e) {\(-x_c\)};
                    \diagram*{
                        (a) -- [scalar] (b),
                        (a) -- [scalar] (c),
                        (a) -- [scalar] (d),
                        (a) -- [scalar] (e),
                    };
                    \end{feynman}
                    \end{tikzpicture}
                    \nonumber \\
    &= -2 \mathcal S_{\alpha\beta} \left[C_{\phi^2F^2}\right]^{A_\alpha B_\beta}_{cd}\agl{3}{4}^2 
    \nonumber \\ 
    & \quad \times 
    \left[
    \sum_{\gamma=1}^{N_G} g_\gamma^2 \left(
    \frac{s_{1y}-s_{1x}}{s_{12}}T^\gamma_{abcd}- \frac{s_{12}+s_{1y}}{s_{1x}}T^\gamma_{acbd}- \frac{s_{12}+s_{1x}}{s_{1y}}T^\gamma_{adbc}
    \right)
    -\lambda_{abcd} 
    \right]\,,
\end{align}
with $T^\alpha_{abcd} = \theta^{A_\alpha}_{ab}\theta^{A_\alpha}_{cd}$.
After parametrizing $(x,y)$ in terms of $(1,2)$ one finds
\begin{align}
    \mathcal{A}_2(z,\bar z) &= 2 \mathcal S_{\alpha\beta} \left[C_{\phi^2F^2}\right]^{A_\alpha B_\beta}_{cd}\agl{3}{4}^2
    \nonumber \\ 
    & \quad  \times
    \left[
    \sum_{\gamma=1}^{N_G} g_\gamma^2
    \left(
    \frac{-1+z\bar z}{1+z\bar z}T^\gamma_{abcd}
    +
    \frac{2+z\bar z}{z\bar z}T^\gamma_{acbd}
    +
    (1+2z\bar z)T^\gamma_{adbc}
    \right)
    +
    \lambda_{abcd}
    \right]\,.
\end{align}
Finally, the third term of Eq.~\eqref{eq:diag-master-formula} is given by
\begin{align}
    \mathcal{A}_3 &=
    \sum_{h_y=\pm}
                \begin{tikzpicture}[baseline = (a)]
                    \begin{feynman}[small]
                    \node [blue, crossed dot, thick] (a);
                    \vertex [below right=1.3cm of a] (b) {\(y_{C_\gamma}^{h_y}\)};
                    \vertex [above right=1.3cm of a] (c) {\(x_c\)} ;
                    \vertex [below left=1.3cm of a] (d) {\(3_{A_\alpha}^-\)} ;
                    \vertex [above left=1.3cm of a] (e) {\(1_a\)};
                    \diagram*{
                        (a) -- [photon] (b),
                        (a) -- [scalar] (c),
                        (a) -- [photon] (d),
                        (a) -- [scalar] (e),
                    };
                    \end{feynman}
                    \end{tikzpicture}
                    \times
                    \begin{tikzpicture}[baseline = (a)]
                        \begin{feynman}[small]
                        \node [blob] (a);
                        \vertex [below right=1.3cm of a] (b) {\(4_{B_\beta}^-\)};
                        \vertex [above right=1.3cm of a] (c) {\(2_b\)} ;
                        \vertex [below left=1.3cm of a] (d) {\(-y_{C_\gamma}^{-h_y}\)} ;
                        \vertex [above left=1.3cm of a] (e) {\(-x_c\)};
                        \diagram*{
                            (a) -- [photon] (b),
                            (a) -- [scalar] (c),
                            (a) -- [photon] (d),
                            (a) -- [scalar] (e),
                        };
                        \end{feynman}
                        \end{tikzpicture}
                        =
                        \begin{tikzpicture}[baseline = (a)]
                    \begin{feynman}[small]
                    \node [blue, crossed dot, thick] (a);
                    \vertex [below right=1.3cm of a] (b) {\(y_{C_\gamma}^{-}\)};
                    \vertex [above right=1.3cm of a] (c) {\(x_c\)} ;
                    \vertex [below left=1.3cm of a] (d) {\(3_{A_\alpha}^-\)} ;
                    \vertex [above left=1.3cm of a] (e) {\(1_a\)};
                    \diagram*{
                        (a) -- [photon] (b),
                        (a) -- [scalar] (c),
                        (a) -- [photon] (d),
                        (a) -- [scalar] (e),
                    };
                    \end{feynman}
                    \end{tikzpicture}
                    \times
                    \begin{tikzpicture}[baseline = (a)]
                        \begin{feynman}[small]
                        \node [blob] (a);
                        \vertex [below right=1.3cm of a] (b) {\(4_{B_\beta}^-\)};
                        \vertex [above right=1.3cm of a] (c) {\(2_b\)} ;
                        \vertex [below left=1.3cm of a] (d) {\(-y_{C_\gamma}^{+}\)} ;
                        \vertex [above left=1.3cm of a] (e) {\(-x_c\)};
                        \diagram*{
                            (a) -- [photon] (b),
                            (a) -- [scalar] (c),
                            (a) -- [photon] (d),
                            (a) -- [scalar] (e),
                        };
                        \end{feynman}
                        \end{tikzpicture}
                        \nonumber\\
                        &=
    4 \mathcal S_{\gamma\alpha} \left[C_{\phi^2F^2}\right]^{A_\alpha C_\gamma}_{ac} \frac{\agl{3}{y}^2 \agl{4}{x}\sqr{y}{x}}{\agl{2}{x}\agl{2}{y}\sqr{4}{2}\sqr{x}{2}}  g_\beta
    \bigg[
    -(1-\delta_{\beta\gamma})g_\gamma s_{2x}Y^{B_\beta C_\gamma}_{bc}
    \nonumber \\
    &\quad
    +
    \delta_{\beta\gamma}g_\beta \left(
    s_{2y}Y^{B_\beta C_\beta}_{bc}
    -
    s_{24} Y^{C_\beta B_\beta}_{bc}
    \right)
    \bigg]\,,
\end{align}
with $Y^{A_\alpha B_\beta}_{ab} = \theta^{A_\alpha}_{ac}\theta^{B_\beta}_{cb}$.
Again, only $h_y = -1$ gives a non-vanishing contribution.
The parametrization of $(x,y)$ in terms of $(2,4)$ yields the following rational function:
\begin{align}
    \mathcal{A}_3 (z,\bar z)
    &=
    4 \mathcal S_{\gamma\alpha} \left[C_{\phi^2F^2}\right]^{A_\alpha C_\gamma}_{ac}
    \frac{(z\agl{2}{3}+\agl{3}{4})^2}{z\bar z (1+z\bar z)}g_\beta 
    \bigg[
    (1-\delta_{\beta\gamma})g_\gamma z\bar z  Y^{B_\beta C_\gamma}_{bc}
    \nonumber \\
    &\quad
    +
    \delta_{\beta\gamma }g_\beta
    \left(
    -
    Y^{B_\beta C_\beta}_{bc}
    +
    (1+z\bar z)Y^{C_\beta B_\beta}_{bc}
    \right)
    \bigg]\,.
\end{align}
The evaluation of the RHS of Eq.~\eqref{eq:diag-master-formula} is achieved by applying the Stokes integration procedure to the function
\begin{equation}
    \mathcal{A}(z,\bar z) = \frac{1}{2}\delta_{\alpha\beta} \mathcal{A}_1 (z,\bar z) + \frac{1}{2} \mathcal{A}_2 (z,\bar z) + \sum_{\sigma(\{a,b\}\times \{A_\alpha,B_\beta\})}\sum_{\gamma=1}^{N_G} \mathcal{A}_3 (z,\bar z)\,,
\end{equation}
where the factors of $1/2$ have been included to account for the integration of indistinguishable particles.
The integral over $\bar z$ generates a rational contribution given by
\begin{align}
    G(z,\bar z) &= \text{Rational}\left[ \int \frac{\text{d}z}{(1+z\bar z)^2}\mathcal{A}(z,\bar z)\right] \nonumber \\
    &= \mathcal S_{\alpha\beta} \left[C_{\phi^2F^2}\right]^{A_\alpha B_\beta}_{cd}\agl{3}{4}^2 \frac{1}{z(1+z\bar z)^2}\bigg[
    \sum_{\gamma=1}^{N_G} g_\gamma^2 [-z\bar z T^\gamma_{abcd}+(1+z\bar z ) (T^\gamma_{acbd}+T^\gamma_{adbc})]
    \nonumber \\ &\quad
    -(1+z\bar z)\lambda_{abcd}
    \bigg]
    + 
    \sum_{\sigma(\{a,b\}\times \{A_\alpha,B_\beta\})}\sum_{\gamma=1}^{N_G} 
    2 \mathcal S_{\gamma\alpha} \left[C_{\phi^2F^2}\right]^{A_\alpha C_\gamma}_{ac}
    \frac{(z\agl{2}{3}+\agl{3}{4})^2}{z (1+z\bar z)^2 }g_\beta \nonumber \\
    &\quad \times \left [
    -(1-\delta_{\beta\gamma})g_\gamma Y^{B_\beta C_\gamma}_{bc} - \delta_{\beta\gamma}g_\beta 
    \left(
    (3+2z\bar z)Y^{B_\beta C_\beta}_{bc}+2(1+z\bar z)Y^{C_\beta B_\beta}_{bc}
    \right)
    \right]\,.
\end{align}
Lastly, the phase space integration is given by the residue of the function $G(z,\bar z)$ at $z_0=0$:
\begin{align}
\int \text{dLIPS}_2 \, \mathcal{A} &= -\frac{1}{8\pi} \Res_{(z_0,\bar z_0)=(0,0)} G(z,\bar z) \nonumber \\
&= 
\frac{1}{8\pi} \mathcal S_{\alpha\beta} \left[C_{\phi^2F^2}\right]^{A_\alpha B_\beta}_{cd}\agl{3}{4}^2
\bigg[
- \sum_{\gamma=1}^{N_G} g_\gamma^2  \left(T^\gamma_{acbd}+T^\gamma_{adbc}\right)+\lambda_{abcd}
\bigg]\nonumber \\
&\quad +
\frac{1}{4\pi} 
\sum_{\sigma(\{a,b\}\times \{A_\alpha,B_\beta\})}\sum_{\gamma=1}^{N_G} 
\mathcal S_{\gamma\alpha} \left[C_{\phi^2F^2}\right]^{A_\alpha C_\gamma}_{ac}
   g_\beta
   \bigg[
   (1-\delta_{\beta\gamma})g_\gamma Y^{B_\beta C_\gamma}_{bc}
   \nonumber \\ &\quad +
    \delta_{\beta\gamma}g_\beta  \left(
    3 Y^{B_\beta C_\beta}_{bc}- 2Y^{C_\beta B_\beta}_{bc}
    \right)
   \bigg]\,.
\end{align}
The resulting anomalous dimension is reported in Eq.~\eqref{eq:phi2F2_phi2F2}.

\section{Results}
\label{sec:results}
\begin{figure}[tbp]
\centering
\input{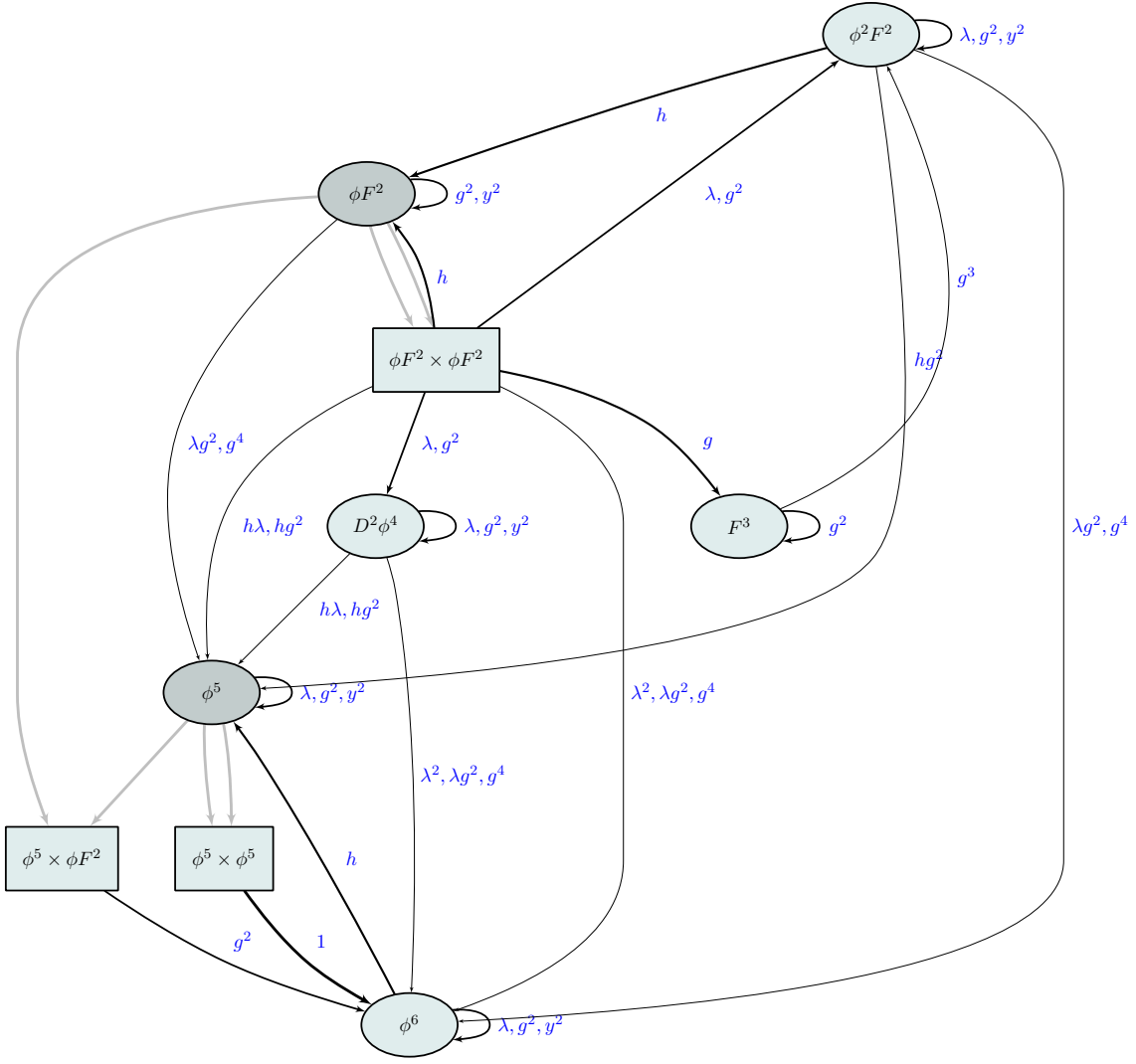}
\caption{Dimension-five and dimension-six operator mixing pattern.\label{fig:graph}}
\end{figure}

In this section, we report the complete set of RGEs for the general bosonic EFT given in Eq.~\eqref{eq:fullEFT}. The contributions are ordered according to their mass dimension, starting from higher-dimensional operators. The operator definitions can be found in Tabs.~\ref{tab:dim5} and \ref{tab:dim6}. A graphical representation of the mixing pattern involving dimension-five and dimension-six operators is depicted in Fig.~\ref{fig:graph}. The figure can also be used to navigate to individual results of the full RGE. 
The collinear ADMs $\gamma_c$ for scalar and vector fields are defined in App.~\ref{sec:coll_ADM}. Lastly, the symmetry factors $\mathcal{S}_{\alpha\beta}$ are defined in App.~\ref{sec:symm_factors}.

\subsection{Running of dimension-6 operators}

\subsubsection{\texorpdfstring{$\phi^6$}{phi6} class}

In addition to the results presented below, one needs to supplement the RGEs of the $\phi^6$-class WCs by the expressions given in Eq.~\eqref{eq:dotCphi6<-Ctildeandbar}.
The explicit expressions for $\dot{\tildee C}_{D^2\phi^4}$ and $\dot{\overline C}_{D^2\phi^4}$ can be obtained using Eq.~\eqref{eq:decomposition} and the RGEs in Sec.~\ref{sec:D2phi4RGE}.

\paragraph{\texorpdfstring{$\phi^6 \leftarrow \phi^6$}{phi6fromphi6}:}
\label{par:phi6<-phi6}

\begin{align}
    \left[\dot C_{\phi^6}\right]_{abcdef} &= \frac{1}{2!4!}\sum_{\sigma(\{a,b,c,d,e,f\})}\bigg[\lambda_{efgh}+2\sum_\alpha g_\alpha^2 T^\alpha_{eghf}\bigg]\left[ C_{\phi^6}\right]_{abcdgh}\nonumber \\
    & +\frac{1}{5!}\sum_{\sigma(\{a,b,c,d,e,f\})} \gamma_{c,s}^{ag}\left[ C_{\phi^6}\right]_{gbcdef}
    \label{eq:phi6<-phi6}\,.
\end{align}

\paragraph{\texorpdfstring{$\phi^6 \leftarrow D^2\phi^4$:}{phi6fromD2phi4}}
\label{par:phi6<-D2phi4}

\begin{align}
    \left[\dot C_{\phi^6}\right]_{abcdef} &= \frac{1}{2}\frac{1}{6!}\sum_{\sigma(\{a,b,c,d,e,f\})}\lambda_{ghab}\bigg[\lambda_{ihcd}\left[ C_{D^2\phi^4}\right]_{gief}\nonumber\\
    &+\frac{1}{3}\lambda_{icde}\left(\left[C_{D^2\phi^4}\right]_{ihgf}+\left[C_{D^2\phi^4}\right]_{ighf}-\left[C_{D^2\phi^4}\right]_{hgif}\right)\bigg]\nonumber\\
    &- \frac{6}{6!} \sum_{\sigma(\{a,b,c,d,e,f\})}\sum_{\alpha=1}^{N_G}\sum_{\beta=1}^{N_G} g_\alpha^2 g_\beta^2 T^\alpha_{fhei} T^\beta_{dicg}\left[C_{D^2\phi^4}\right]_{abgh}\nonumber\\
    &-\frac{3}{6!} \sum_{\sigma(\{a,b,c,d,e,f\})}\sum_{\alpha=1}^{N_G} g_\alpha^2 \lambda_{abhi} T^\alpha_{cidg} \left[C_{D^2\phi^4}\right]_{ghef}\,.
    \label{eq:phi6<-D2phi4}
\end{align}

\paragraph{\texorpdfstring{$\phi^6 \leftarrow \phi^5\times\phi^5$:}{phi6fromphi5xphi5}}
\label{par:phi6<-phi5xphi5}

\begin{equation}
    \left[\dot C_{\phi^6}\right]_{abcdef} = -\frac{1}{2}\frac{(5!)^2}{6!(3!)^2}\sum_{\sigma(\{a,b,c,d,e,f\})}\left[ C_{\phi^5}\right]_{abcgh}\left[ C_{\phi^5}\right]_{ghdef}\,.
\end{equation}

\paragraph{\texorpdfstring{$\phi^6 \leftarrow \phi F^2\times\phi^5$:}{phi6fromphiF2xphi5}}
\label{par:phi6<-phiF2xphi5}

\begin{equation}
    \left[\dot C_{\phi^6}\right]_{abcdef} = -\frac{5!}{6!}\sum_{\sigma(\{a,b,c,d,e,f\})} \sum_{\alpha=1}^{N_G} \sum_{\beta=1}^{\alpha} g_\alpha g_\beta \left[C_{\phi^5}\right]_{abcdg} Y_{eg}^{A_\alpha B_\beta} \left[C_{\phi F^2}\right]_{f}^{A_\alpha B_\beta}\,.
\end{equation}

\paragraph{\texorpdfstring{$\phi^6 \leftarrow \phi^2 F^2$}{phifromphi2F2}:}
\label{par:phi6<-phi2F2}

\begin{align}
    \left[\dot C_{\phi^6}\right]_{abcdef} &=
    \frac{2}{6!}\sum_{\sigma(\{a,b,c,d,e,f\})} \Bigg[\sum_{\alpha=1}^{N_G} \sum_{\beta=1}^{\alpha} g_\alpha g_\beta \lambda_{abcg} Y_{dg}^{A_\alpha B_\beta} \left[C_{\phi^2 F^2}\right]_{ef}^{A_\alpha B_\beta}\nonumber\\
    &- 6 \sum_{\alpha=1}^{N_G}\sum_{\beta=1}^{\alpha}\sum_{\gamma=1}^{N_G} g_\alpha g_\beta g_\gamma^2 \,Y_{ce}^{A_\alpha C_\gamma} Y_{df}^{B_\beta C_\gamma} \left[C_{\phi^2 F^2}\right]_{ab}^{A_\alpha B_\beta}\Bigg]\,.
    \label{eq:phi6<-phi2F2}
\end{align}

\paragraph{\texorpdfstring{$\phi^6 \leftarrow \phi F^2 \times \phi F^2$}{phi6fromphiF2xphiF2}:}
\label{par:phi6<-phiF2xphiF2}

\begin{align}
    \left[\dot C_{\phi^6}\right]_{abcdef} &= \frac{1}{6!} \sum_{\sigma(\{a,b,c,d,e,f\})} \Bigg[-\frac{1}{18}\lambda_{abcg}\lambda_{defh} \sum_{\alpha=1}^{N_G}\sum_{\beta=1}^\alpha \mathcal{S}_{\alpha\beta} \left[C_{\phi F^2}\right]_{g}^{A_\alpha B_\beta} \left[C_{\phi F^2}\right]_{h}^{A_\alpha B_\beta}\nonumber\\
    &-24 \sum_{\alpha=1}^{N_G}\sum_{\beta=1}^{\alpha}\sum_{\gamma=1}^{\beta}\sum_{\delta=1}^{N_G} g_\alpha g_\gamma g_\delta^2 \left[C_{\phi F^2}\right]_{a}^{A_\alpha B_\beta} \left[C_{\phi F^2}\right]_{b}^{B_\beta C_\gamma} Y_{cd}^{C_\gamma D_\delta} Y_{ef}^{D_\delta A_\alpha}\nonumber\\
    &-24 \sum_{\beta=1}^{N_G}\sum_{\alpha=1}^{\beta}\sum_{\gamma=1}^{\alpha}\sum_{\delta=1}^{N_G} g_\beta g_\gamma g_\delta^2 \left[C_{\phi F^2}\right]_{a}^{A_\alpha B_\beta} \left[C_{\phi F^2}\right]_{b}^{A_\alpha C_\gamma} Y_{cd}^{C_\gamma D_\delta} Y_{ef}^{D_\delta B_\beta}\nonumber\\
    &-24 \sum_{\alpha=1}^{N_G}\sum_{\beta=1}^{\alpha}\sum_{\gamma=1}^{N_G}\sum_{\delta=1}^{\gamma} g_\alpha g_\beta g_\gamma g_\delta \left[C_{\phi F^2}\right]_{a}^{A_\alpha B_\beta} Y_{bc}^{B_\beta C_\gamma} \left[C_{\phi F^2}\right]_{d}^{C_\gamma D_\delta}  Y_{ef}^{A_\alpha D_\delta}\Bigg]\nonumber \\
    &+ \left(\left[C_{\phi F^2}\right]\to \left[C_{\phi \tildee F^2}\right]\right)\,.    
\end{align}

\subsubsection{\texorpdfstring{$D^2\phi^4$}{D2phi4} class}
\label{sec:D2phi4RGE}
\paragraph{\texorpdfstring{$D^2\phi^4\leftarrow D^2\phi^4$}{D2phi4fromD2phi4}:}
\label{par:D2phi4<-D2phi4}
~
\newline The running of the $\widehat C_{D^2\phi^4}$ combination defined in Eq.~\eqref{eq:hatCD2phi4} is given by 
\begin{align}
\left[\dot{\widehat C}_{D^2\phi^4}\right]_{abcd} &=
-\frac{1}{6}\sum_{\alpha=1}^{N_G}g_\alpha^2 
\bigg\{
\bigg[
\left(T^\alpha_{cdef}+24T^\alpha_{cedf}
\right)\left[\widehat C_{D^2\phi^4}\right]_{aebf} \nonumber \\
&+ 6\left(T^\alpha_{cedf}+T^\alpha_{cfde}\right)\left[\widehat C_{D^2\phi^4}\right]_{abef} - (a\leftrightarrow c) - (b\leftrightarrow d) + 
\binom{a\leftrightarrow c}{b\leftrightarrow d}
\bigg] \nonumber \\
&+ \bigg[
2\left(T^\alpha_{bdef}+24T^\alpha_{bedf}
\right)\left[\widehat C_{D^2\phi^4}\right]_{aecf} + 
\binom{a\leftrightarrow c}{b\leftrightarrow d}
\bigg]
\bigg\}
\label{eq:D2phi4<-D2phi4}\\
&+ \bigg[
\lambda_{cdef}
\left[\widehat C_{D^2\phi^4}\right]_{abef}
- (a\leftrightarrow c) - (b\leftrightarrow d) + 
\binom{a\leftrightarrow c}{b\leftrightarrow d}
\bigg] \nonumber \\
&+\gamma_{c,s}^{ae}\left[\widehat C_{D^2\phi^4}\right]_{ebcd}+\gamma_{c,s}^{be}\left[\widehat C_{D^2\phi^4}\right]_{aecd}+\gamma_{c,s}^{ce}\left[\widehat C_{D^2\phi^4}\right]_{abed}+\gamma_{c,s}^{de}\left[\widehat C_{D^2\phi^4}\right]_{abce}\,.\nonumber
\end{align}
The running of the full operator (including off-shell contributions) is given by
\begin{align}
\left[\dot{C}_{D^2\phi^4}\right]_{abcd} &=
\frac{1}{6}\sum_{\alpha} g_{\alpha}^2 \Big[
-6\Bigl( T^\alpha_{cedf} + T^\alpha_{cfde} \Bigr)
\left[C_{D^2\phi^4}\right]_{abef} -6\Bigl( T^\alpha_{aebf} + T^\alpha_{afbe} \Bigr)
\left[C_{D^2\phi^4}\right]_{efcd}\nonumber \\
&-\Bigl( T^\alpha_{bdef} + 15\,T^\alpha_{bedf} - 9\,T^\alpha_{bfde} \Bigr)
\left[C_{D^2\phi^4}\right]_{aecf} \nonumber\\
&-\Bigl( T^\alpha_{bcef} + 15\,T^\alpha_{becf} - 9\,T^\alpha_{bfce} \Bigr)
\left[C_{D^2\phi^4}\right]_{aedf} \nonumber\\
&+\Bigl( T^\alpha_{bdef} + 9\,T^\alpha_{bedf} - 15\,T^\alpha_{bfde} \Bigr)
\left[C_{D^2\phi^4}\right]_{afce} \nonumber\\
&+\Bigl( T^\alpha_{bcef} + 9\,T^\alpha_{becf} - 15\,T^\alpha_{bfce} \Bigr)
\left[C_{D^2\phi^4}\right]_{afde} \nonumber\\
&-\Bigl( T^\alpha_{adef} + 15\,T^\alpha_{aedf} - 9\,T^\alpha_{afde} \Bigr)
\left[C_{D^2\phi^4}\right]_{becf} \label{eq:D2phi4<-D2phi4_offshell}\\
&-\Bigl( T^\alpha_{acef} + 15\,T^\alpha_{aecf} - 9\,T^\alpha_{afce} \Bigr)
\left[C_{D^2\phi^4}\right]_{bedf}\nonumber \\
&+\Bigl( T^\alpha_{adef} + 9\,T^\alpha_{aedf} - 15\,T^\alpha_{afde} \Bigr)
\left[C_{D^2\phi^4}\right]_{bfce}\nonumber \\
&+\Bigl( T^\alpha_{acef} + 9\,T^\alpha_{aecf} - 15\,T^\alpha_{afce} \Bigr)
\left[C_{D^2\phi^4}\right]_{bfde} 
\Big]\nonumber \\
&-2\lambda_{efcd}\left(\left[C_{D^2\phi^4}\right]_{ebaf}-\left[C_{D^2\phi^4}\right]_{efab}\right)\nonumber\\
&+ \gamma_{s,c}^{ae}\left[C_{D^2\phi^4}\right]_{ebcd}
+
\gamma_{s,c}^{be}\left[C_{D^2\phi^4}\right]_{aecd}
+
\gamma_{s,c}^{ce}\left[C_{D^2\phi^4}\right]_{abed}
+
\gamma_{s,c}^{de}\left[C_{D^2\phi^4}\right]_{abce}\,.\nonumber
\end{align}

\paragraph{\texorpdfstring{$D^2\phi^4\leftarrow \phi F^2\times \phi F^2$}{D2phi4fromphiF2xphiF2}:}
\label{par:D2phi4<-phiF2xphiF2}
~
\newline The running of the $\widehat C_{D^2\phi^4}$ combination defined in Eq.~\eqref{eq:hatCD2phi4} is given by 
\begin{align}
    \left[\dot{\widehat{C}}_{D^2\phi^4}\right]_{abcd} &
    = 8 \sum_{\alpha=1}^{N_G}\sum_{\gamma=1}^{\alpha}\sum_{\beta=1}^{\gamma}\Bigg\{ g_\alpha g_\beta \Bigg[Y_{ab}^{A_\alpha B_\beta} \left(\left[C_{\phi F^2}\right]^{A_\alpha C_\gamma}_c\left[C_{\phi F^2}\right]^{B_\beta C_\gamma}_d + \left[C_{\phi F^2}\right]^{A_\alpha C_\gamma}_d\left[C_{\phi F^2}\right]^{B_\beta C_\gamma}_c\right)\nonumber\\
    & \qquad\qquad +\frac{1}{2}\theta^{A_\alpha}_{ab} \left[C_{\phi F^2}\right]^{C_\gamma A_\alpha}_e \left( \theta^{B_\beta}_{de} \left[C_{\phi F^2}\right]^{C_\gamma B_\beta}_c + \theta^{B_\beta}_{ce} \left[C_{\phi F^2}\right]^{C_\gamma B_\beta}_d\right)\nonumber\\
    & \qquad\qquad - (a \leftrightarrow c) - (b\leftrightarrow d) + 
    \binom{a\leftrightarrow c}{b\leftrightarrow d} \Bigg]\nonumber\\
    &\qquad\qquad - \frac{1}{3}  \left(\theta^{A_\alpha}_{ab} \theta^{B_\beta}_{cd} + 2\theta^{A_\alpha}_{ac} \theta^{B_\beta}_{bd}+\theta^{A_\alpha}_{ad} \theta^{B_\beta}_{bc}\right)\left[C_{\phi F^2}\right]^{A_\alpha C_\gamma}_e \left[C_{\phi F^2}\right]^{C_\gamma B_\beta}_e \Bigg\}\nonumber \\
    &\qquad\qquad+\left(\left[C_{\phi F^2}\right]\to \left[C_{\phi \tildee F^2}\right]\right)\,.
\end{align}
The running of the full operator (including off-shell contributions) is given by
\begin{align}
    \left[\dot{C}_{D^2\phi^4}\right]_{abcd} &
    = 8 \sum_{\alpha=1}^{N_G}\sum_{\gamma=1}^{\alpha}\sum_{\beta=1}^{\gamma}\Bigg\{ g_\alpha g_\beta \Bigg[Y_{ab}^{A_\alpha B_\beta} \left(\left[C_{\phi F^2}\right]^{A_\alpha C_\gamma}_c\left[C_{\phi F^2}\right]^{B_\beta C_\gamma}_d + \left[C_{\phi F^2}\right]^{A_\alpha C_\gamma}_d\left[C_{\phi F^2}\right]^{B_\beta C_\gamma}_c\right)\nonumber\\
    & \qquad\qquad\qquad +\frac{1}{2}\theta^{A_\alpha}_{ab} \left[C_{\phi F^2}\right]^{C_\gamma A_\alpha}_e \left( \theta^{B_\beta}_{de} \left[C_{\phi F^2}\right]^{C_\gamma B_\beta}_c + \theta^{B_\beta}_{ce} \left[C_{\phi F^2}\right]^{C_\gamma B_\beta}_d\right)\nonumber\\
    & \qquad\qquad\qquad  -\frac{1}{6}\left(\theta^{A_\alpha}_{ad}\theta^{B_\beta}_{bc}+\theta^{A_\alpha}_{ac}\theta^{B_\beta}_{bd}\right)
    \left[C_{\phi F^2}\right]^{A_\alpha C_\gamma }_e \left[C_{\phi F^2}\right]^{B_\beta C_\gamma }_e \Bigg\}\nonumber\\
    &\qquad\qquad\qquad+\left(\left[C_{\phi F^2}\right]\to \left[C_{\phi \tildee F^2}\right]\right)\,.
    \label{eq:D2phi4_phiF22}
\end{align}

\subsubsection{\texorpdfstring{$\phi^2F^2$}{phi2F2} class}
\paragraph{\texorpdfstring{$\phi^2F^2 \leftarrow \phi^2F^2$:}{phi2F2fromphi2F2}}
\label{par:phi2F2<-phi2F2}

\begin{align}
     \left[\dot{C}_{\phi^2 F^2}\right]_{ab}^{A_\alpha B_\beta} &= \bigg[\lambda_{abcd}+2\sum_{\gamma=1}^{N_G} g_\gamma^2T^\gamma_{a c d b} \bigg]\left[C_{\phi^2 F^2}\right]_{c d}^{A_\alpha B_\beta}\nonumber\\
    & +2\sum_{\sigma(\{A_\alpha ,B_\beta\}\times \{a , b\})}\sum_{\gamma=1}^{N_G}g_\alpha g_\gamma\Big[ \delta_{\alpha\gamma}\Big(3Y_{ac}^{A_\alpha C_\gamma}-2Y_{ac}^{C_\gamma A_\alpha}\Big) \nonumber\\
    & +(1-\delta_{\alpha\gamma}) \mathcal S_{\alpha\beta}^{-1} \mathcal{S}_{\gamma\beta}  Y_{ac }^{A_\alpha C_\gamma}   \Big]\left[C_{\phi^2 F^2}\right]_{cb}^{C_\gamma B_\beta}\nonumber\\
    & +\gamma_{c,s}^{ac}\left[ C_{\phi^2  F^2}\right]_{cb}^{A_\alpha B_\beta} +\gamma_{c,s}^{bc}\left[ C_{\phi^2  F^2}\right]_{ac}^{A_\alpha B_\beta}\nonumber\\
    &+\gamma^{A_\alpha C_\alpha}_{c,v}\left[ C_{\phi^2  F^2}\right]_{ab}^{C_\alpha B_\beta} + \gamma^{C_\beta B_\beta}_{c,v}\left[ C_{\phi^2  F^2}\right]_{ab}^{A_\alpha C_\beta}\,,
    \label{eq:phi2F2_phi2F2}
\end{align}
\begin{align}
     \left[\dot{ C}_{\phi^2 \widetilde F^2}\right]_{ab}^{A_\alpha B_\beta} &= \bigg[\lambda_{abcd}+2\sum_{\gamma=1}^{N_G} g_\gamma^2T^\gamma_{a c d b}\bigg]\left[ C_{\phi^2 \widetilde F^2}\right]_{c d}^{A_\alpha B_\beta}\nonumber\\
    & +2\sum_{\sigma(\{A_\alpha ,B_\beta\}\times \{a , b\})}\sum_{\gamma=1}^{N_G}g_\alpha g_\gamma\Big[ \delta_{\alpha\gamma}\Big(3Y_{ac}^{A_\alpha C_\gamma}-2Y_{ac}^{C_\gamma A_\alpha}\Big) \nonumber\\
    & +(1-\delta_{\alpha\gamma}) \mathcal S_{\alpha\beta}^{-1} \mathcal{S}_{\gamma\beta}  Y_{ac }^{A_\alpha C_\gamma}   \Big]\left[ C_{\phi^2 \widetilde F^2}\right]_{cb}^{C_\gamma B_\beta}\nonumber\\
    & +\gamma_{c,s}^{ac}\left[ C_{\phi^2  \tildee F^2}\right]_{cb}^{A_\alpha B_\beta} +\gamma_{c,s}^{bc}\left[ C_{\phi^2  \tildee F^2}\right]_{ac}^{A_\alpha B_\beta}\nonumber\\
    &+\gamma^{A_\alpha C_\alpha}_{c,v}\left[ C_{\phi^2  \tildee F^2}\right]_{ab}^{C_\alpha B_\beta} + \gamma^{C_\beta B_\beta}_{c,v}\left[ C_{\phi^2  \tildee F^2}\right]_{ab}^{A_\alpha C_\beta}\,.
\end{align}

\paragraph{\texorpdfstring{$\phi^2F^2\leftarrow F^3$}{phiF2fromF3}:}
\label{par:phi2F2<-F3}

\begin{align}
    \left[\dot{C}_{\phi^2 F^2}\right]_{ab}^{A_\alpha B_\beta} &= \frac{3}{2}i\delta_{\alpha\beta}g_\alpha^3 
    \sum_{\sigma(\{A_\alpha,B_\alpha\}\times\{a,b\})}
    \Big[
    \theta^{D_\alpha}_{bc}
    \left(Y_{ac}^{A_\alpha C_\alpha}-2 Y_{ac}^{C_\alpha A_\alpha}\right)
    \left[ C_{F^3}\right]^{B_\alpha C_\alpha D_\alpha}\nonumber\\
    &-i Y_{ab}^{C_\alpha D_\alpha}
    f^{B_\alpha D_\alpha E_\alpha}\left[ C_{F^3}\right]^{A_\alpha C_\alpha E_\alpha}
    \Big]\nonumber \\
    &- 3i(1-\delta_{\alpha\beta}) 
    \sum_{\sigma(\{A_\alpha,B_\beta\}\times\{a,b\})}
    g^2_\alpha g_\beta 
    \theta^{D_\alpha}_{bc}Y^{B_\beta C_\alpha}_{ac}
    \left[ C_{F^3}\right]^{A_\alpha C_\alpha D_\alpha}\,,\label{eq:CF3_to_Cphi2F2}
\end{align}
\begin{align}
    \left[\dot{C}_{\phi^2 \widetilde F^2}\right]_{ab}^{A_\alpha B_\beta} &= \frac{3}{2}i\delta_{\alpha\beta}g_\alpha^3 
    \sum_{\sigma(\{A_\alpha,B_\alpha\}\times\{a,b\})}
    \Big[
    \theta^{D_\alpha}_{bc}
    \left(Y_{ac}^{A_\alpha C_\alpha}-2 Y_{ac}^{C_\alpha A_\alpha}\right)
    \left[C_{\widetilde F^3}\right]^{B_\alpha C_\alpha D_\alpha}\nonumber\\
    &-i Y_{ab}^{C_\alpha D_\alpha}
    f^{B_\alpha D_\alpha E_\alpha}\left[  C_{\widetilde F^3}\right]^{A_\alpha C_\alpha E_\alpha}
    \Big]\nonumber \\
    &- 3i(1-\delta_{\alpha\beta}) 
    \sum_{\sigma(\{A_\alpha,B_\beta\}\times\{a,b\})}
    g^2_\alpha g_\beta 
    \theta^{D_\alpha}_{bc}Y^{B_\beta C_\alpha}_{ac}
    \left[  C_{\widetilde F^3}\right]^{A_\alpha C_\alpha D_\alpha}\,.
\end{align}

\paragraph{\texorpdfstring{$\phi^2F^2\leftarrow \phi F^2\times\phi F^2$}{phiF2fromphiF2xphiF2}:}
\label{par:phi2F2<-phiF2xphiF2}

\begin{align}
    \left[\dot C_{\phi^2 F^2}\right]_{ab}^{A_\alpha B_\beta} &= 
    \frac{1}{4} \mathcal S_{\alpha\beta}^{-1} \sum_{\sigma(\{A_\alpha,B_\beta\}\times \{a,b\})} \sum_{\gamma=1}^{N_G}\mathcal S_{\alpha\gamma}\mathcal S_{\beta\gamma} g_\gamma^2 \left\{\frac{1}{3}[S_2(R_\gamma)]_{ac}\left[C_{\phi F^2}\right]_{c}^{A_\alpha C_\gamma} \left[C_{\phi F^2}\right]_{b}^{B_\beta C_\gamma}\right.\nonumber \\
    &\left.+ Y_{ac}^{C_\gamma D_\gamma} \left[C_{\phi F^2}\right]_{c}^{A_\alpha C_\gamma} \left[C_{\phi F^2}\right]_{b}^{B_\beta D_\gamma} + 2 Y_{cb}^{C_\gamma D_\gamma} \left[C_{\phi F^2}\right]_{a}^{A_\alpha C_\gamma} \left[C_{\phi F^2}\right]_{c}^{B_\beta D_\gamma} \right\}\nonumber\\
    &+\mathcal S_{\alpha\beta}^{-1}\sum_{\sigma(\{A_\alpha,B_\beta\}\times \{a,b\})}\sum_{\gamma=1}^{N_G}\mathcal S_{\alpha\gamma}\mathcal S_{\beta\gamma} g_\gamma^2\bigg\{-\frac{11}{3} [C_2(G_\gamma)]^{A_\alpha D_\alpha}\left[C_{\phi F^2}\right]_{a}^{D_\alpha C_\gamma} \left[C_{\phi F^2}\right]_{b}^{B_\beta C_\gamma}\nonumber \\
    &+ 3i f^{C_\gamma D_\gamma E_\gamma} \theta^{C_\gamma}_{bc} \left[C_{\phi F^2}\right]_{a}^{A_\alpha D_\gamma} \left[C_{\phi F^2}\right]_{c}^{B_\beta E_\gamma}\bigg\}\nonumber\\
    &+\frac{1}{4}\mathcal S_{\alpha\beta}^{-1}\bigg\{\sum_{\gamma=1}^{N_G}\mathcal S_{\alpha\gamma}\mathcal S_{\beta\gamma}\bigg(\left[C_{\phi F^2}\right]_{c}^{A_\alpha C_\gamma} \left[C_{\phi F^2}\right]_{d}^{B_\beta C_\gamma}+\left[C_{\phi F^2}\right]_{c}^{B_\beta C_\gamma} \left[C_{\phi F^2}\right]_{d}^{A_\alpha C_\gamma}\bigg)\nonumber \\
    &\times \bigg(\lambda_{abcd}-\sum_{\delta=1}^{N_G}g_\delta^2\left(T^\delta_{acbd}+T^\delta_{adbc}\right)\bigg)\bigg\}\nonumber\\
    & + \frac{1}{2}\mathcal S_{\alpha\beta}^{-1} \sum_{\sigma(\{A_\alpha,B_\beta\}\times \{a,b\})}\sum_{\gamma=1}^{N_G}\sum_{\delta=1}^{N_G}\mathcal S_{\alpha\gamma}\mathcal S_{\beta\delta} g_\gamma g_\delta Y_{ab}^{C_\gamma D_\delta} \left[C_{\phi F^2}\right]_{c}^{A_\alpha C_\gamma} \left[C_{\phi F^2}\right]_{c}^{B_\beta D_\delta}\nonumber\\
    & - \left(\left[C_{\phi F^2}\right]\to \left[C_{\phi \tildee F^2}\right]\right)\,,
    \label{eq:phi2F2-phiF2xphiF2}
\end{align}
\begin{align}
    \left[\dot C_{\phi^2 \tildee F^2}\right]_{ab}^{A_\alpha B_\beta} &= 
    \frac{1}{4} \mathcal S_{\alpha\beta}^{-1} \sum_{\sigma(\{A_\alpha,B_\beta\}\times \{a,b\})} \sum_{\gamma=1}^{N_G}\mathcal S_{\alpha\gamma}\mathcal S_{\beta\gamma} g_\gamma^2 \left\{\frac{1}{3}[S_2(R_\gamma)]_{ac}\left[C_{\phi F^2}\right]_{c}^{A_\alpha C_\gamma} \left[C_{\phi \tildee F^2}\right]_{b}^{B_\beta C_\gamma}\right.\nonumber \\
    &\left.+ Y_{ac}^{C_\gamma D_\gamma} \left[C_{\phi F^2}\right]_{c}^{A_\alpha C_\gamma} \left[C_{\phi \tildee F^2}\right]_{b}^{B_\beta D_\gamma} + 2 Y_{cb}^{C_\gamma D_\gamma} \left[C_{\phi F^2}\right]_{a}^{A_\alpha C_\gamma} \left[C_{\phi \tildee F^2}\right]_{c}^{B_\beta D_\gamma} \right\}\nonumber\\
    &+\mathcal S_{\alpha\beta}^{-1}\sum_{\sigma(\{A_\alpha,B_\beta\}\times \{a,b\})}\sum_{\gamma=1}^{N_G}\mathcal S_{\alpha\gamma}\mathcal S_{\beta\gamma} g_\gamma^2\bigg\{-\frac{11}{3} [C_2(G_\gamma)]^{A_\alpha D_\alpha}\left[C_{\phi F^2}\right]_{a}^{D_\alpha C_\gamma} \left[C_{\phi \tildee F^2}\right]_{b}^{B_\beta C_\gamma}\nonumber \\
    &+ 3i f^{C_\gamma D_\gamma E_\gamma} \theta^{C_\gamma}_{bc} \left[C_{\phi F^2}\right]_{a}^{A_\alpha D_\gamma} \left[C_{\phi \tildee F^2}\right]_{c}^{B_\beta E_\gamma}\bigg\}\nonumber\\
    &+\frac{1}{4}\mathcal S_{\alpha\beta}^{-1}\bigg\{\sum_{\gamma=1}^{N_G}\mathcal S_{\alpha\gamma}\mathcal S_{\beta\gamma}\bigg(\left[C_{\phi F^2}\right]_{c}^{A_\alpha C_\gamma} \left[C_{\phi \tildee F^2}\right]_{d}^{B_\beta C_\gamma}+\left[C_{\phi F^2}\right]_{c}^{B_\beta C_\gamma} \left[C_{\phi \tildee F^2}\right]_{d}^{A_\alpha C_\gamma}\bigg)\nonumber \\
    &\times \bigg(\lambda_{abcd}-\sum_{\delta=1}^{N_G}g_\delta^2\left(T^\delta_{acbd}+T^\delta_{adbc}\right)\bigg)\bigg\}\nonumber\\
    & + \frac{1}{2}\mathcal S_{\alpha\beta}^{-1} \sum_{\sigma(\{A_\alpha,B_\beta\}\times \{a,b\})}\sum_{\gamma=1}^{N_G}\sum_{\delta=1}^{N_G}\mathcal S_{\alpha\gamma}\mathcal S_{\beta\delta}g_\gamma g_\delta Y_{ab}^{C_\gamma D_\delta} \left[C_{\phi F^2}\right]_{c}^{A_\alpha C_\gamma} \left[C_{\phi \tildee F^2}\right]_{c}^{B_\beta D_\delta}\nonumber\\
    & + \left(\left[C_{\phi F^2}\right]\leftrightarrow \left[C_{\phi \tildee F^2}\right]\right)\,.
    \label{eq:phi2F2-phiF2xphiF2_CPV}
\end{align}

\subsubsection{\texorpdfstring{$F^3$}{F3} class}
\paragraph{\texorpdfstring{$F^3 \leftarrow F^3$}{F3fromF3}:}
\label{par:F3<-F3}

\begin{align}
    \left[\dot C_{F^3}\right]^{A_\alpha B_\alpha C_\alpha} &= 4 g_\alpha^2 \Big(F^{B_\alpha C_\alpha D_\alpha F_\alpha} \left[ C_{F^3}\right]^{A_\alpha D_\alpha F_\alpha} + F^{C_\alpha A_\alpha D_\alpha F_\alpha} \left[ C_{F^3}\right]^{B_\alpha D_\alpha F_\alpha}\nonumber \\& +F^{A_\alpha B_\alpha D_\alpha F_\alpha} \left[ C_{F^3}\right]^{C_\alpha D_\alpha F_\alpha}  \Big) + \gamma_{c,v}^{A_\alpha D_\alpha}\left[ C_{ F^3}\right]^{D_\alpha B_\alpha C_\alpha}\nonumber\\ 
    &
    + \gamma_{c,v}^{B_\alpha D_\alpha}\left[ C_{F^3}\right]^{A_\alpha D_\alpha C_\alpha} + \gamma_{c,v}^{C_\alpha D_\alpha}\left[ C_{F^3}\right]^{A_\alpha B_\alpha D_\alpha}\,,
    \label{eq:F3_to_F3}
\end{align}
\begin{align}
    \left[\dot{ C}_{\widetilde F^3}\right]^{A_\alpha B_\alpha C_\alpha} &= 4 g_\alpha^2 \Big(F^{B_\alpha C_\alpha D_\alpha F_\alpha} \left[  C_{\widetilde F^3}\right]^{A_\alpha D_\alpha F_\alpha} + F^{C_\alpha A_\alpha D_\alpha F_\alpha} \left[  C_{\widetilde F^3}\right]^{B_\alpha D_\alpha F_\alpha}\nonumber \\& +F^{A_\alpha B_\alpha D_\alpha F_\alpha} \left[  C_{\widetilde F^3}\right]^{C_\alpha D_\alpha F_\alpha}  \Big)+ \gamma_{c,v}^{A_\alpha D_\alpha}\left[ C_{\widetilde F^3}\right]^{D_\alpha B_\alpha C_\alpha}\nonumber\\ 
    &
    + \gamma_{c,v}^{B_\alpha D_\alpha}\left[ C_{\widetilde F^3}\right]^{A_\alpha D_\alpha C_\alpha} + \gamma_{c,v}^{C_\alpha D_\alpha}\left[ C_{\widetilde F^3}\right]^{A_\alpha B_\alpha D_\alpha}\,.
    \label{eq:F3t_to_F3t}
\end{align}

\paragraph{\texorpdfstring{$F^3\leftarrow \phi F^2 \times \phi F^2$}{F3fromphiF2xphiF2}:}
\label{par:F3<-phiF2xphiF2}

\begin{align}
    \left[\dot C_{F^3}\right]^{A_\alpha B_\alpha C_\alpha} &= -\frac{8}{3}g_\alpha \bigg[ 
    f^{C_\alpha D_\alpha E_\alpha}\left( \left[C_{\phi F^2}\right]^{A_\alpha D_\alpha}_a \left[C_{\phi F^2}\right]^{B_\alpha E_\alpha}_a - \left[ C_{\phi \widetilde F^2}\right]^{A_\alpha D_\alpha}_a \left[ C_{\phi \widetilde F^2}\right]^{B_\alpha E_\alpha}_a \right) \nonumber \\
    &+
    f^{B_\alpha D_\alpha E_\alpha}\left( \left[C_{\phi F^2}\right]^{C_\alpha D_\alpha}_a \left[C_{\phi F^2}\right]^{A_\alpha E_\alpha}_a - \left[ C_{\phi \widetilde F^2}\right]^{C_\alpha D_\alpha}_a \left[ C_{\phi \widetilde F^2}\right]^{A_\alpha E_\alpha}_a \right)\label{eq:F3<-phiF2xphiF2}\\
    &+
    f^{A_\alpha D_\alpha E_\alpha}\left( \left[C_{\phi F^2}\right]^{B_\alpha D_\alpha}_a \left[C_{\phi F^2}\right]^{C_\alpha E_\alpha}_a - \left[ C_{\phi \widetilde F^2}\right]^{B_\alpha D_\alpha}_a \left[ C_{\phi \widetilde F^2}\right]^{C_\alpha E_\alpha}_a \right)
    \bigg]\nonumber \,,
\end{align}
\begin{align}
\left[\dot{ C}_{\widetilde F^3}\right]^{A_\alpha B_\alpha C_\alpha} &= -\frac{16}{3}g_\alpha \bigg( 
f^{C_\alpha D_\alpha E_\alpha} \left[C_{\phi F^2}\right]^{A_\alpha D_\alpha}_a \left[ C_{\phi \widetilde F^2}\right]^{B_\alpha E_\alpha}_a +f^{B_\alpha D_\alpha E_\alpha} \left[C_{\phi F^2}\right]^{C_\alpha D_\alpha}_a \left[ C_{\phi \widetilde F^2}\right]^{A_\alpha E_\alpha}_a \nonumber   \\
&+
f^{A_\alpha D_\alpha E_\alpha} \left[C_{\phi F^2}\right]^{B_\alpha D_\alpha}_a \left[ C_{\phi \widetilde F^2}\right]^{C_\alpha E_\alpha}_a 
\bigg)\,. 
\label{eq:F3tilde<-phiF2xphiF2}
\end{align}

\subsection{Running of dimension-5 operators}

\subsubsection{\texorpdfstring{$\phi^5$}{phi5} class}

In addition to the results presented below, one needs to supplement the RGEs of the $\phi^5$-class WCs by the expression given in Eq.~\eqref{eq:dotCphi5<-Ctildeandbar}.
The explicit expressions for $\dot{\tildee C}_{D^2\phi^4}$ and $\dot{\overline C}_{D^2\phi^4}$ can be obtained using Eq.~\eqref{eq:decomposition} and the RGEs in Sec.~\ref{sec:D2phi4RGE}.

\paragraph{\texorpdfstring{$\phi^5 \leftarrow \phi^5$}{phi5fromphi5}:}
\label{par:phi5<-phi5}

\begin{align}
    \left[\dot C_{\phi^5}\right]_{abcde} &= \frac{1}{2!3!}\sum_{\sigma(\{a,b,c,d,e\})}\bigg[\lambda_{defg}+2\sum_{\alpha=1}^{N_G} g_\alpha^2 T^\alpha_{dfge}\bigg]\left[ C_{\phi^5}\right]_{abcfg} \nonumber \\
    & + \frac{1}{4!}\sum_{\sigma(\{a,b,c,d,e\})}\gamma_{c,s}^{af}\left[C_{\phi^5}\right]_{fbcde}\,.
\end{align}

\paragraph{\texorpdfstring{$\phi^5 \leftarrow \phi^6$}{phi5fromphi6}:}
\label{par:phi5<-phi6}

\begin{equation}
    \left[\dot C_{\phi^5}\right]_{abcde} = \frac{6!}{5!4!}\sum_{\sigma(\{a,b,c,d,e\})}h_{efg}\left[ C_{\phi^6}\right]_{abcdfg}\,.
\end{equation}

\paragraph{\texorpdfstring{$\phi^5 \leftarrow D^2\phi^4$}{phi5fromD2phi4}:}
\label{par:phi5<-D2phi4}

\begin{align}
    \left[\dot C_{\phi^5}\right]_{abcde} &=  \frac{1}{5!}\sum_{(\{a,b,c,d,e\})}\Bigg[2h_{fga}\lambda_{hgbc}\left[C_{D^2\phi^4}\right]_{fhde}\nonumber\\ \nonumber
    &+\left(\frac{1}{3}h_{fga}\lambda_{hbcd}+\frac{1}{2}h_{hcd}\lambda_{fgab}\right)\left(\left[C_{D^2\phi^4}\right]_{hgfe}+\left[C_{D^2\phi^4}\right]_{hfge}-\left[C_{D^2\phi^4}\right]_{gfhe}\right)\\
    &-6\sum_{\alpha=1}^{N_G} g_\alpha^2 h_{agh} T^\alpha_{bhcf} \left[C_{D^2\phi^4}\right]_{fgde}\Bigg]\,.
\end{align}

\paragraph{\texorpdfstring{$\phi^5 \leftarrow \phi^2 F^2$}{phi5fromphi2F2}:}
\label{par:phi5<-phi2F2}

\begin{align}
    \left[\dot C_{\phi^5}\right]_{abcde} &= \frac{6}{5!} \sum_{\sigma(\{a,b,c,d,e\})} \sum_{\alpha=1}^{N_G} \sum_{\beta=1}^{\alpha} g_\alpha g_\beta h_{abf} Y^{A_\alpha B_\beta}_{cf} \left[C_{\phi^2 F^2}\right]^{A_\alpha B_\beta}_{de}\,.
\end{align}

\paragraph{\texorpdfstring{$\phi^5\leftarrow \phi F^2\times\phi F^2$}{phi5fromphiF2xphiF2}:}
\label{par:phi5<-phiF2xphiF2}

\begin{align}
    \left[\dot C_{\phi^5}\right]_{abcde} &= -\frac{1}{5!} \sum_{\sigma(\{a,b,c,d,e\})} \frac{1}{3} h_{abf}\lambda_{cdeg} \sum_{\alpha=1}^{N_G}\sum_{\beta=1}^\alpha \mathcal{S}_{\alpha\beta} \left[C_{\phi F^2}\right]_{f}^{A_\alpha B_\beta} \left[C_{\phi F^2}\right]_{g}^{A_\alpha B_\beta}\nonumber \\
    &+\left(\left[C_{\phi F^2}\right]\to \left[C_{\phi \tildee F^2}\right]\right)\,.
\end{align}

\paragraph{\texorpdfstring{$\phi^5\leftarrow \phi F^2$}{phi5fromphiF2}:}
\label{par:phi5<-phiF2}

\begin{align}
    \left[\dot C_{\phi^5}\right]_{abcde} &=
    \frac{4}{5!}\sum_{\sigma(\{a,b,c,d,e\})} \Bigg[\sum_{\alpha=1}^{N_G} \sum_{\beta=1}^{\alpha} g_\alpha g_\beta \lambda_{abcf} Y_{df}^{A_\alpha B_\beta} \left[C_{\phi F^2}\right]_{e}^{A_\alpha B_\beta}\nonumber\\
    &- 6 \sum_{\alpha=1}^{N_G}\sum_{\beta=1}^{\alpha}\sum_{\gamma=1}^{N_G} g_\alpha g_\beta g_\gamma^2 \,Y_{bd}^{A_\alpha C_\gamma} Y_{ce}^{B_\beta C_\gamma} \left[C_{\phi F^2}\right]_{a}^{A_\alpha B_\beta}\Bigg]\,.
    \label{eq:phi5<-phiF2}
\end{align}

\subsubsection{\texorpdfstring{$\phi F^2$}{phiF2} class}

\paragraph{\texorpdfstring{$\phi F^2 \leftarrow \phi F^2$}{phiF2fromphiF2}:}
\label{par:phiF2<-phiF2}

\begin{align}
    \left[\dot C_{\phi F^2}\right]_a^{A_\alpha B_\beta} &= 
    2\sum_{\sigma(\{A_\alpha ,B_\beta\})}\sum_{\gamma=1}^{N_G}g_\alpha g_\gamma\Big[ \delta_{\alpha\gamma}\Big(3Y_{ab}^{A_\alpha C_\gamma}-2Y_{ab}^{C_\gamma A_\alpha}\Big) \nonumber\\
    & +(1-\delta_{\alpha\gamma}) \mathcal S_{\alpha\beta}^{-1} \mathcal{S}_{\gamma\beta}  Y_{ab }^{A_\alpha C_\gamma}   \Big]\left[C_{\phi F^2}\right]_{b}^{C_\gamma B_\beta}\label{eq:phiF2-phiF2} \\
    &+\gamma_{c,s}^{ac}\left[ C_{\phi  F^2}\right]_{c}^{A_\alpha B_\beta}+\gamma^{A_\alpha C_\alpha}_{c,v}\left[ C_{\phi  F^2}\right]_{a}^{C_\alpha B_\beta} + \gamma^{C_\beta B_\beta}_{c,v}\left[ C_{\phi F^2}\right]_{a}^{A_\alpha C_\beta}\,,\nonumber
\end{align}
\begin{align}
    \left[\dot{ C}_{\phi \widetilde F^2}\right]_a^{A_\alpha B_\beta} &= 
    2\sum_{\sigma(\{A_\alpha ,B_\beta\})}\sum_{\gamma=1}^{N_G}g_\alpha g_\gamma\Big[ \delta_{\alpha\gamma}\Big(3Y_{ab}^{A_\alpha C_\gamma}-2Y_{ab}^{C_\gamma A_\alpha}\Big)\nonumber\\
    & +(1-\delta_{\alpha\gamma}) \mathcal S_{\alpha\beta}^{-1} \mathcal{S}_{\gamma\beta}  Y_{ab }^{A_\alpha C_\gamma}   \Big]\left[ C_{\phi \widetilde F^2}\right]_{b}^{C_\gamma B_\beta}\label{eq:phiFt2-phiFt2}\\
    &+\gamma_{c,s}^{ac}\left[ C_{\phi  \tildee F^2}\right]_{c}^{A_\alpha B_\beta}+\gamma^{A_\alpha C_\alpha}_{c,v}\left[ C_{\phi  \tildee F^2}\right]_{a}^{C_\alpha B_\beta} + \gamma^{C_\beta B_\beta}_{c,v}\left[ C_{\phi \tildee F^2}\right]_{a}^{A_\alpha C_\beta}\,.\nonumber
\end{align}

\paragraph{\texorpdfstring{$\phi F^2 \leftarrow \phi^2 F^2$}{phiF2fromphi2F2}:}
\label{par:phiF2<-phi2F2}

\begin{align}
    \left[\dot C_{\phi F^2}\right]_a^{A_\alpha B_\beta} &=
    2 h_{abc}\left[C_{\phi^2 F^2}\right]_{bc}^{A_\alpha B_\beta}\,,
\\
    \left[\dot C_{\phi \widetilde F^2}\right]_a^{A_\alpha B_\beta} &=
    2 h_{abc}\left[C_{\phi^2 \widetilde F^2}\right]_{bc}^{A_\alpha B_\beta}\,.
\end{align}

\paragraph{\texorpdfstring{$\phi F^2\leftarrow \phi F^2\times\phi F^2$}{phiF2fromphiF2xphiF2}:}
\label{par:phiF2<-phiF2xphiF2}

\begin{align}
    \left[\dot C_{\phi F^2}\right]^{A_\alpha B_\beta}_a &= 
    2 \mathcal S_{\alpha\beta}^{-1}h_{abc}\sum_{\gamma=1}^{N_G}\mathcal S_{\alpha\gamma}\mathcal S_{\beta\gamma}
    \left(
    \left[ C_{\phi F^2}\right]^{A_\alpha C_\gamma}_b  \left[ C_{\phi F^2}\right]^{B_\beta C_\gamma}_c
    - \left[ C_{\phi \widetilde F^2}\right]^{A_\alpha C_\gamma}_b  \left[ C_{\phi \widetilde F^2}\right]^{B_\beta C_\gamma}_c
    \right)\,,\label{eq:phiF2-phiF2xphiF2}
    \\
    \left[\dot C_{\phi \widetilde F^2}\right]^{A_\alpha B_\beta}_a &= 
    2 \mathcal S_{\alpha\beta}^{-1}h_{abc}\sum_{\gamma=1}^{N_G}\mathcal S_{\alpha\gamma}\mathcal S_{\beta\gamma}
    \left(
    \left[ C_{\phi F^2}\right]^{A_\alpha C_\gamma}_b  \left[ C_{\phi \widetilde F^2}\right]^{B_\beta C_\gamma}_c
    + \left[ C_{\phi \widetilde F^2}\right]^{A_\alpha C_\gamma}_b  \left[ C_{\phi  F^2}\right]^{B_\beta C_\gamma}_c
    \right)\,.
\end{align}

\subsection{Running of renormalizable couplings}

\subsubsection{Gauge couplings and topological angles} 

\paragraph{\texorpdfstring{$F^2\leftarrow \phi^2 F^2$}{F2from phi2F2}:}

\begin{align}
    \dot g_\alpha \delta^{A_\alpha B_\alpha} &= 4 g_\alpha  m^2_{ab} \left[C_{\phi^2F^2}\right]_{ab}^{A_\alpha B_\alpha}\,,
    \label{eq:g_phi2F2}
\\
    \dot\vartheta_\alpha\delta^{A_\alpha B_\alpha} &= \frac{32\pi^2}{g_\alpha^2} 2 m^2_{ab} \left[C_{\phi^2\tildee F^2}\right]_{ab}^{A_\alpha B_\alpha}\,.
    \label{eq:theta_phi2F2}
\end{align}

\paragraph{\texorpdfstring{$F^2\leftarrow \phi F^2 \times \phi F^2$}{F2fromphiF2xphiF2}:}

\begin{align}
    \dot g_\alpha \delta^{A_\alpha B_\alpha} &= 2 g_\alpha m^2_{ab}\sum_{\beta=1}^{N_G} \mathcal S_{\alpha\beta}^2
    \left(
    \left[C_{\phi F^2} \right]^{A_\alpha C_\beta }_a \left[C_{\phi F^2} \right]^{C_\beta B_\alpha }_b
    -
    \left[C_{\phi \widetilde F^2} \right]^{A_\alpha C_\beta }_a \left[C_{\phi \widetilde F^2} \right]^{C_\beta B_\alpha }_b
    \right)\,,
    \label{eq:F2<-phiF2xphiF2}
\\
    \dot\vartheta_\alpha \delta^{A_\alpha B_\alpha} &= \frac{32\pi^2}{g_\alpha^2}
     m^2_{ab} \sum_{\beta=1}^{N_G} \mathcal S_{\alpha\beta}^2
    \left(
    \left[C_{\phi F^2} \right]^{A_\alpha C_\beta }_a \left[C_{\phi \tildee F^2} \right]^{C_\beta B_\alpha }_b
    +
    \left[C_{\phi \widetilde F^2} \right]^{A_\alpha C_\beta }_a \left[C_{\phi F^2} \right]^{C_\beta B_\alpha }_b
    \right)\,.
    \label{eq:Ftilde2<-phiF2xphiF2}
\end{align}

\subsubsection{Scalar quartic}

In addition to the results presented below, one needs to supplement the RGEs for quartic scalar couplings by the terms in Eq.~\eqref{eq:dotlambda<-Ctildeandbar}.
The explicit expressions for $\dot{\tildee C}_{D^2\phi^4}$ and $\dot{\overline C}_{D^2\phi^4}$ can be obtained using Eq.~\eqref{eq:decomposition} and the RGEs in Sec.~\ref{sec:D2phi4RGE}.

\paragraph{\texorpdfstring{$\phi^{4} \leftarrow \phi^6$}{phi4fromphi6}:}

\begin{equation}
    \dot \lambda_{abcd} = -6!\,m^2_{ef}\left[C_{\phi^6}\right]_{abcdef}\,.
    \label{eq:lambda_CH6}
\end{equation}

\paragraph{\texorpdfstring{$\phi^{4} \leftarrow \phi^2 F^2$}{phi4fromphi2F2}:}

\begin{align}
    \dot \lambda_{abcd} &= -12 \sum_{\sigma(\{a,b,c,d\})} \sum_{\alpha=1}^{N_G} \sum_{\beta=1}^{\alpha} g_\alpha g_\beta m^2_{ae} Y^{A_\alpha B_\beta}_{be} \left[C_{\phi^2 F^2}\right]^{A_\alpha B_\beta}_{cd}\,.
    \label{eq:phi2F2_to_phi4}
\end{align}

\paragraph{\texorpdfstring{$\phi^{4} \leftarrow D^2\phi^4$}{phi4fromD2phi4}:}

\begin{align}
    \dot \lambda_{abcd} &= - \sum_{\sigma(\{a,b,c,d\})} \Bigg[2(m^2_{ef}\lambda_{gfab}+h_{efa}h_{gfb})\left[C_{D^2\phi^4}\right]_{egcd}\label{eq:phi4<-D2phi4}\\\nonumber 
    &+\left(\frac{1}{3}m^2_{ef}\lambda_{gcab}+h_{efa}h_{gbc}+m^2_{gc}\lambda_{efab}\right)\left(\left[C_{D^2\phi^4}\right]_{gfed}+\left[C_{D^2\phi^4}\right]_{gefd}-\left[C_{D^2\phi^4}\right]_{fegd}\right)\\
    &-6\sum_{\alpha=1}^{N_G} g_\alpha^2 m^2_{fg} T^\alpha_{agbe} \left[C_{D^2\phi^4}\right]_{efcd}
    -\frac{4}{3!}\lambda_{ebcd}m^2_{fg}\left(
    \left[C_{D^2\phi^4}\right]_{feag}-\left[C_{D^2\phi^4}\right]_{fgae}
    \right)
    \Bigg]\,.
    \nonumber
\end{align}

\paragraph{\texorpdfstring{$\phi^4 \leftarrow \phi^5$}{phi4fromphi5}:}
\begin{equation}
    \dot \lambda_{abcd} = - \frac{5!}{3!}\sum_{\sigma(\{a,b,c,d\})} h_{def}\left[C_{\phi^5}\right]_{abcef}\,.
\end{equation}

\paragraph{\texorpdfstring{$\phi^4 \leftarrow \phi F^2$}{phi4fromphiF2}:}

\begin{align}
    \dot\lambda_{abcd} &=
    -12\sum_{\sigma(\{a,b,c,d\})} \sum_{\alpha=1}^{N_G} \sum_{\beta=1}^{\alpha} g_\alpha g_\beta h_{abe} Y_{ce}^{A_\alpha B_\beta} \left[C_{\phi F^2}\right]_{d}^{A_\alpha B_\beta}\,.
    \label{eq:phi4<-phiF2}
\end{align}

\paragraph{\texorpdfstring{$\phi^4 \leftarrow \phi F^2 \times \phi F^2$}{phi4fromphiF2xphiF2}:}

\begin{align}
    \dot \lambda_{abcd} &= \sum_{\sigma(\{a,b,c,d\})} \left(\frac{2}{3} m^2_{ae}\lambda_{bcdf}+\frac{1}{2}h_{abe} h_{cdf}\right) \sum_{\alpha=1}^{N_G}\sum_{\beta=1}^\alpha \mathcal{S}_{\alpha\beta} \left[C_{\phi F^2}\right]_{e}^{A_\alpha B_\beta} \left[C_{\phi F^2}\right]_{f}^{A_\alpha B_\beta}\nonumber \\
    &+ \left(\left[C_{\phi F^2}\right]\to \left[C_{\phi \tildee F^2}\right]\right)\,.
\end{align}

\subsubsection{Scalar trilinear}

In addition to the results presented below, one needs to supplement the RGEs for scalar trilinear couplings by the expression in Eq.~\eqref{eq:dotCphi3<-Ctildeandbar}.
The explicit expressions for $\dot{\tildee C}_{D^2\phi^4}$ and $\dot{\overline C}_{D^2\phi^4}$ can be obtained using Eq.~\eqref{eq:decomposition} and the RGEs in Sec.~\ref{sec:D2phi4RGE}.

\paragraph{\texorpdfstring{$\phi^{3} \leftarrow \phi^2 F^2$}{phi3fromphi2F2}:}

\begin{align}
    \dot h_{abc} &= -12 \sum_{\sigma(\{a,b,c\})} \sum_{\alpha=1}^{N_G} \sum_{\beta=1}^{\alpha} g_\alpha g_\beta t_{e} Y^{A_\alpha B_\beta}_{ae} \left[C_{\phi^2 F^2}\right]^{A_\alpha B_\beta}_{bc}\,.
    \label{eq:phi2F2_to_phi3}
\end{align}

\paragraph{\texorpdfstring{$\phi^{3} \leftarrow D^2\phi^4$}{phi3fromD2phi4}:}

\begin{align}
    \dot h_{abc} &= - \sum_{\sigma(\{a,b,c\})} \Bigg[4 m_{de}^2 h_{fea}\left[C_{D^2\phi^4}\right]_{dfbc}\nonumber \\
    &+\left(m^2_{de} h_{fab}+2m_{fa}^2 h_{deb}+t_f \lambda_{deab}\right)\left(\left[C_{D^2\phi^4}\right]_{hedc}+\left[C_{D^2\phi^4}\right]_{hdec}-\left[C_{D^2\phi^4}\right]_{edfc}\right)\nonumber \\
    &
    -\frac{4}{2!}h_{dbc}
    m^2_{ef}\left(
    \left[C_{D^2\phi^4}\right]_{edaf}-\left[C_{D^2\phi^4}\right]_{efad}
    \right)
    \Bigg]\,.
    \label{eq:phi3<-D2phi4}
\end{align}

\paragraph{\texorpdfstring{$\phi^3 \leftarrow \phi^5$}{phi3fromphi5}:}

\begin{equation}
    \dot h_{abc} = -5!\,m^2_{de}\left[C_{\phi^5}\right]_{abcde}\,.
\end{equation}

\paragraph{\texorpdfstring{$\phi^3 \leftarrow \phi F^2$}{phi3fromphiF2}:}

\begin{align}
    \dot h_{abc} &=
    -24\sum_{\sigma(\{a,b,c\})} \sum_{\alpha=1}^{N_G} \sum_{\beta=1}^{\alpha} g_\alpha g_\beta m^2_{ad} Y_{bd}^{A_\alpha B_\beta} \left[C_{\phi F^2}\right]_{c}^{A_\alpha B_\beta}\,.
    \label{eq:phi3<-phiF2}
\end{align}

\paragraph{\texorpdfstring{$\phi^3 \leftarrow \phi F^2 \times \phi F^2$}{phi3fromphiF2xphiF2}:}

\begin{align}
    \dot h_{abc} &= \sum_{\sigma(\{a,b,c\})} \left(\frac{2}{3} t_{d}\lambda_{abce}+2 m^2_{ad} h_{bce}\right) \sum_{\alpha=1}^{N_G}\sum_{\beta=1}^\alpha \mathcal{S}_{\alpha\beta} \left[C_{\phi F^2}\right]_{d}^{A_\alpha B_\beta} \left[C_{\phi F^2}\right]_{e}^{A_\alpha B_\beta}\nonumber \\
    &+ \left(\left[C_{\phi F^2}\right]\to \left[C_{\phi \tildee F^2}\right]\right)\,.
\end{align}

\subsubsection{Scalar mass}

\paragraph{\texorpdfstring{$\phi^{2} \leftarrow D^2\phi^4$}{phi2fromD2phi4}:}

\begin{align}
    \dot{(m^2)}_{ab} &= -\sum_{\sigma(\{a,b\})}\Bigg[2 m_{cd}^2 m_{ed}^2\left[C_{D^2\phi^4}\right]_{ceab}
    \nonumber \\
    &+2\left(m_{cd}^2 m_{ea}^2 + t_e h_{cda}\right)\left(\left[C_{D^2\phi^4}\right]_{edcb}+\left[C_{D^2\phi^4}\right]_{ecdb}-\left[C_{D^2\phi^4}\right]_{dceb}\right)\nonumber \\
    &-4m^2_{cb} m^2_{de}
    \left(
    \left[C_{D^2\phi^4}\right]_{dcae}-\left[C_{D^2\phi^4}\right]_{deac}
    \right)
    \Bigg]\,.\label{eq:phi2<-D2phi4}
\end{align}

\paragraph{\texorpdfstring{$\phi^2 \leftarrow \phi F^2 \times \phi F^2$}{phi2fromphiF2xphiF2}:}

\begin{align}
    \dot{(m^2)}_{ab} &= 4\left(t_{c}h_{abd}+ m^2_{ac} m^2_{bd}\right) \sum_{\alpha=1}^{N_G}\sum_{\beta=1}^\alpha \mathcal{S}_{\alpha\beta} \left[C_{\phi F^2}\right]_{c}^{A_\alpha B_\beta} \left[C_{\phi F^2}\right]_{d}^{A_\alpha B_\beta}\nonumber \\
    &+ \left(\left[C_{\phi F^2}\right]\to \left[C_{\phi \tildee F^2}\right]\right)\,.
    \label{eq:phi2<-phiF2xphiF2}
\end{align}

\paragraph{\texorpdfstring{$\phi^2 \leftarrow \phi F^2$}{phi2fromphiF2}:}

\begin{align}
    \dot{(m^2)}_{ab} &=
    -24\sum_{\sigma(\{a,b\})} \sum_{\alpha=1}^{N_G} \sum_{\beta=1}^{\alpha} g_\alpha g_\beta t_{c} Y_{ac}^{A_\alpha B_\beta} \left[C_{\phi F^2}\right]_{b}^{A_\alpha B_\beta}\,.
    \label{eq:phi2<-phiF2}
\end{align}

\subsubsection{Tadpole}

\paragraph{\texorpdfstring{$\phi \leftarrow D^2\phi^4$}{phifromD2phi4}:}

\begin{align}
    \dot{t}_{a} &= 
    2 m^2_{bc} t_d \left(
    \left[C_{D^2\phi^4}\right]_{cdab}-2\left[C_{D^2\phi^4}\right]_{cbad}
    \right)\,.
\end{align}

\paragraph{\texorpdfstring{$\phi \leftarrow \phi F^2 \times \phi F^2$}{phifromphiF2xphiF2}:}

\begin{align}
    \dot{t}_{a} &= 4 t_{b } m^2_{ac} \sum_{\alpha=1}^{N_G}\sum_{\beta=1}^\alpha \mathcal{S}_{\alpha\beta}\left( \left[C_{\phi F^2}\right]_{b}^{A_\alpha B_\beta} \left[C_{\phi F^2}\right]_{c}^{A_\alpha B_\beta} + \left[C_{\phi \tildee F^2}\right]_{b}^{A_\alpha B_\beta} \left[C_{\phi \tildee F^2}\right]_{c}^{A_\alpha B_\beta}\right)\,.
\end{align}

\subsubsection{Vacuum energy}

\paragraph{\texorpdfstring{$\Lambda \leftarrow \phi F^2 \times \phi F^2$}{LambdafromphiF2xphiF2}:}

\begin{align}
    \dot{\Lambda} &= 2 t_{a} t_b \sum_{\alpha=1}^{N_G}\sum_{\beta=1}^\alpha \mathcal{S}_{\alpha\beta} \left( \left[C_{\phi F^2}\right]_{b}^{A_\alpha B_\beta} \left[C_{\phi F^2}\right]_{c}^{A_\alpha B_\beta} + \left[C_{\phi \tildee F^2}\right]_{b}^{A_\alpha B_\beta} \left[C_{\phi \tildee F^2}\right]_{c}^{A_\alpha B_\beta}\right)\,.
\end{align}

\section{Comparison with existing literature}
\label{sec:comp}

This section applies the results of the generic EFT to the specific cases of different prominent examples. First, we rederive the RGEs for the SMEFT bosonic sector. Second, we cross-check numerous results collected in the literature for the SMEFT extended by an additional axion-like particle ($a+$SMEFT). Our purpose is twofold: We cross-check our results against the ones from Alonso, Jenkins, Manohar, and Trott~\cite{Jenkins:2013zja,Alonso:2013hga} and secondly, we illustrate how to use our results to derive RGEs of other EFTs with arbitrary symmetry and particle content.   

To use the derived RGEs, we first specify how to obtain the generators acting on the real scalar field constructed out of degrees of freedom contained in a complex scalar field. 
Let us start with a complex scalar $\varphi$ transforming as 
\begin{equation}
    \varphi \rightarrow \exp(i\sum_{\alpha=1}^{N_G}\varepsilon^{A_{\alpha}} t^{A_\alpha})\varphi\,,
\end{equation}
under the gauge factor $G_\alpha$. Next, we identify the transformation of the \emph{real} scalar field 
\begin{equation}
    \phi = \begin{pmatrix}
        \Re (\varphi) \\
        \Im (\varphi)
    \end{pmatrix}\,,\quad \phi \rightarrow \exp(i\sum_{\alpha=1}^{N_G}\varepsilon^{A_{\alpha}} \theta^{A_\alpha})\phi\,,
\end{equation}
implying the following form of the generators $\theta^{A_\alpha}$ 
\begin{equation}
    \theta^{A_\alpha} = i\begin{pmatrix}
        \Im (t^{A_\alpha}) & \Re(t^{A_\alpha}) \\
        -\Re (t^{A_\alpha}) & \Im(t^{A_\alpha})
    \end{pmatrix}\,.
    \label{eq:real_generators}
\end{equation}
The generators $\theta^{A_\alpha}$ are therefore purely imaginary and antisymmetric.

\subsection{SMEFT}
\label{subsec:SMEFT_RGEs}

In the case of the Standard Model with the gauge group 
\begin{equation}
    G_3\times G_2\times G_1 = SU(3)_c\times SU(2)_L\times U(1)_Y\,,
\end{equation}
the only scalar particle is the SM Higgs 
\begin{equation}
    H = \frac{1}{\sqrt 2}\begin{pmatrix}
        \phi_1 + i \phi_3 \\
        \phi_2 + i \phi_4
    \end{pmatrix}\,,
    \label{eq:cmplx_Higgs}
\end{equation}
which transforms as $(\textbf{1}, \textbf{2}, 1/2)$ under $G_3\times G_2\times G_1$. The Lagrangian of the bosonic sector reads
\begin{equation}
    \mathcal{L}_{\rm SM} =  -\frac{1}{4} G_{\mu\nu}^A G^{A\mu\nu} -\frac{1}{4} W_{\mu\nu}^I W^{I\mu\nu} -\frac{1}{4} B_{\mu\nu} B^{\mu\nu} + (D_\mu H)^\dagger (D^\mu H) - \lambda \left(H^\dagger H - \frac{v^2}{2}\right)^2\,,
\end{equation}
and the covariant derivative is defined as 
\begin{equation}
    D_\mu H = \left(\partial_\mu - i g_2 \frac{\tau^I}{2} W_{\mu}^I - i g_1 y_h B_\mu\right) H\,,
\end{equation}
where $\tau^I$ are the Pauli matrices, and $2y_h = \mathbb{1}$. Using Eq.~\eqref{eq:real_generators}, the generators acting on the real scalar field $\phi = (\phi_1\,, \phi_2\,, \phi_3\,, \phi_4)^\intercal$ constructed out of four degrees of freedom contained in the SM Higgs field in Eq.~\eqref{eq:cmplx_Higgs} read
\begin{equation}
    \theta^h = \frac{i}{2} \begin{pmatrix}
        0 & \mathbb{1} \\
        -\mathbb{1} & 0
    \end{pmatrix}\,,\quad \theta^{I} = \frac{i}{2} \begin{pmatrix}
         \Im(\tau^I) & \Re (\tau^I) \\
         - \Re(\tau^I) & \Im (\tau^I)
    \end{pmatrix}\,,
    \label{eq:theta_h_and_I}
\end{equation} 
with $I=1,2,3$, and $h$ being the label of the hypercharge generator. The covariant derivative of $\phi$ is
\begin{equation}
    (D_\mu \phi)_a = \left(\partial_\mu \delta_{ab} - i g_2 \theta^I_{ab} W_{\mu}^I - i g_1 \theta^h_{ab} B_\mu\right) \phi_b\,,
\end{equation}
and the Lagrangian expressed in terms of the $\phi$ field reads
\begin{equation}
    \mathcal{L}_{\rm SM} \supset  \frac{1}{2} (D_\mu \phi)_a(D^\mu \phi)_a + \frac{\lambda v^2}{2} \delta_{ab}\, \phi_a \phi_b - \frac{\lambda}{4} \delta_{(ab}\delta_{cd)}\, \phi_a\phi_b\phi_c\phi_d\,.
\end{equation}
Hence, comparison with the Lagrangian parameters in Eq.~\eqref{eq:L4} leads to the relations
\begin{align}
    m_{ab}^2 & = - \lambda v^2 \delta_{ab} = -\frac{1}{2}m_H^2 \delta_{ab}\label{eq:mass_SM}\,,\\
    \lambda_{abcd} &= \frac{4!}{4} \lambda\delta_{(ab}\delta_{cd)} = 2\lambda (\delta_{ab}\delta_{cd}+\delta_{ad}\delta_{bc}+\delta_{ac}\delta_{bd})\,,\label{eq:quartic_SM}
\end{align}
with the real scalar indices $a,b,c,d$ running from 1 to 4. 

We write the dimension-six SMEFT Lagrangian in the Warsaw basis~\cite{Grzadkowski:2010es} as
\begin{equation}
    \mathcal{L}^{(6)} = \sum_i C_i \mathcal{O}_i\,,
\end{equation}
where the bosonic operators $\mathcal{O}_i$ are collected in Tab.~\ref{tab:SMEFT_bosonic_op}. After expanding the Higgs field in terms of its real degrees of freedom, one can perform the translation to the dimension-six operators in Sec.~\ref{subsec:dim6_general} to obtain the appropriate Clebsch-Gordan coefficients that multiply the SMEFT Wilson coefficients. These Clebsch-Gordan coefficients are then used in combination with the results in Sec.~\ref{sec:results} to obtain the beta functions for the SMEFT Wilson coefficients.
\renewcommand{\arraystretch}{1.4}
\begin{table}[t]
\begin{centering}
\begin{tabular}{|c|c|c|c|}
\hline 
\multicolumn{2}{|c|}{\textbf{\large{}$\boldsymbol{F^{3}}$}} & \multicolumn{2}{c|}{{\large{}$\text{\ensuremath{\boldsymbol{\phi^{2} F^{2}}}}$}}\tabularnewline
\hline 
$\mathcal{O}_{G}$  & $f^{ABC}G_{\mu}^{A\nu}G_{\nu}^{B\rho}G_{\rho}^{C\mu}$  & $\mathcal{O}_{HG}$  & $G_{\mu\nu}^{A}G^{A\mu\nu}(H^{\dagger}H)$\tabularnewline
$\mathcal{O}_{\widetilde{G}}$  & $f^{ABC}\widetilde{G}_{\mu}^{A\nu}G_{\nu}^{B\rho}G_{\rho}^{C\mu}$  & $\mathcal{O}_{H\widetilde{G}}$  & $\widetilde{G}_{\mu\nu}^{A}G^{A\mu\nu}(H^{\dagger}H)$ \tabularnewline
$\mathcal{O}_{W}$  & $\epsilon^{IJK}W_{\mu}^{I\nu}W_{\nu}^{J\rho}W_{\rho}^{K\mu}$  & $\mathcal{O}_{HW}$  & $W_{\mu\nu}^{I}W^{I\mu\nu}(H^{\dagger}H)$ \tabularnewline
$\mathcal{O}_{\widetilde{W}}$  & $\epsilon^{IJK}\widetilde{W}_{\mu}^{I\nu}W_{\nu}^{J\rho}W_{\rho}^{K\mu}$  & $\mathcal{O}_{H\widetilde{W}}$  & $\widetilde{W}_{\mu\nu}^{I}W^{I\mu\nu}(H^{\dagger}H)$\tabularnewline
\cline{1-2} \cline{2-2} 
\multicolumn{2}{|c|}{{\large{}$\boldsymbol{D^{2}\phi^{4}}$}} & $\mathcal{O}_{HB}$  & $B_{\mu\nu}B^{\mu\nu}(H^{\dagger}H)$\tabularnewline
\cline{1-2} \cline{2-2} 
$\mathcal{O}_{H\Box}$  & $(H^\dagger H)\Box (H^\dagger H)$ & $\mathcal{O}_{H\widetilde{B}}$  & $\widetilde{B}_{\mu\nu}B^{\mu\nu}(H^{\dagger}H)$\tabularnewline
$\mathcal{O}_{HD}$  & $|H^\dagger D_\mu H|^2$ & $\mathcal{O}_{HWB}$  & $W_{\mu\nu}^I B^{\mu\nu}(H^{\dagger}\tau^I H)$
\tabularnewline
\cline{1-2} \cline{2-2} 
\multicolumn{2}{|c|}{{\large{}$\boldsymbol{\phi^{6}}$}} & $\mathcal{O}_{H\widetilde{W}B}$  & $\widetilde{W}_{\mu\nu}^I B^{\mu\nu}(H^{\dagger}\tau^I H)$\tabularnewline
\cline{1-2} \cline{2-2} 
$\mathcal{O}_{H}$  & $(H^\dagger H)^3$ &  & \tabularnewline
\hline
\end{tabular}
\caption{The bosonic dimension-six SMEFT operators in the Warsaw basis.}
\label{tab:SMEFT_bosonic_op}
\end{centering}
\end{table}
\renewcommand{\arraystretch}{1.0}

For example, consider the operator $O_{HWB}$ after the Higgs field has been expanded in terms of real fields $\phi_a$. The relevant term in the Lagrangian reads
\begin{equation}
    C_{HWB} W_{\mu\nu}^I B^{\mu\nu} (H^\dagger \tau^I H) = \frac{1}{2}C_{HWB} \Sigma_{ab}^I  \phi_a \phi_b W_{\mu\nu}^I B^{\mu\nu}\,,
\end{equation}
such that
\begin{equation}
\left[C_{\phi^2 F^2}\right]_{ab}^{A_\alpha B_\beta} = \left[C_{HWB}\right]_{ab}^I = \frac{1}{2} C_{HWB} \Sigma_{ab}^I\,,
\end{equation}
where $G_\alpha = G_2 = SU(2)_L$ and $G_\beta=G_1 = U(1)_Y$. The resulting Clebsch-Gordan coefficients, with the Higgs field defined in Eq.~\eqref{eq:cmplx_Higgs}, read
\begin{equation}
    \Sigma^1 = \begin{pmatrix}
        0 & 1 & 0 & 0\\
        1 & 0 & 0 & 0\\
        0 & 0 & 0 & 1\\
        0 & 0 & 1 & 0
    \end{pmatrix}\,,\quad  \Sigma^2 = \begin{pmatrix}
        0 & 0 & 0 & 1\\
        0 & 0 & -1 & 0\\
        0 & -1 & 0 & 0\\
        1 & 0 & 0 & 0
    \end{pmatrix}\,,\quad  \Sigma^3 = \begin{pmatrix}
        1 & 0 & 0 & 0\\
        0 & -1 & 0 & 0\\
        0 & 0 & 1 & 0\\
        0 & 0 & 0 & -1
    \end{pmatrix}\,.
    \label{eq:Sigma_matrices}
\end{equation}
Once the Clebsch-Gordan coefficients for all operators are known, one can use the generic EFT anomalous dimensions to obtain the running. 

The common ingredient for all RGEs are the collinear anomalous dimensions. Using Eq.~\eqref{eq:coll_anom_dim},
one finds the results for the SM vector and scalar fields at lowest order in the EFT expansion
\begin{align}
    \gamma_{c,G} &= - g_3^2\, b_{0,3}\,,
    \label{eq:gammaG}\\
    \gamma_{c,W} &= - g_2^2\, b_{0,2}\,,
    \label{eq:gammaW}\\
    \gamma_{c,B} &= - g_1^2\, b_{0,1}\,,
    \label{eq:gammaB}\\
    \gamma_{c,H} &= - g_1^2 - 3g_2^2\,,
    \label{eq:gammaH}
\end{align}
where $\theta^h_{ac}\theta^h_{cb}=\delta_{ab}/4$ and $\theta^I_{ac}\theta^I_{cb}=3\delta_{ab}/4$ was used.
Dimension-six terms in the collinear anomalous dimensions are present, but they only affect the running of the renormalizable couplings, and we have already included them in the corresponding RGEs.

In the following subsections, the beta functions for the Wilson coefficients of the operators in Tab.~\ref{tab:SMEFT_bosonic_op} will be derived. To make the results more transparent, we study each operator class independently. 

\subsubsection{\texorpdfstring{$D^2\phi^4$}{D2phi4} class}
In the Warsaw basis, the two independent $D^2\phi^4$-class operators are
\begin{equation}
\mathcal O_{HD} = (H^\dagger D_\mu H)^*(H^\dagger D^\mu H)\,,
\qquad
    \mathcal O_{H\Box} = (H^\dagger H)\Box (H^\dagger H)\,.
\end{equation}
The former, using the identity $\delta_{i\ell}\delta_{kj}=(\tau^I_{ij}\tau^I_{k\ell}+\delta_{ij}\delta_{k\ell})/2$, can be written as
\begin{equation}\label{eq:OHD}
    \mathcal O_{HD} = \frac{1}{2}\mathcal O_\parallel+\frac{1}{2}\mathcal O_\perp\,,
\end{equation}
where we defined
\begin{equation}
    \mathcal O_\parallel = (H^\dagger H)[(D_\mu H)^\dagger (D^\mu H)] \,,
    \qquad
    \mathcal O_\perp = (H^\dagger \tau^I H)[(D_\mu H)^\dagger \tau^I (D^\mu H)]\,.
\end{equation}
After a field redefinition of the Higgs field, the box operator can be recast as
\begin{equation}\label{eq:OHBox}
    \mathcal O_{H\Box} = 2 \mathcal O_\parallel + \dotsc\,,
\end{equation}
where the dots indicate operators that are not included in the $D^2\phi^4$ class and do not affect this cross-check. 
We can use $\mathcal O_\parallel$ and $\mathcal O_\perp$ as a basis since they naturally arise from the definition of $\mathcal O_{D^2\phi^4}$ via
\begin{equation}
    \left[C_{D^2\phi^4}\right]_{abcd} = \frac{1}{4}C_\parallel \delta_{ab}\delta_{cd}+\frac{1}{4}C_\perp \Sigma^I_{ab}\Sigma^I_{cd}\,,
\end{equation}
or, using the definition in Eq.~\eqref{eq:hatCD2phi4}, 
\begin{equation}
    \left[\widehat C_{D^2\phi^4}\right]_{abcd} = \frac{1}{2}C_\parallel\left( \delta_{ab}\delta_{cd}-\delta_{ad}\delta_{bc}\right)+\frac{1}{2}C_\perp \left( \Sigma^I_{ab}\Sigma^I_{cd}-\Sigma^I_{ad}\Sigma^I_{bc}\right)\,.
\end{equation}
The results obtained in this basis are then related to the ones in the Warsaw basis via the following rotation
\begin{equation}
    \begin{pmatrix}
    C_\parallel \\ C_\perp
\end{pmatrix} = \mathbf{C} \begin{pmatrix}
    C_{H\Box} \\ C_{HD}
\end{pmatrix}\,, \qquad
\mathbf{C}=\begin{pmatrix}
    2 & \frac{1}{2} \\ 0 & \frac{1}{2}
\end{pmatrix}\,,
\end{equation}
which follows from Eqs.~\eqref{eq:OHD} and \eqref{eq:OHBox}. 

\paragraph{\texorpdfstring{$D^2\phi^4 \leftarrow D^2\phi^4$}{D2phi4fromD2phi4}:}
The self-renormalization of $\mathcal O_{D^2\phi^4}$ described in Eq.~\eqref{eq:D2phi4<-D2phi4} receives contributions which are proportional to $\lambda$ and $g^2$.
We analyze them separately in the following.

\subparagraph{Lambda dependence:}
As stated previously, we have $\lambda_{abcd}=\frac{4!}{4}\lambda \delta_{(ab}\delta_{cd)}= 2\lambda (\delta_{ab}\delta_{cd}+\delta_{ad}\delta_{bc}+\delta_{ac}\delta_{bd})$.
Using $\Tr(\Sigma^I)=0$ and $\Sigma^I \Sigma^I = 3 \mathbb{1}$, the terms of Eq.~\eqref{eq:D2phi4<-D2phi4} proportional to $\lambda$ are
\begin{align}
    \left[\dot{\widehat C}_{D^2\phi^4}\right]_{abcd} 
    &=\lambda_{cdef}
    \left[\widehat C_{D^2\phi^4}\right]_{abef}
    - (a\leftrightarrow c) - (b\leftrightarrow d) + 
    \binom{a\leftrightarrow c}{b\leftrightarrow d}\nonumber \\
    &= 6\lambda \Big[
    \left(2C_\parallel -  C_\perp\right) \left(\delta_{ab}\delta_{cd}-\delta_{ad}\delta_{bc}\right)
    + C_\perp \left(\Sigma^I_{ab}\Sigma^I_{cd}-\Sigma^I_{ad}\Sigma^I_{bc}\right)
    \Big]\,,
\end{align}
from which we can read
\begin{equation}
     \begin{pmatrix}
    \dot C_\parallel \\ \dot C_\perp
\end{pmatrix}
=
 \begin{pmatrix}
    24\lambda & -12\lambda \\
    0 & 12\lambda
\end{pmatrix}
 \begin{pmatrix}
    C_\parallel \\ C_\perp
\end{pmatrix}\,.
\end{equation}
In terms of $C_{H\Box}$ and $C_{HD}$, this is equivalent to
\begin{align}
    \begin{pmatrix}
    \dot C_{H\Box} \\ \dot C_{HD}
\end{pmatrix}
    &= \mathbf{C}^{-1}
    \begin{pmatrix}
    24\lambda & -12\lambda \\
    0 & 12\lambda
\end{pmatrix}
    \mathbf{C}
    \begin{pmatrix}
     C_{H\Box} \\  C_{HD}
    \end{pmatrix}
    = 
    \begin{pmatrix}
    24\lambda & 0\\
    0 & 12\lambda
    \end{pmatrix}
    \begin{pmatrix}
     C_{H\Box} \\  C_{HD}
    \end{pmatrix}\,.\label{eq:dotCHBoxandCHDlambda}
\end{align}

\subparagraph{Gauge coupling dependence:}
As stated previously, we have $T^{(1)}_{abcd}=\theta^h_{ab}\theta^h_{cd}$ and $T^{(2)}_{abcd}=\theta^I_{ab}\theta^I_{cd}$.
Exploiting the identities associated with the multiplication among these matrices collected in Eqs.~\eqref{eq:matrixid1} and \eqref{eq:matrixid2} and using the expression of the collinear anomalous dimension of the Higgs field $\gamma_{c,H} = -g_1^2-3g_2^2$, the terms of Eq.~\eqref{eq:D2phi4<-D2phi4} proportional to $g^2$ are
\begin{align}
    \left[\dot{\widehat C}_{D^2\phi^4}\right]_{abcd}
    &=
    -\frac{1}{6}\sum_{\alpha=1}^{2}g_\alpha^2 
    \bigg\{
    \bigg[
    \left(T^\alpha_{cdef}+24T^\alpha_{cedf}
    \right)\left[\widehat C_{D^2\phi^4}\right]_{aebf} \nonumber \\
    &\quad+ 6\left(T^\alpha_{cedf}+T^\alpha_{cfde}\right)\left[\widehat C_{D^2\phi^4}\right]_{abef} - (a\leftrightarrow c) - (b\leftrightarrow d) + 
    \binom{a\leftrightarrow c}{b\leftrightarrow d}
    \bigg] \nonumber \\
    &\quad+ \bigg[
    2\left(T^\alpha_{bdef}+24T^\alpha_{bedf}
    \right)\left[\widehat C_{D^2\phi^4}\right]_{aecf} + 
    \binom{a\leftrightarrow c}{b\leftrightarrow d}
    \bigg]
    \bigg\} \nonumber \\
    &\quad+4 \gamma_{c,H}\left[\widehat C_{D^2\phi^4}\right]_{abcd}\nonumber \\
    &=
    \frac{1}{6}g_1^2 \Big[
    C_\parallel \Big(
    9(\delta_{ad}\delta_{bc}-\delta_{ab}\delta_{cd})-20(\theta^h_{ad}\theta^h_{bc}+2\theta^h_{ac}\theta^h_{bd}+\theta^h_{ab}\theta^h_{cd} )
    \Big)  \\
    &\quad-3 C_\perp \Big(
    6(\theta^I_{ad}\theta^I_{bc}+2\theta^I_{ac}\theta^I_{bd}+\theta^I_{ab}\theta^I_{cd} )
    +2 (\theta^h_{ad}\theta^h_{bc}+2\theta^h_{ac}\theta^h_{bd}+\theta^h_{ab}\theta^h_{cd} )\nonumber \\
    &\quad-3(\Sigma^I_{ad}\Sigma^I_{bc}-\Sigma^I_{ab}\Sigma^I_{cd})
    \Big)
    \Big]\nonumber \\
    &\quad +\frac{1}{6}g_2^2 \Big[
    C_\parallel \Big(
    27(\delta_{ad}\delta_{bc}-\delta_{ab}\delta_{cd})
    -20 (\theta^I_{ad}\theta^I_{bc}+2\theta^I_{ac}\theta^I_{bd}+\theta^I_{ab}\theta^I_{cd} )
    \Big) \nonumber \\
    &\quad+2 C_\perp \Big(
    (\theta^I_{ad}\theta^I_{bc}+2\theta^I_{ac}\theta^I_{bd}+\theta^I_{ab}\theta^I_{cd} )
    -27 (\theta^h_{ad}\theta^h_{bc}+2\theta^h_{ac}\theta^h_{bd}+\theta^h_{ab}\theta^h_{cd} )
    \Big)
    \Big]\nonumber\,.
\end{align}
Then, we can use the following two identities
\begin{align}
    \theta^h_{ad}\theta^h_{bc}+2\theta^h_{ac}\theta^h_{bd}+\theta^h_{ab}\theta^h_{cd}&=\frac{1}{2}\left(\delta_{ad}\delta_{bc}-\delta_{ab}\delta_{cd}\right)+\frac{1}{4}\left(\Sigma^I_{ad}\Sigma^I_{bc}-\Sigma^I_{ab}\Sigma^I_{cd}\right) \,,\\
    \theta^I_{ad}\theta^I_{bc}+2\theta^I_{ac}\theta^I_{bd}+\theta^I_{ab}\theta^I_{cd} &= \frac{3}{4} \left(\delta_{ad}\delta_{bc}-\delta_{ab}\delta_{cd}\right)\,,
\end{align}
in order to further simplify the expression:
\begin{align}
    \left[\dot{\widehat C}_{D^2\phi^4}\right]_{abcd} &= \frac{1}{12}g_1^2 \Big[
    \big(2C_\parallel + 33 C_\perp\big) \big(\delta_{ab}\delta_{cd}-\delta_{ad}\delta_{bc}\big)
    +5 \big(2C_\parallel - 3 C_\perp\big)\big(\Sigma^I_{ab}\Sigma^I_{cd}-\Sigma^I_{ad}\Sigma^I_{bc}\big)
    \Big]\nonumber \\
    &+\frac{1}{4}g_2^2 \Big[
    \big( -8C_\parallel + 17 C_\perp\big) \big(\delta_{ab}\delta_{cd}-\delta_{ad}\delta_{bc}\big)
    +9 C_\perp \big(\Sigma^I_{ab}\Sigma^I_{cd}-\Sigma^I_{ad}\Sigma^I_{bc}\big)
    \Big]\,,
\end{align}
from which one finds
\begin{equation}
     \begin{pmatrix}
    \dot C_\parallel \\ \dot C_\perp
\end{pmatrix}
=
 \begin{pmatrix}
    \frac{1}{3}g_1^2 - 4g_2^2 & \frac{11}{2}g_1^2+\frac{17}{2}g_2^2 \\
    \frac{5}{3}g_1^2 & -\frac{5}{2}g_1^2+\frac{9}{2}g_2^2
\end{pmatrix}
 \begin{pmatrix}
    C_\parallel \\ C_\perp
\end{pmatrix}\,.
\end{equation}
In terms of $C_{H\Box}$ and $C_{HD}$, this is equivalent to
\begin{align}
    \begin{pmatrix}
    \dot C_{H\Box} \\ \dot C_{HD}
\end{pmatrix}
    &= \mathbf{C}^{-1}
    \begin{pmatrix}
    \frac{1}{3}g_1^2 - 4g_2^2 & \frac{11}{2}g_1^2+\frac{17}{2}g_2^2 \\
    \frac{5}{3}g_1^2 & -\frac{5}{2}g_1^2+\frac{9}{2}g_2^2
\end{pmatrix}
    \mathbf{C}
    \begin{pmatrix}
     C_{H\Box} \\  C_{HD}
    \end{pmatrix}
    \nonumber \\
    &=
    \begin{pmatrix}
    -\frac{4}{3}g_1^2 - 4g_2^2 & \frac{5}{3}g_1^2\\
    \frac{20}{3}g_1^2 & -\frac{5}{6}g_1^2+\frac{9}{2}g_2^2
    \end{pmatrix}
    \begin{pmatrix}
     C_{H\Box} \\  C_{HD}
    \end{pmatrix}\,.\label{eq:dotCHBoxandCHDgc}
\end{align}
The same result is obtained using Eq.~\eqref{eq:D2phi4<-D2phi4_offshell} for the running of the full WC, including the off-shell contributions.
Indeed, $[C_{D^2\phi^4}]_{abcd}$ in the Warsaw basis is given by
\begin{equation}
    \left[C_{D^2\phi^4}\right]_{abcd} = -\frac{1}{2}C_{H\Box} (\delta_{ac}\delta_{bd}+\delta_{ad}\delta_{bc})+\frac{1}{8}C_{HD}(\delta_{ab}\delta_{cd}+\Sigma^I_{ab}\Sigma^I_{cd})\,.
\end{equation}
The RHS of Eq.~\eqref{eq:D2phi4<-D2phi4_offshell} gives then
\begin{align}
    \frac{1}{48}\Big[&
    2\Big(
    8C_{H\Box}(2g_1^2+11g_2^2-6\lambda)-C_{HD}(11g_1^2+27g_2^2+24\lambda)
    \Big)(\delta_{ac}\delta_{bd}+\delta_{ad}\delta_{bc})\nonumber \\
    &+ \Big(40 C_{H\Box}(g_1^2+2g_2^2+12\lambda)
    +C_{HD}(13g_1^2-27g_2^2+24\lambda)
    \Big)\delta_{ab}\delta_{cd}\nonumber \\
    &+
    \Big(
    40 C_{H\Box}g_1^2
    +C_{HD}(-5g_1^2+27g_2^2+72\lambda)
    \Big)\Sigma^I_{ab}\Sigma^I_{cd}
    \Big]\,,
\end{align}
which generates the correct running of $C_{H\Box}$ and $C_{HD}$ in Eqs.~\eqref{eq:dotCHBoxandCHDlambda} and \eqref{eq:dotCHBoxandCHDgc}, plus
\begin{align}
    \left[\dot{\tildee C}_{D^2\phi^4}\right]_{abcd} &= 
    \frac{1}{24}\Big[
    40 C_{H\Box}(g_2^2+6\lambda)+3C_{HD}(3g_1^2 -9g_2^2-8\lambda)
    \Big](\delta_{ab}\delta_{cd}+\delta_{ac}\delta_{bd}+\delta_{ad}\delta_{bc})\,,\label{eq:dotCtildeSMEFT}\\
    \left[\dot{\overline C}_{D^2\phi^4}\right]_{abcd} &=0\label{eq:dotCbarSMEFT}\,,
\end{align}
which modify the running of $C_H$ and $\lambda$.

\subsubsection{\texorpdfstring{$\phi^6$}{phi6} class}

The only $\phi^6$-class operator in the SMEFT is $\mathcal O_H = (H^\dagger H)^3$, whose WC is recast in the real scalar basis as
\begin{equation}
    \left[C_{\phi^6}\right]_{abcdef}=\frac{1}{8}C_H \delta_{(ab}\delta_{cd}\delta_{ef)}=\frac{1}{8}C_H  \frac{1}{6!}\sum_{\sigma(\{a,b,c,d,e,f\})}\delta_{ab}\delta_{cd}\delta_{ef}\,.
\end{equation}

\paragraph{\texorpdfstring{$\phi^6\leftarrow \phi^6$}{phi6tophi6}:}
This anomalous dimension, reported in Eq.~\eqref{eq:phi6<-phi6}, receives contributions which are proportional to $\lambda$ and $g^2$.
They are analyzed separately as follows.

\subparagraph{Lambda dependence:}
The Higgs quartic $\lambda$ is recast in the real basis in Eq.~\eqref{eq:quartic_SM}. From Eq.~\eqref{eq:phi6<-phi6} it follows that
\begin{align}
    \left[\dot C_{\phi^6}\right]_{abcdef} &= 
    \frac{1}{2!4!}\sum_{\sigma(\{a,b,c,d,e,f\})}\lambda_{efgh} \left[C_{\phi^6}\right]_{abcdgh}\nonumber \\
    &= \frac{27}{2} \lambda  C_H \delta_{(ab}\delta_{cd}\delta_{ef)}\,,
\end{align}
where we used the identity
\begin{equation}
    \sum_{\sigma(\{a,b,c,d,e,f\})}\delta_{(ef}\delta_{gh)}\delta_{(ab}\delta_{cd}\delta_{gh)} = 48(14+N)\delta_{(ab}\delta_{cd}\delta_{ef)} = 864 \,\delta_{(ab}\delta_{cd}\delta_{ef)}\,,
    \label{eq:identity_permutations}
\end{equation}
for $N=\delta_{gg}=4$.
Therefore,
\begin{equation}
    \dot C_H = 108 \lambda C_H\,.
\end{equation}

\subparagraph{Gauge coupling dependence:}
For the generators $\,\theta^h$ and $\,\theta^I$ associated with the gauge groups $U(1)_Y$ and $SU(2)_L$, respectively, one finds $T^{(1)}_{abcd} = \theta^h_{ab}\theta^h_{cd}$ and $T^{(2)}_{abcd}=\theta^I_{ab}\theta^I_{cd}$.
The terms in Eq.~\eqref{eq:phi6<-phi6} proportional to the gauge couplings are
\begin{align}
    \left[\dot C_{\phi^6}\right]_{abcdef} &=\frac{1}{2!4!}\sum_{\sigma(\{a,b,c,d,e,f\})}\sum_{\alpha=1}^2 2 g_\alpha^2 T^\alpha_{eghf}\left[ C_{\phi^6}\right]_{abcdgh} + 6 \gamma_{c,H}\left[ C_{\phi^6}\right]_{abcdef}\nonumber\\
    &= \frac{3}{16}(g_1^2+3g_2^2)C_H \delta_{(ab}\delta_{cd}\delta_{ef)} + 6 (-g_1^2-3g_2^2)\frac{1}{8}C_H  \delta_{(ab}\delta_{cd}\delta_{ef)} \nonumber\\
    &= -\frac{9}{16}(g_1^2+3g_2^2)C_H \delta_{(ab}\delta_{cd}\delta_{ef)}\,,
\end{align}
where $\theta^h_{ac}\theta^h_{cb} = \delta_{ab}/4$, $\theta^I_{ac}\theta^I_{cb} = 3\delta_{ab}/4$, and $\gamma_{c,H}=-g_1^2-3g_2^2$.
Therefore, one finds
\begin{equation}
    \dot C_H =  -\frac{9}{2}(g_1^2+3g_2^2)C_H \,.
\end{equation}

\paragraph{\texorpdfstring{$\phi^6 \leftarrow D^2\phi^4$}{phi6fromD2phi4}:}

The first contribution proportional to $g^4$ from Eq.~\eqref{eq:phi6<-D2phi4} is given by
\begin{align}
    \left[\dot C_{\phi^6}\right]_{abcdef} &= - \frac{6}{6!} \sum_{\sigma(\{a,b,c,d,e,f\})}[g_1^4 T^{(1)}_{fhei}T^{(1)}_{dicg} + g_1^2 g_2^2 (T^{(1)}_{fhei}T^{(2)}_{dicg}+T^{(2)}_{fhei}T^{(1)}_{dicg}) +g_2^4 T^{(2)}_{fhei}T^{(2)}_{dicg}]\nonumber \\
    &\quad \times 
    \left[C_{D^2\phi^4}\right]_{abgh}\,.
\end{align}
The pieces proportional to $g_1^4$, $g_1^2g_2^2$, and $g_2^4$  give, before symmetrization, the following results
    \begin{align}
    T^{(1)}_{fhei}T^{(1)}_{dicg} [C_{D^2\phi^4}]_{abgh}&=\frac{1}{128} \delta_{de} \left[C_{HD}\delta_{ab}\delta_{cf}+ C_{HD}\Sigma_{ab}^I\Sigma_{cf}^I\right.\\
    &\left.\qquad+ 16 C_{H\Box}\left(\theta^h_{af}\theta^h_{bc}+\theta^h_{ac}\theta^h_{bf}\right)\right]\,,\nonumber\\
    \left(T^{(1)}_{fhei}T^{(2)}_{dicg} +T^{(2)}_{fhei}T^{(1)}_{dicg}\right)[C_{D^2\phi^4}]_{abgh} &= \frac{1}{64} \Sigma_{de}^I \left[C_{HD}\left(\delta_{cf}\Sigma_{ab}^I + \delta_{ab}\Sigma_{cf}^I\right)\right.\\ &\left.+8C_{H\Box}\left(\theta_{bf}^I\theta_{ac}^h+\theta_{af}^I\theta_{bc}^h+\theta_{bc}^I\theta_{af}^h+\theta_{ac}^I\theta_{bf}^h\right)\right]\,,\nonumber\\
    T^{(2)}_{fhei}T^{(2)}_{dicg} [C_{D^2\phi^4}]_{abgh}&=\frac{1}{128}\Big[C_{HD} \delta _{a b} \left(3 \delta _{c f} \delta _{d e}-8 \theta _{c f}^{I} \theta _{d e}^{I}\right)\\ 
    &+\delta _{d e} \Big( 16 C_{H\Box} \left(\theta _{a f}^{I} \theta _{b c}^{I}+\theta _{a c}^{I} \theta _{b f}^{I}\right) 
    + C_{HD} \Sigma _{a b}^{I}\Sigma _{c f}^{I} \nonumber\\
    & -2C_{HD}\Sigma _{a b}^{I}\Sigma _{c f}^{I} \Big) 
    + 2i\theta _{d e}^{K} \Big( 4iC_{HD} \theta _{c f}^h\Sigma _{a b}^{K}\nonumber\\
    &+ \epsilon ^{J K L} C_{HD}  \Sigma _{a b}^{L} \Sigma _{c f}^{J} + \epsilon ^{I J K} \Big(16 C_{H\Box} \left(\theta _{a f}^{I} \theta _{b c}^{J}+\theta _{a c}^{J} \theta _{b f}^{I}\right)\nonumber\\ 
    &+ \epsilon ^{I L M} \epsilon^{J M N} C_{HD} \Sigma _{a b}^{L} \Sigma _{c f}^{N} \Big) \Big) \Big]\,.\nonumber
    \end{align}
After symmetrizing the full expression, all terms involving at least one $\theta^h$ or $\theta^I$ matrix vanish since they are antisymmetric. This in turn implies that all terms proportional to $C_{H\Box}$ vanish. After exploiting the identity in Eq.~\eqref{eq:thetaI_thetaJ} the remaining contribution is given by
\begin{align}
    \left[\dot C_{\phi^6}\right]_{abcdef} = -\frac{3}{32}C_{HD}(g_1^2+g_2^2)^2 \delta_{(ab}\delta_{cd}\delta_{ef)}\,,
\end{align}
which yields
\begin{equation}
    \dot C_H = -\frac{3}{4}C_{HD}(g_1^2+g_2^2)^2\,.\label{eq:dotCH<-D2phi41}
\end{equation}
On the other hand, the terms of this anomalous dimension that are proportional to $\lambda^2$ and $\lambda g^2$ are affected by $[\dot{\widehat C}_{D^2\phi^4}]_{abcd}$ and $[\dot{\overline C}_{D^2\phi^4}]_{abcd}$, reported in Eqs.~\eqref{eq:dotCtildeSMEFT} and \eqref{eq:dotCbarSMEFT}, respectively.
From Eq.~\eqref{eq:dotCphi6<-Ctildeandbar} we then find
\begin{align}
    \left[C_{\phi^6}\right]_{abcdef} &= \frac{1}{6!}\frac{1}{3!}\sum_{\sigma(\{a,b,c,d,e,f\})}\lambda_{abcg}\left(\frac{1}{3}
    \left[\dot{\widehat C}_{D^2\phi^4}\right]_{gdef}+\left[\dot{\overline C}_{D^2\phi^4}\right]_{gdef}
    \right) \nonumber \\
    &= \left[
    \frac{5}{3}\lambda (g_2^2+6\lambda)C_{H\Box} + \frac{1}{8}\lambda (3g_1^2-9g_2^2-8\lambda)C_{HD}
    \right] \delta_{(ab}\delta_{cd}\delta_{ef)}\label{eq:dotCH<-D2phi42}\,.
\end{align}
This contribution needs to be added to the one that can be obtained from Eq.~\eqref{eq:phi6<-D2phi4}. Its RHS proportional to $\lambda^2$ and $\lambda g^2$ reads 
\begin{align}
    &\frac{1}{2}\frac{1}{6!}\sum_{\sigma(\{a,b,c,d,e,f\})}\lambda_{ghab}\bigg[\lambda_{ihcd}\left[ C_{D^2\phi^4}\right]_{gief}\nonumber \\
    &+\frac{1}{3}\lambda_{icde}\left(\left[C_{D^2\phi^4}\right]_{ihgf}+\left[C_{D^2\phi^4}\right]_{ighf}-\left[C_{D^2\phi^4}\right]_{hgif}\right)\bigg]\nonumber \\
    &-\frac{3}{6!} \sum_{\sigma(\{a,b,c,d,e,f\})}\sum_{\alpha=1}^{3} g_\alpha^2 \lambda_{abhi} T^\alpha_{cidg} \left[C_{D^2\phi^4}\right]_{ghef}\nonumber \\
    = & \left[\lambda^2(-30C_{H\Box}+7C_{HD})+\frac{3}{8}\lambda (g_1^2 +g_2^2)C_{HD}\right]\delta_{(ab}\delta_{cd}\delta_{ef)}\label{eq:dotCH<-D2phi43}\,.
\end{align}
Therefore, summing the contributions in Eqs.~\eqref{eq:dotCH<-D2phi41}, \eqref{eq:dotCH<-D2phi42}, and \eqref{eq:dotCH<-D2phi43} leads to the anomalous dimension 
\begin{equation}
    \dot C_H = -\frac{3}{4}(g_1^2+g_2^2)^2C_{HD} -16 \lambda^2(10 C_{H\Box }-3C_{HD}) + \frac{40}{3}\lambda g_2^2 C_{H\Box} + 6\lambda (g_1^2-g_2^2)C_{HD} \,.
\end{equation}

\paragraph{\texorpdfstring{$\phi^6 \leftarrow \phi^2 F^2$}{phi6fromphi2F2}:}
First, let us focus on the contributions proportional to $\lambda g^2$.
From Eq.~\eqref{eq:phi6<-phi2F2} one finds
\begin{align}
    \left[\dot C_{\phi^6}\right]_{abcdef} &= \frac{2}{6!}\sum_{\sigma(\{a,b,c,d,e,f\})}\lambda_{abcg} (g_1^2 \theta^h_{dh}\theta^h_{hg}\frac{1}{2}C_{HB}\delta_{ef}
    + g_1 g_2 \theta^I_{dh}\theta^h_{hg}\frac{1}{2}C_{HWB}\Sigma^I_{ef}\nonumber \\
    &\quad
    + g_2^2 \theta^I_{dh}\theta^I_{hg}\frac{1}{2}C_{HW}\delta_{ef})\nonumber \\
    &= \frac{\lambda}{1440}\sum_{\sigma(\{a,b,c,d,e,f\})}[(g_1^2 C_{HB}+3g_2^2C_{HW})(\delta_{ad}\delta_{bc}+\delta_{ac}\delta_{bd}+\delta_{ab}\delta_{cd})\delta_{ef} \nonumber \\
    &\quad+ g_1 g_2 C_{HWB}(\Sigma^I_{ad}\delta_{bc}+\delta_{ac}\Sigma^I_{bd}+\delta_{ab}\Sigma^I_{cd})\Sigma^I_{ef}]\nonumber \\
    &= \frac{3}{2}\lambda(g_1^2 C_{HB} + 3 g_2^2 C_{HW} + g_1 g_2 C_{HWB})\delta_{(ab}\delta_{cd}\delta_{ef)}\,,
\end{align}
where we used the identity in Eq.~\eqref{eq:thetaI_thetaJ} to eliminate the dependence on the $\Sigma^I$ matrices.
Thus,
\begin{equation}
    \dot C_H = 12 \lambda(g_1^2 C_{HB} + 3 g_2^2 C_{HW} + g_1 g_2 C_{HWB})\,.
\end{equation}
Regarding the contributions proportional to $g^4$, they are captured by the following piece of Eq.~\eqref{eq:phi6<-phi2F2}:
\begin{equation}
    \left[\dot C_{\phi^6}\right]_{abcdef} = - \frac{12}{6!}\sum_{\sigma(\{a,b,c,d,e,f\})} \sum_{\alpha=1}^{3}\sum_{\beta=1}^{\alpha}\sum_{\gamma=1}^{3} g_\alpha g_\beta g_\gamma^2 \,Y_{ce}^{A_\alpha C_\gamma} Y_{df}^{B_\beta C_\gamma} \left[C_{\phi^2 F^2}\right]_{ab}^{A_\alpha B_\beta}\,.
\end{equation}
This yields
\begin{align}
    \left[\dot C_{\phi^6}\right]_{abcdef} &=
    - \frac{12}{6!}\sum_{\sigma(\{a,b,c,d,e,f\})}
    \bigg[
    \frac{g_1^2}{2}C_{HB}\delta_{ab}(g_1^2 \theta^h_{cg}\theta^h_{ge}\theta^h_{dh}\theta^h_{hf}
    +
    g_2^2 \theta^h_{cg}\theta^I_{ge}\theta^h_{dh}\theta^I_{hf}) \\
    &\quad + 
    \frac{g_2^2}{2}C_{HW}\delta_{ab}\delta^{IJ}
    (g_1^2 \theta^I_{cg}\theta^h_{ge}\theta^J_{dh}\theta^h_{hf}
    +g_2^2 
    \theta^I_{cg}\theta^K_{ge}\theta^J_{dh}\theta^K_{hf}
    )\nonumber \\
    &\quad + \frac{g_1 g_2}{2}C_{HWB}\Sigma^I_{ab}(g_1^2 
    \theta^I_{cg}\theta^h_{ge}\theta^h_{dh}\theta^h_{hf}
    + g_2^2
    \theta^I_{cg}\theta^J_{ge}\theta^h_{dh}\theta^J_{hf}
    )
    \bigg] \nonumber \\
    &= 
    - \frac{12}{6!}\sum_{\sigma(\{a,b,c,d,e,f\})}
    \bigg[
    \frac{g_1^2}{32}C_{HB}\delta_{ab}(g_1^2 \delta_{ce}\delta_{df}+g_2^2 \Sigma^I_{ce}\Sigma^I_{df})\nonumber \\
    &\quad + \frac{g_2^2}{32}
    C_{HW}\delta_{ab}
    (g_1^2 \Sigma^I_{ce}\Sigma^I_{df}
    +g_2^2 (3\delta_{ce}\delta_{df}
    -8 \theta^I_{ce}\theta^I_{df}
    )
    )\nonumber \\
    &\quad +
    \frac{g_1 g_2}{32}C_{HWB}\Sigma^I_{ab}
    (
    g_1^2 \Sigma^I_{ce}\delta_{df}
    +
    g_2^2  (
    \delta_{ce}
    +
    2i \epsilon^{IJK}\theta^K_{ce}
    )\Sigma^J_{df}
    )
    \bigg]\nonumber \\
    &=-\frac{3}{8}\left[g_1^2(g_1^2+g_2^2)C_{HB} + g_2^2(g_1^2+3g_2^2)C_{HW} + g_1g_2 (g_1^2+g_2^2) C_{HWB}\right]\delta_{(ab}\delta_{cd}\delta_{ef)}\,,\nonumber
\end{align}
where we used the identities in Eqs.~\eqref{eq:matrixid1}, \eqref{eq:matrixid2}, and \eqref{eq:thetaI_thetaJ}.
The anomalous dimension of $C_H$ is then
\begin{equation}
    \dot C_H = -3 \left[g_1^2(g_1^2+g_2^2)C_{HB} + g_2^2(g_1^2+3g_2^2)C_{HW} + g_1g_2 (g_1^2+g_2^2) C_{HWB}\right]\,.
\end{equation}

\subsubsection{\texorpdfstring{$\phi^2 F^2$}{phi2F2} class}

The beta function for the operators in this class is given in Eq.~\eqref{eq:phi2F2_phi2F2}. As can be seen, the RGEs for the operators in this class receive contributions from themselves and the $F^3$ class. The associated Clebsch-Gordan coefficients are
\begin{align}
    \left[C_{HG}\right]_{ab}^{AB} &= \frac{1}{2} C_{HG}\, \delta_{ab}\,\delta^{AB}\,,&
    \left[C_{HW}\right]_{ab}^{IJ} &=\frac{1}{2} C_{HW} \,\delta_{ab}\,\delta^{IJ}\,,\\
    \left[C_{HB}\right]_{ab} &= \frac{1}{2} C_{HB} \,\delta_{ab}\,,&
    \left[C_{HWB}\right]_{ab}^I &= \frac{1}{2} C_{HWB}\, \Sigma_{ab}^I\,,
\end{align}
and similarly for the CP-odd counterparts
\begin{align}
    \left[C_{H\tildee G}\right]_{ab}^{AB} &= \frac{1}{2} C_{H\tildee G}\, \delta_{ab}\,\delta^{AB}\,,&
    \left[C_{H\tildee W}\right]_{ab}^{IJ} &=\frac{1}{2} C_{H\tildee W} \,\delta_{ab}\,\delta^{IJ}\,,\\
    \left[C_{H\tildee B}\right]_{ab} &= \frac{1}{2} C_{H\tildee B} \,\delta_{ab}\,,&
    \left[C_{H\tildee WB}\right]_{ab}^I &= \frac{1}{2} C_{H\tildee WB}\, \Sigma_{ab}^I\,.
\end{align}

\paragraph{\texorpdfstring{$\phi^2 F^2 \leftarrow \phi^2 F^2$}{phi2F2fromphi2F2}:}

We start from the running of the operators in the $\phi^2 F^2$ class into themselves. The corresponding anomalous dimension is given in Eq.~\eqref{eq:phi2F2_phi2F2}, where the first line leads to
\begin{equation}
    \left[\dot{C}_{\phi^2 F^2}\right]_{ab}^{A_\alpha B_\beta} = \bigg[\lambda_{abcd}+2\sum_\gamma g_\gamma^2T^\gamma_{a c d b}\bigg]\left[C_{\phi^2 F^2}\right]_{c d}^{A_\alpha B_\beta}\,.
\end{equation}
For the operators listed in Tab.~\ref{tab:SMEFT_bosonic_op} except $\mathcal{O}_{HWB}$ and $\mathcal{O}_{H\widetilde{W}B}$, the $\lambda$-dependence can be extracted using
\begin{equation}
    \lambda_{abcd}\delta_{cd} = 2(2+N) \lambda \delta_{ab} = 12 \delta_{ab}\,,
    \label{eq:phi2F2_lambda_delta}
\end{equation}
while for the case of $\mathcal{O}_{HWB}$ and $\mathcal{O}_{H\widetilde{W}B}$  we have
\begin{equation}
    \lambda_{abcd}\Sigma^I_{cd} = 4 \lambda \Sigma^I_{ab}\,.
    \label{eq:phi2F2_lambda_sigma}
\end{equation}
Similarly, the part proportional to the gauge coupling is 
\begin{align}
    2\sum_\gamma g_\gamma^2 T^\gamma_{a c d b}\delta_{cd} &= 2 g_1^2 \theta^h_{ac}\theta^h_{db}\delta_{cd} +  2g_2^2 \theta^I_{ac}\theta^I_{db} \delta_{cd}= \left(\frac{1}{2}g_1^2 + \frac{3}{2} g_2^2\right) \delta_{ab}\,,
    \label{eq:phi2F2_g_delta}
\end{align}
for the operators with the Clebsch-Gordan coefficients proportional to $\delta_{ab}$, while 
\begin{align}
    2\sum_\gamma g_\gamma^2 T^\gamma_{a c d b}\Sigma_{cd}^I = 2g_1^2 \theta^h_{ac}\theta^h_{db} \Sigma_{cd}^I +  2g_2^2 \theta^J_{ac}\theta^J_{db} \Sigma_{cd}^I= \left(\frac{1}{2}g_1^2 - \frac{1}{2} g_2^2\right) \Sigma_{ab}^I\,,
    \label{eq:phi2F2_g_sigma}
\end{align}
for the operators with the Clebsch-Gordan coefficients proportional to $\Sigma^I_{ab}$. The second and third line in Eq.~\eqref{eq:phi2F2_phi2F2} 
\begin{align}
     \left[\dot{C}_{\phi^2 F^2}\right]_{ab}^{A_\alpha B_\beta} &= 2\sum_{\sigma(\{A_\alpha ,B_\beta\}\times \{a , b\})}\sum_{\gamma}g_\alpha g_\gamma\Big[ \delta_{\alpha\gamma}\Big(3Y_{ac}^{A_\alpha C_\gamma}-2Y_{ac}^{C_\gamma A_\alpha}\Big) \nonumber\\
    & +(1-\delta_{\alpha\gamma}) \mathcal S_{\alpha\beta}^{-1} \mathcal{S}_{\gamma\beta}  Y_{ac }^{A_\alpha C_\gamma}   \Big]\left[C_{\phi^2 F^2}\right]_{cb}^{C_\gamma B_\beta}\,,
    \label{eq:phi2F2_phi2F2_23lines}
\end{align}
differ for each operator and we evaluate them next.
\subparagraph{\texorpdfstring{$\mathcal{O}_{HG}$}{OHG}:}

For this operator, all $Y^{AB}_{ab}$ are trivially zero since the Higgs is not charged under $SU(3)_c$. Therefore, combining Eqs.~\eqref{eq:phi2F2_lambda_delta}, \eqref{eq:phi2F2_g_delta}, and the results for the collinear anomalous dimensions $\gamma_{c,G}$ and $\gamma_{c,H}$ one finds
\begin{equation}
    \dot{C}_{HG} = \left(12\lambda -\frac{3}{2} g_1^2 -\frac{9}{2} g_2^2  - 2 b_{0,3} g_3^2\right) C_{HG}\,.
\end{equation}

\subparagraph{\texorpdfstring{$\mathcal{O}_{HW}$}{OHW}:} In this case, the Clebsch-Gordan coefficient on the left-hand side in Eq.~\eqref{eq:phi2F2_phi2F2_23lines} has two $G_2$ adjoint indices such that $\alpha=\beta=2$. There are two options for the gauge factor $G_\gamma$ on the right-hand side, $\gamma=2$ or $\gamma=1$. Let us start with $\alpha=\gamma=2$ when only the first line in Eq.~\eqref{eq:phi2F2_phi2F2_23lines} contributes. In this case, we have the following contribution 
\begin{equation}
    \dot{C}_{HW}\, \delta_{ab} \delta^{IJ} = 2\sum_{\sigma(\{I,J\}\times \{a , b\})} g_2^2\, \frac{1}{4}\, C_{HW}\, \delta_{ab} \delta^{IJ}\,,
\end{equation}
which results in 
\begin{equation}
    \dot{C}_{HW} = 2 g_2^2 C_{HW}\,.
\end{equation}
On the other hand, if $\alpha\ne\gamma=1$, only the second line in Eq.~\eqref{eq:phi2F2_phi2F2_23lines} contributes as
\begin{align}
     \dot{C}_{HW}\, \delta_{ab} \delta^{IJ} &= 2\sum_{\sigma(\{I,J\}\times \{a , b\})} g_2 g_1\, \mathcal{S}_{22}^{-1} \mathcal{S}_{12}\,\theta_{ad}^{I}\theta^h_{dc} C_{HWB} \Sigma_{cb}^J\nonumber\\
     &= \frac{1}{4} g_1 g_2 \sum_{\sigma(\{I,J\}\times \{a , b\})} \delta_{ab} \delta^{IJ}\,,
\end{align}
where we used $\theta_{ad}^I\,\theta^h_{dc} \Sigma_{cb}^J = \frac{1}{4} \delta_{ab} \delta^{IJ}$ and that $\mathcal{S}_{22} =2$, $\mathcal{S}_{12} =1$, as shown in Eq.~\eqref{eq:symm_fact}.
Summing all contributions, together with the collinear anomalous dimensions $\gamma_{c,W}$ and $\gamma_{c,H}$ one obtains
\begin{equation}
     \dot{C}_{HW} = \left(12\lambda-\frac{3}{2} g_1^2 - \frac{5}{2}g_2^2 -2\,b_{0,2}g_2^2\right) C_{HW} + g_1 g_2 \,C_{HWB}\,.
\end{equation}

\subparagraph{\texorpdfstring{$\mathcal{O}_{HB}$}{OHB}:} This operator is analogous to the previous one, where now we have $\alpha=\beta=1$, and the Clebsch-Gordan coefficient on the left-hand side of Eq.~\eqref{eq:phi2F2_phi2F2_23lines} carries no adjoint indices. The sum over gauge factors $\gamma$ includes two cases. First, when $\alpha=\gamma=1$, which leads to
\begin{equation}
    \dot{C}_{HB}\, \delta_{ab} = 2\times 2\sum_{\sigma(\{a , b\})} g_1^2\, \frac{1}{4}\, C_{HB}\, \delta_{ab} \,.
\end{equation}
The additional factor of 2 comes from permuting $A_2=I$ and $B_2=J$ on the right-hand side of Eq.~\eqref{eq:phi2F2_phi2F2_23lines} and we used $\theta_{ac}^h\theta_{cb}^h=\delta_{ab}/4$.
Similarly, if $\alpha\ne\gamma=2$, one finds
\begin{align}
    \dot{C}_{HB}\, \delta_{ab} & = 2\times 2 \sum_{\sigma(\{a , b\})} g_1 g_2\, \frac{1}{2} \theta^h_{ad}\theta^{I}_{dc} C_{HWB} \Sigma_{cb}^I\, C_{HB}\, \delta_{ab} \nonumber\\
    & = 3 g_1 g_2 C_{HWB}  \delta_{ab} \,,
\end{align}
where we used
$\theta^h_{ad}\theta^{I}_{dc} \Sigma_{cb}^I = \frac{3}{4} \delta_{ab}$.
Summing all contributions, together with the collinear anomalous dimensions $\gamma_{c,B}$ and $\gamma_{c,H}$ results in
\begin{equation}
     \dot{C}_{HB} = \left(12\lambda+\frac{1}{2} g_1^2 -2\,b_{0,1}g_1^2 - \frac{9}{2}g_2^2 \right) C_{HB} + 3 g_1 g_2 \,C_{HWB}\,.
\end{equation}

\subparagraph{\texorpdfstring{$\mathcal{O}_{HWB}$}{OHWB}:} For this operator the associated Clebsch-Gordan coefficient has one $G_2$ adjoint index. Let us take $\alpha=2$ and $\beta=1$. Again, let us start with the case $\gamma=\alpha=2$ in Eq.~\eqref{eq:phi2F2_phi2F2_23lines}, obtaining
\begin{align}
    \dot{C}_{HWB} \Sigma_{ab}^I & = 2 \sum_{\sigma(\{a,b\})} g_2 ^2 \left(3 \theta^I_{ad}\theta^J_{dc} - 2\theta^J_{ad}\theta^I_{dc}\right) C_{HWB} \Sigma_{cb}^J\nonumber\\
    & = 11 g_2^2 C_{HWB} \Sigma_{ab}^I\,,
\end{align}
where we used
\begin{equation}
    3 \theta^I_{ad}\theta^J_{dc}\Sigma_{cb}^J - 2\theta^J_{ad}\theta^I_{dc}\Sigma_{cb}^J = \frac{11}{4} \Sigma_{ab}^I\,.
\end{equation}
For $\gamma=1\ne\alpha$ one obtains
\begin{align}
    \dot{C}_{HWB} \Sigma_{ab}^I & = 2 \sum_{\sigma(\{a,b\})} g_2 g_1 \mathcal{S}_{21}^{-1} \mathcal{S}_{11} \theta^{I}_{ad}\theta^h_{dc} C_{HB} \delta_{cb}\nonumber\\
    & = 2 g_1 g_2 C_{HB} \Sigma_{ab}^I\,.
\end{align}
We now need to permute $A_\alpha$ and $B_\beta$ by taking $\alpha=1$ and $\beta=2$. Likewise, let us first take $\gamma=\alpha=1$, obtaining 
\begin{align}
    \dot{C}_{HWB} \Sigma_{ab}^I & = 2 \sum_{\sigma(\{a,b\})} g_1 ^2 \left(3 \theta^h_{ad}\theta^h_{dc} - 2\theta^h_{ad}\theta^h_{dc}\right) C_{HWB} \Sigma_{cb}^I\nonumber\\
    & = g_1^2 C_{HWB} \Sigma_{ab}^I\,,
\end{align}
where we used
\begin{equation}
    3 \theta^h_{ad}\theta^h_{dc}\Sigma_{cb}^I - 2\theta^h_{ad}\theta^h_{dc}\Sigma_{cb}^I = \frac{1}{4} \Sigma_{ab}^I\,.
\end{equation}
Finally, $\gamma=2\ne\alpha$ leads to
\begin{align}
    \dot{C}_{HWB} \Sigma_{ab}^I & = 2 \sum_{\sigma(\{a,b\})} g_1 g_2 \mathcal{S}_{21}^{-1} \mathcal{S}_{22} \theta^{h}_{ad}\theta^J_{dc} C_{HW} \delta_{cb} \delta^{JI} \nonumber\\
    & = 2 g_1 g_2 C_{HW} \Sigma_{ab}^I\,.
\end{align}
Summing all contributions, together with the collinear anomalous dimensions $\gamma_{c,W}$, $\gamma_{c,B}$, and $\gamma_{c,H}$, results in
\begin{equation}
     \dot{C}_{HWB} = \left(4\lambda-\frac{1}{2} g_1^2 - b_{0,1}g_1^2 + \frac{9}{2}g_2^2 -b_{0,2}g_2^2\right) C_{HWB} + 2 g_1 g_2 \,C_{HB}+ 2 g_1 g_2 \,C_{HW}\,.
\end{equation}

\paragraph{\texorpdfstring{$\phi^2 F^2 \leftarrow F^3$:}{F3tophi2F2}}

Next, we study the running of the $\phi^2 F^2$ operators induced by the $F^3$ class. Since the only scalar field in the SM is a color singlet, it is easy to see, using Eq.~\eqref{eq:CF3_to_Cphi2F2}, that the only running contributions arise from $\mathcal{O}_{W}$ into $\mathcal{O}_{HW}$ and $\mathcal{O}_{HWB}$. 

\subparagraph{\texorpdfstring{$\mathcal{O}_{HW}$:}{OHW}} For this operator, both adjoint indices correspond to $SU(2)_L$ and only the first two lines of Eq.~\eqref{eq:CF3_to_Cphi2F2} contribute
\begin{align}
    \frac{1}{2} \dot{C}_{HW} \delta_{ab} \delta^{IJ} &= \frac{3}{2}i g_2^3 \sum_{\sigma(\{I,J\}\times\{a,b\})} \left[\theta^L_{bc}\left(\theta^I_{ae}\theta^K_{ec}-2\theta^K_{ae}\theta^I_{ec}\right)\epsilon^{JKL} C_{W}-i\theta^K_{ac}\theta^L_{cb}\epsilon^{JLM}\epsilon^{IKM}C_W\right]\nonumber\\
    &= \frac{3}{2}i g_2^3 \sum_{\sigma(\{I,J\}\times\{a,b\})} \left[\frac{1}{2}\left(\epsilon^{IJK}\theta^K_{ab}-\frac{3i}{2}\delta_{ab}\delta^{IJ}\right)-\frac{i}{2}\left(\delta_{ab}\delta^{IJ}-i\epsilon^{IJK}\theta^K_{ab}\right)\right]C_W\nonumber
    \\
    &=\frac{3}{2}i g_2^3 \sum_{\sigma(\{I,J\}\times\{a,b\})} -\frac{5i}{4} C_W\delta_{ab}\delta^{IJ}\,.
    \label{eq:CW_to_CHW}
\end{align}
One finds
\begin{equation}
    \dot{C}_{HW} = 15 g_2^3 C_W\,,
\end{equation}
which agrees with the result in~\cite{Alonso:2013hga} up to a different convention for the sign used in the covariant derivative. In the second line of Eq.~\eqref{eq:CW_to_CHW} we explicitly write the results of the first two terms in the first line obtained by repeated use of the identity in Eq.~\eqref{eq:thetaI_thetaJ}.

\subparagraph{\texorpdfstring{$\mathcal{O}_{HWB}$}{OHWB}:} In this case, there is only one adjoint index corresponding to the $SU(2)_L$ gauge factor, such that only the last lines of Eq.~\eqref{eq:CF3_to_Cphi2F2} enters 
\begin{align}
    \frac{1}{2}\dot{C}_{HWB} \Sigma^I_{ab} &= -3i\sum_{\sigma(\{a,b\})} g_2^2 g_1 \theta^K_{bc} \theta^h_{ae} \theta^J_{ec} \epsilon^{IJK} C_W\nonumber\\
    &= - 3ig_1 g_2^2 \sum_{\sigma(\{a,b\})} -i \theta^h_{ae} \theta^I_{eb}C_W\,,
\end{align}
which results in
\begin{equation}
    \dot{C}_{HWB} = -3g_1 g_2^2 C_W\,.
\end{equation}
Since the RGE formula for the CP-odd operators is the same, their running is also given by the above results with the substitution $F\to \tildee F$.

\subsubsection{\texorpdfstring{$F^3$}{F3} class}

\paragraph{\texorpdfstring{$F^3\leftarrow F^3$}{F3fromF3}:}

The relevant RGEs for this sector are given in Eqs.~\eqref{eq:F3_to_F3} and~\eqref{eq:F3t_to_F3t}. To obtain the RGEs of the SMEFT WCs the following relation is used
\begin{equation}
    F^{B_\alpha C_\alpha D_\alpha F_\alpha} \left[ C_{F^3}\right]^{A_\alpha D_\alpha F_\alpha} = C_2(G_\alpha) \left[ C_{F^3}\right]^{A_\alpha B_\alpha C_\alpha}\,,
\end{equation}
where $C_2(G_\alpha)$ is the quadratic Casimir of the adjoint representation. In combination with the collinear anomalous dimensions in Eqs.~\eqref{eq:gammaG} and~\eqref{eq:gammaW} one finds
\begin{alignat}{2}
    \dot C_{G} &= 3(12 - b_{0,3})\,g_3^2\, C_{G} &&= g_3^2\, C_{G}\,,\\
    \dot C_{W} &= 3(8 - b_{0,2})\,g_2^2\, C_{W} &&= \frac{5}{2} g_2^2\, C_{W} \,.
\end{alignat}
The RGEs for the CP-violating operators are the same, and the result can be obtained by replacing $F\to\tildee F$.

\subsubsection{Gauge couplings and topological angles} The running of the gauge couplings induced by dimension-6 terms is computed in Eq.~\eqref{eq:g_phi2F2}. For each simple gauge group $G_\alpha$ $(\alpha=1,2,3)$ of the SM, we obtain the same result  
\begin{equation}
    \dot{g}_\alpha = 4 g_\alpha \left(-\lambda v^2 \delta_{ab}\right) \frac{1}{2} C_{HX} \delta_{ab}\,,
\end{equation}
where we used Eq.~\eqref{eq:mass_SM} and $X=G,W,B$. The results read
\begin{equation}
     \dot{g}_3 =  - 4\, m_H^2\, g_3\, C_{HG}\,,\quad  \dot{g}_2 =  - 4\, m_H^2\, g_2\, C_{HW}\,,\quad
     \dot{g}_1 =  - 4\, m_H^2\, g_1\, C_{HB}\,.
\end{equation}
We also used the Higgs boson mass defined as $m_H^2=2\lambda v^2$ and $\delta_{aa}=4$.

The running of the topological angles induced by dimension-six terms is computed in Eq.~\eqref{eq:theta_phi2F2}. Again, for each SM gauge factor $G_\alpha$ $(\alpha=1,2,3)$ one finds the same result  
\begin{equation}
    \dot{\vartheta}_\alpha = \frac{32\pi^2}{g_\alpha^2} 2\left(-\lambda v^2 \delta_{ab}\right) \frac{1}{2} C_{H\widetilde{X}} \delta_{ab}\,,
\end{equation}
where we used Eq.~\eqref{eq:mass_SM} and $X=G,W,B$. The results read
\begin{equation}
     \dot{\vartheta}_3 = -16\pi^2 \frac{4m_H^2}{g_3^2} C_{H\widetilde{G}}\,,\quad \dot{\vartheta}_2 = -16\pi^2  \frac{4m_H^2}{g_2^2} C_{H\widetilde{W}}\,,\quad \dot{\vartheta}_1 = -16\pi^2  \frac{4m_H^2}{g_1^2} C_{H\widetilde{B}}\,,
\end{equation}
after the theta terms are normalized such that $\mathcal{L}_{\vartheta_\alpha} = \frac{\vartheta_\alpha g_\alpha^2}{32\pi^2} F_{\mu\nu}^{A_\alpha} \widetilde{F}^{A_\alpha\mu\nu}$ for each simple gauge group $G_\alpha$.

\subsubsection{Higgs quartic}
\paragraph{\texorpdfstring{$\phi^4\leftarrow \phi^6$}{phi4fromphi6}:} The running of the Higgs quartic coupling induced by $\mathcal{O}_{H}$ is computed in Eq.~\eqref{eq:lambda_CH6}, which gives
\begin{equation}
    \frac{4!}{4}\dot{\lambda}\delta_{(ab}\delta_{cd)} = - 6! \left(-\lambda v^2 \delta_{ef}\right) \times \frac{1}{8} C_{H} \delta_{(ab}\delta_{cd}\delta_{ef)}\,.
\end{equation}
We can use 
\begin{equation}
    \delta_{ef} \delta_{(ab}\delta_{cd}\delta_{ef)} = \frac{4!}{6!} 6(4+\delta_{ee})\delta_{(ab}\delta_{cd)}\,,
\end{equation}
which results in
\begin{equation}
    \dot{\lambda} = 12\, m_H^2\, C_H\,.
\end{equation}

\paragraph{\texorpdfstring{$\phi^4\leftarrow \phi^2 F^2$}{phi4fromphi2F2}:} Operators in the class $\phi^2 F^2$ run into the scalar quartic coupling according to Eq.~\eqref{eq:phi2F2_to_phi4}. In SMEFT, the relevant operators are $\mathcal{O}_{HB}$, $\mathcal{O}_{HW}$, and $\mathcal{O}_{HWB}$. Applying Eq.~\eqref{eq:phi2F2_to_phi4} for the case of $\mathcal{O}_{HB}$ gives
\begin{align}
    \frac{4!}{4}\dot{\lambda}\delta_{(ab}\delta_{cd)} &= - 12 \sum_{\sigma(\{a,b,c,d\})} \left(-\frac{1}{2}m_H^2\delta_{af}\right)\left( g_1^2\theta_{bg}^h\theta_{gf}^h\frac{1}{2}C_{HB}\delta_{cd}+g_2^2\theta_{bg}^I\theta_{gf}^J\frac{1}{2}C_{HW}\delta_{cd}\delta^{IJ}\right)\nonumber\\
    &= 18 m_H^2 \left(C_{HB} +3 C_{HW}\right)\delta_{(ab}\delta_{cd)}\,,
\end{align}
which gives
\begin{equation}
    \dot{\lambda} = 3 g_1^2 m_H^2 C_{HB} + 9 g_2^2 m_H^2 C_{HW}\,.
\end{equation}
Similarly, the case of $\mathcal{O}_{HWB}$ can be obtained as
\begin{align}
    \frac{4!}{4}\dot{\lambda}\delta_{(ab}\delta_{cd)} &= - 12 \sum_{\sigma(\{a,b,c,d\})} g_2 g_1 \left(-\frac{1}{2}m_H^2\delta_{ae}\right) \theta_{bf}^I\theta^h_{fe} \frac{1}{2} C_{HWB} \Sigma_{cd}^I\nonumber\\
    &= \frac{3}{4}g_1 g_2 m_H^2 C_{HWB} \sum_{\sigma(\{a,b,c,d\})} \Sigma_{ab}^I \Sigma_{cd}^I\nonumber\\
    &= \frac{3}{4}g_1 g_2 m_H^2 C_{HWB}\, 4!\, \delta_{(ab}\delta_{cd)}\,,
\end{align}
which finally gives
\begin{equation}
    \dot\lambda = 3 g_1 g_2 m_{H}^2 C_{HWB}\,.
\end{equation}

\paragraph{\texorpdfstring{$\phi^4\leftarrow D^2 \phi^4$}{phi4fromD2phi4}:}

This anomalous dimension is reported in Eq.~\eqref{eq:phi4<-D2phi4}. 
The relevant piece of the RHS of this equation is
\begin{align}
    &\quad - \sum_{\sigma(\{a,b,c,d\})} \Bigg[2m^2_{ef}\lambda_{gfab}\left[C_{D^2\phi^4}\right]_{egcd}\nonumber\\\nonumber 
    &\quad +\left(\frac{1}{3}m^2_{ef}\lambda_{gcab}+m^2_{gc}\lambda_{efab}\right)\left(\left[C_{D^2\phi^4}\right]_{gfed}+\left[C_{D^2\phi^4}\right]_{gefd}-\left[C_{D^2\phi^4}\right]_{fegd}\right)\\
    &\quad -6\sum_{\alpha=1}^{3} g_\alpha^2 m^2_{fg} T^\alpha_{agbe} \left[C_{D^2\phi^4}\right]_{efcd}
    -\frac{4}{3!}\lambda_{ebcd}m^2_{fg}\left(
    \left[C_{D^2\phi^4}\right]_{feag}-\left[C_{D^2\phi^4}\right]_{fgae}
    \right)
    \Bigg]\nonumber \\
    &= m_H^2\bigg[  4  \lambda \left(-26 C_{H\Box}+7C_{HD}\right)
    +\frac{3}{2}(g_1^2+g_2^2) C_{HD}
    \bigg](\delta_{ab}\delta_{cd}+\delta_{ac}\delta_{bd}+\delta_{ad}\delta_{bc})\,.\label{eq:lambdadotSMEFT1}
\end{align}
This contribution needs to be added to the one in Eq.~\eqref{eq:dotlambda<-Ctildeandbar} generated by $[\dot{\tildee C}_{D^2\phi^4}]_{abcd}$ and $[\dot{\overline C}_{D^2\phi^4}]_{abcd}$, which are reported in Eqs.~\eqref{eq:dotCtildeSMEFT} and \eqref{eq:dotCbarSMEFT}, respectively:
\begin{align}
    &\quad - \sum_{\sigma(\{a,b,c,d\})}m^2_{ae}\left(
    \frac{1}{3}\left[\dot{\tildee C}_{D^2\phi^4}\right]_{ebcd}+\left[\dot{\overline C}_{D^2\phi^4}\right]_{ebcd}
    \right)\label{eq:lambdadotSMEFT2} \\
    &= m_H^2\bigg[  4\lambda (10 C_{H\Box}-C_{HD})+\frac{3}{2}g_1^2C_{HD}+\frac{1}{6}g_2^2(40 C_{H\Box }-27 C_{HD})
    \bigg](\delta_{ab}\delta_{cd}+\delta_{ac}\delta_{bd}+\delta_{ad}\delta_{bc})\,.\nonumber
\end{align}
By summing Eqs.~\eqref{eq:lambdadotSMEFT1} and \eqref{eq:lambdadotSMEFT2} one obtains $\dot \lambda_{abcd}$, which gives
\begin{equation}
    \dot\lambda =\frac{1}{6}m_H^2\left[4\left(5g_2^2-48\lambda\right)C_{H\Box}+9\left(
    g_1^2-g_2^2+8\lambda
    \right)C_{HD}\right] \,.
\end{equation}

\subsubsection{Higgs mass}

The running of the Higgs mass $m_H^2$ only receives contributions from $D^2\phi^4$-class operators.
The corresponding anomalous dimension is reported in Eq.~\eqref{eq:phi2<-D2phi4}, which gives 
\begin{align}
    -\frac{1}{2}\dot{(m_H^2)}\delta_{ab} &= -\sum_{\sigma(\{a,b\})}\Bigg[2 m_{cd}^2 m_{ed}^2\left[C_{D^2\phi^4}\right]_{ceab}
    \nonumber \\
    &\quad +2m_{cd}^2 m_{ea}^2 \left(\left[C_{D^2\phi^4}\right]_{edcb}+\left[C_{D^2\phi^4}\right]_{ecdb}-\left[C_{D^2\phi^4}\right]_{dceb}\right)\nonumber \\
    &\quad-4m^2_{cb} m^2_{de}
    \left(
    \left[C_{D^2\phi^4}\right]_{dcae}-\left[C_{D^2\phi^4}\right]_{deac}
    \right)
    \Bigg] \nonumber \\
    &= m_H^4\left(2 C_{H\Box}-C_{HD}\right)\delta_{ab}\,,
\end{align}
and therefore one finds
\begin{equation}
    \dot{(m_H^2)} = m_H^4 \left(-4C_{H\Box}+2C_{HD}\right)\,.
\end{equation}

\subsection{CP-violating ALP --- SMEFT interference}

Models that involve axion-like particles (ALPs) are well-motivated extensions of the SM with a light scalar gauge singlet. ALPs have the potential to solve the strong CP problem \cite{Peccei:1977hh,Weinberg:1977ma,Wilczek:1977pj} and arise naturally in various extensions of the SM \cite{Gripaios:2009pe,Gripaios:2016mmi,Chala:2017sjk,Ema:2016ops,Calibbi:2016hwq,deGiorgi:2024elx}. In this subsection, we will derive the RGEs resulting from a CP-violating ALP field $a$ that couples to the SM. The corresponding bosonic Lagrangian up to dimension-5 reads~\cite{Georgi:1986df,DiLuzio:2023cuk,DiLuzio:2023lmd} 
\begin{align}
    \mathcal L_{\text{ALP}} &= \frac{1}{2}(\partial_\mu a)(\partial^\mu a)-\frac{1}{2}m^2 a^2 + C_{aG}\, a\, G^A_{\mu\nu} G^{A\,\mu\nu} + C_{a\tildee G}\, a\, G^A_{\mu\nu} \tildee G^{A\,\mu\nu} \\ & \quad+ 
    C_{aW}\, a\, W^I_{\mu\nu} W^{I\,\mu\nu} + C_{a\tildee W}\, a\, W^I_{\mu\nu} \tildee W^{I\,\mu\nu} +
    C_{aB}\, a\, B_{\mu\nu} B^{\mu\nu} + C_{a\tildee B}\, a\, B_{\mu\nu} \tildee B^{\mu\nu}\,,\nonumber 
\end{align}
where we also allow for couplings of the ALP to CP-even operators.\footnote{The same Lagrangian can also be adopted for generic CP-even and CP-odd scalars. Their distinction from axion-like particles and possible ways to discriminate them have been discussed in~\cite{Alda:2024cxn}.}

\subsubsection{\texorpdfstring{$D^2\phi^4$}{D2phi4} class}
Since the ALP is a gauge singlet, we are interested only in the following contractions from the RHS of Eq.~\eqref{eq:D2phi4_phiF22}
\begin{align}
     &\quad -\frac{4}{3}\sum_{\alpha=1}^3 \sum_{\gamma=1}^\alpha \sum_{\beta=1}^\gamma   g_\alpha g_\beta \left(\theta^{A_\alpha}_{ad}\theta^{B_\beta}_{bc}+\theta^{A_\alpha}_{ac}\theta^{B_\beta}_{bd}\right)\nonumber \\
    &\quad \times \left(
    \left[C_{\phi F^2}\right]^{A_\alpha C_\gamma }_e \left[C_{\phi F^2}\right]^{B_\beta C_\gamma }_e 
    +
    \left[C_{\phi \tildee F^2}\right]^{A_\alpha C_\gamma }_e \left[C_{\phi \tildee F^2}\right]^{B_\beta C_\gamma }_e
    \right) \nonumber \\
    &=
    -\frac{4}{3}
    g_1^2 \left(\theta^h_{ad}\theta^h_{bc} + \theta^h_{ac}\theta^h_{bd}\right)
    \left(C_{aB}^2+C_{a\tildee B}^2\right) -\frac{4}{3}
    g_2^2 \left(\theta^I_{ad}\theta^I_{bc} + \theta^I_{ac}\theta^I_{bd}\right)
    \left(C_{aW}^2+C_{a\tildee W}^2\right) \nonumber \\
    &= 
    -\frac{1}{3}
    g_1^2 \left(
    \delta_{ac}\delta_{bd}+\delta_{ad}\delta_{bc}-\delta_{ab}\delta_{cd}-\Sigma^I_{ab}\Sigma^I_{cd}
    \right)
    \left(C_{aB}^2+C_{a\tildee B}^2\right)\nonumber \\
    &\quad +\frac{1}{3}
    g_2^2 \left(
    2 \delta_{ab}\delta_{cd}-\delta_{ac}\delta_{bd}-\delta_{ad}\delta_{bc}
    \right)
    \left(C_{aW}^2+C_{a\tildee W}^2\right)\,.
\end{align}
From this expression one can obtain the running of both $C_{H\Box}$ and $C_{HD}$, but also $[\dot{\tildee C}_{D^2\phi^4}]_{abcd}$ and $[\dot{\overline C}_{D^2\phi^4}]_{abcd}$:
\begin{align}
    \dot C_{H\Box} &= \frac{2}{3}\,g_1^2 \left(C_{aB}^2+C_{a\tildee B}^2\right) + 2\,g_2^2  \left(C_{aW}^2+C_{a\tildee W}^2\right)\,,
    \\
    \dot C_{HD} &= \frac{8}{3}\,g_1^2 \left(C_{aB}^2+C_{a\tildee B}^2\right)\,,\\
    \left[\dot {\tildee C}_{D^2\phi^4}\right]_{abcd} &=\frac{2}{3}\,g_2^2  \left(C_{aW}^2+C_{a\tildee W}^2\right) (\delta_{ac}\delta_{bd}+\delta_{ad}\delta_{bc}+\delta_{ab}\delta_{cd})\,,\label{eq:CD2phi4tildedotALP}\\
    \left[\dot {\overline C}_{D^2\phi^4}\right]_{abcd} &=0\label{eq:CD2phi4bardotALP}\,.
\end{align}
These last two expressions modify the RGEs corresponding to $C_H$ and $\lambda$, as induced by the ALP interactions with gauge bosons.

\subsubsection{\texorpdfstring{$\phi^6$}{phi6} class}
The only ALP contribution to the running of $C_H$ results from the term
\begin{equation}
    \left[C_{\phi^6}\right]_{abcdef} = \frac{1}{6!}\sum_{\sigma(\{a,b,c,d,e,f\})}\frac{1}{3!}\lambda_{abcg}\left(\frac{1}{3}\left[\dot {\tildee C}_{D^2\phi^4}\right]_{gdef}+\left[\dot {\overline C}_{D^2\phi^4}\right]_{gdef}\right)\,,
\end{equation}
where $[\dot {\tildee C}_{D^2\phi^4}]_{abcd}$ is provided in Eq.~\eqref{eq:CD2phi4tildedotALP}, while $[\dot {\overline C}_{D^2\phi^4}]_{abcd}$ is vanishing.
The explicit calculation yields
\begin{equation}
    \dot C_H = \frac{16}{3}\,\lambda\, g_2^2 \left(C_{aW}^2+C_{a\tildee W}^2\right) \,.
\end{equation}

\subsubsection{\texorpdfstring{$\phi^2F^2$}{phi2F2} class}
The anomalous dimensions of the SMEFT parity-even $\phi^2F^2$-class, induced by pairs of ALP interactions, are determined by the term in Eq.~\eqref{eq:phi2F2-phiF2xphiF2} for which the ALP index is contracted among products of $C_{\phi F^2}$ or $C_{\phi \tildee F^2}$ WCs:
\begin{align}
\left[\dot C_{\phi^2F^2}\right]_{ab}^{A_\alpha B_\beta} &=
     \frac{1}{2}\mathcal S_{\alpha\beta}^{-1} \sum_{\sigma(\{A_\alpha,B_\beta\}\times \{a,b\})}\sum_{\gamma=1}^{3}\sum_{\delta=1}^{3}\mathcal S_{\alpha\gamma}\mathcal S_{\beta\delta} g_\gamma g_\delta Y_{ab}^{C_\gamma D_\delta} \Big(\left[C_{\phi F^2}\right]_{c}^{A_\alpha C_\gamma} \left[C_{\phi F^2}\right]_{c}^{B_\beta D_\delta}\nonumber \\
     &\quad - \left[C_{\phi \tildee F^2}\right]_{c}^{A_\alpha C_\gamma} \left[C_{\phi \tildee F^2}\right]_{c}^{B_\beta D_\delta}\Big)\,,
\end{align}
where $a,b$ are Higgs boson indices and $c$ denotes the ALP index.
This expression implies $\dot C_{HG} = 0$ since the Higgs boson is a $SU(3)_c$ singlet.
For $\dot C_{HB}$ one finds
\begin{align}
    \frac{1}{2}\dot C_{HB} \delta_{ab} = 2 \sum_{\sigma(\{a,b\})} g_1^2 \theta^h_{ad}\theta^h_{db}\left(C_{aB}^2-C_{a\tildee B}^2 \right) =  g_1^2 \delta_{ab}\left(C_{aB}^2-C_{a\tildee B}^2 \right)\,,
\end{align}
and therefore 
\begin{align}
    \dot C_{HB} = 2\, g_1^2 \left(C_{aB}^2-C_{a\tildee B}^2 \right)\,.
\end{align}
Similarly, for $\dot C_{HW}$ one finds
\begin{align}
    \frac{1}{2}\dot C_{HW} \delta_{ab} \delta^{IJ} &= \sum_{\sigma(\{I,J\}\times\{a,b\})} g_2^2 \theta^K_{ad}\theta^L_{db}\left(C_{aW}^2-C_{a\tildee W}^2 \right)\delta^{IK}\delta^{JL}\nonumber \\
    &= \sum_{\sigma(\{I,J\}\times\{a,b\})} g_2^2 \left(\frac{1}{4}\delta_{ab}\delta^{IJ}+\frac{i}{2}\epsilon^{IJK}\theta^K_{ab}\right)\left(C_{aW}^2-C_{a\tildee W}^2 \right)\nonumber \\
    &= g_2^2 \delta_{ab}\delta^{IJ} \left(C_{aW}^2-C_{a\tildee W}^2 \right)\,,
\end{align}
which implies
\begin{equation}
    \dot C_{HW} = 2\, g_2^2 \left(C_{aW}^2-C_{a\tildee W}^2 \right)\,.
\end{equation}
Finally, the expression for $\dot C_{HWB}$ is
\begin{align}
    \frac{1}{2} \dot C_{HWB} \Sigma^I_{ab} &= 4 \sum_{\sigma(\{a,b\})} g_1 g_2 \theta^h_{ad}\theta^I_{db} \left(C_{aB}C_{aW} - C_{ a\tildee B}C_{a\tildee W} \right)\nonumber \\
    &= 2 g_1 g_2 \Sigma^I_{ab} \left(C_{aB}C_{aW} - C_{ a\tildee B}C_{a\tildee W} \right)\,,
\end{align}
where we used $\theta^h\theta^I = \frac{1}{4}\Sigma^I$, implying
\begin{equation}
    \dot C_{HWB} = 4\, g_1\, g_2 \left(C_{aB}C_{aW} - C_{ a\tildee B}C_{a\tildee W} \right)\,.
\end{equation}
Similarly, for the RGEs associated with the parity-odd WCs, the relevant term from Eq.~\eqref{eq:phi2F2-phiF2xphiF2_CPV} is
\begin{align}
\left[\dot C_{\phi^2\tildee F^2}\right]_{ab}^{A_\alpha B_\beta} &=
     \frac{1}{2}\mathcal S_{\alpha\beta}^{-1} \sum_{\sigma(\{A_\alpha,B_\beta\}\times \{a,b\})}\sum_{\gamma=1}^{3}\sum_{\delta=1}^{3}\mathcal S_{\alpha\gamma}\mathcal S_{\beta\delta} g_\gamma g_\delta Y_{ab}^{C_\gamma D_\delta} \Big(\left[C_{\phi F^2}\right]_{c}^{A_\alpha C_\gamma} \left[C_{\phi \tildee F^2}\right]_{c}^{B_\beta D_\delta}\nonumber \\
     &\quad + \left[C_{\phi \tildee F^2}\right]_{c}^{A_\alpha C_\gamma} \left[C_{\phi F^2}\right]_{c}^{B_\beta D_\delta}\Big)\,.
\end{align}
After performing the same group algebra as before, one obtains the following anomalous dimensions:
\begin{gather}
    \dot C_{H\tildee B} = 4\,g_1^2\, C_{aB}\,C_{a\tildee B}\,,
    \quad
    \dot C_{H\tildee W} = 4\, g_2^2\, C_{aW}\,C_{a\tildee W} \,,
    \quad
    \dot C_{H\tildee G} = 0\,,\\
    \dot C_{H\tildee WB} = 4\, g_1\, g_2 \left(C_{a\tildee B}C_{aW} + C_{ a B}C_{a\tildee W} \right)\,.
\end{gather}
These results are a generalization of those obtained in \cite{Galda:2021hbr} and many RGEs are new, to the best of our knowledge.

\subsubsection{\texorpdfstring{$F^3$}{F3} class}

The anomalous dimensions of the SMEFT $C_{X}$ WCs, with $X=G,W$, induced by pairs of ALP interactions can be derived from Eq.~\eqref{eq:F3<-phiF2xphiF2}:
\begin{align}
    \dot C_{X} f^{A_\alpha B_\alpha C_\alpha} &= -\frac{8}{3}g_\alpha \left(C_{aX}^2-C_{a\tildee X}^2\right)\left[f^{C_\alpha D_\alpha E_\alpha} \delta^{A_\alpha D_\alpha} \delta^{B_\alpha E_\alpha} + f^{B_\alpha D_\alpha E_\alpha} \delta^{C_\alpha D_\alpha} \delta^{A_\alpha E_\alpha}\right.\nonumber \\
    &\quad  \left. +
    f^{A_\alpha D_\alpha E_\alpha} \delta^{B_\alpha D_\alpha} \delta^{C_\alpha E_\alpha}\right] \nonumber \\
    &= -8 g_\alpha \left(C_{aX}^2-C_{a\tildee X}^2\right) f^{A_\alpha B_\alpha C_\alpha}\,,
\end{align}
which implies 
\begin{equation}
    \dot C_{G} = -8\, g_3 \left(C_{aG}^2-C_{a\tildee G}^2\right)\,, \quad 
    \dot C_{W} = -8\, g_2 \left(C_{aW}^2-C_{a\tildee W}^2\right)\,.
\end{equation}
These results are in agreement with those found in \cite{Galda:2021hbr,Bresciani:2024shu}. Similarly, using Eq.~\eqref{eq:F3tilde<-phiF2xphiF2} the running of the parity-odd WCs $C_{\tildee X}$ is given by:
\begin{equation}
    \dot C_{\tildee G} = -16\, g_3\, C_{aG}\, C_{a\tildee G} \,, \quad
    \dot C_{\tildee W} = -16\, g_2\, C_{aW}\, C_{a\tildee W}\,,
\end{equation}
which agrees with the results found in \cite{DiLuzio:2020oah, Bresciani:2024shu}.

\subsubsection{Gauge couplings and topological angles}

From Eq.~\eqref{eq:F2<-phiF2xphiF2} and \eqref{eq:Ftilde2<-phiF2xphiF2} the running of the gauge couplings and topological angles is given as follows:
\begin{align}
    \dot g_1 &= 8\, g_1\, m^2 \left(C^2_{aB} - C^2_{a\tildee B}\right)\,, & 
    \dot \vartheta_1 &= \frac{32\pi^2}{g_1^2}\,8\,m^2\, C_{aB}\, C_{a\tildee B}  \,,
    \\
    \dot g_2 &= 8\, g_2\, m^2 \left(C^2_{aW} - C^2_{a\tildee W}\right)\,,
    & 
    \dot \vartheta_2 &= \frac{32\pi^2}{g_2^2}\,8\,m^2\, C_{aW}\, C_{a\tildee W} \,,
    \\
    \dot g_3 &= 8\, g_3\, m^2 \left(C^2_{aG} - C^2_{a\tildee G}\right)\,,
    & 
    \dot \vartheta_3 &= \frac{32\pi^2}{g_3^2}\,8\,m^2\, C_{aG}\, C_{a\tildee G} \,.
\end{align}

\subsubsection{Higgs quartic}
The running of the Higgs quartic coupling only receives a contribution proportional to $[\dot{\tildee C}_{D^2\phi^4}]_{abcd}$, given in Eq.~\eqref{eq:CD2phi4tildedotALP}:
\begin{align}
    \dot \lambda_{abcd} =- \sum_{\sigma(\{a,b,c,d\})}
    m^2_{ae} \left(\frac{1}{3}\left[\dot{\tildee C}_{D^2\phi^4}\right]_{bcde}+\left[\dot{\overline C}_{D^2\phi^4}\right]_{bcde}\right) = 8 g_2^2 m_H^2 \left(C_{aW}^2+C_{a\tildee W}^2\right)\delta_{(ab}\delta_{cd)}\,,
\end{align}
which implies
\begin{equation}
    \dot \lambda = \frac{4}{3}\,g_2^2\, m_H^2 \left(C_{aW}^2+C_{a\tildee W}^2\right)\,.
\end{equation}
This agrees with Eq.~(4.5) in \cite{Galda:2021hbr}, up to a sign.

\subsubsection{ALP mass}
Furthermore, from Eq.~\eqref{eq:phi2<-phiF2xphiF2} one finds the running of the ALP mass $m^2$:
\begin{equation}
    \dot{(m^2)} = 8\,m^4 \left(
    C_{aB}^2+C_{a\tildee B}^2
    + 3\,C_{aW}^2+3\,C_{a\tildee W}^2
    + 8\,C_{aG}^2+8\,C_{a\tildee G}^2
    \right)\,.
\end{equation}

\subsection{\texorpdfstring{$O(n)$}{O(n)} EFT}

Finally, we consider the $O(n)$ scalar EFT. 
This theory includes $n$ real scalar gauge singlets and is invariant under global $O(n)$ transformations.
We consider the following dimension-six Lagrangian\,\footnote{The operator $(\phi \cdot \partial_\mu \phi)(\phi \cdot \partial^\mu \phi)$ is neglected since it is redundant.}
\begin{equation}
    \mathcal L_{O(n)} = \frac{1}{2}(\partial_\mu \phi)\cdot (\partial^\mu \phi) - \frac{1}{2}m^2 \phi \cdot \phi - \frac{\lambda}{4}(\phi \cdot \phi)^2 + C_{E}\phi\cdot \phi (\partial_\mu \phi)\cdot (\partial^\mu \phi) + C_1 (\phi\cdot \phi)^3\,.
\end{equation}
The renormalizable couplings of the general EFT are given by
\begin{equation}
    m^2_{ab} = m^2 \,\delta_{ab}\,, \qquad 
    \lambda_{abcd} = \frac{4!}{4}\,\lambda \,\delta_{(ab}\delta_{cd)}\,,
\end{equation}
while the WCs read
\begin{equation}
    \left[C_{D^2\phi^4}\right]_{abcd} = C_E\, \delta_{ab}\delta_{cd}\,,
    \qquad
    \left[C_{\phi^6}\right]_{abcdef} = C_1 \,\delta_{(ab}\delta_{cd}\delta_{ef)}\,.
\end{equation}
In the following, we derive the RGEs of this theory, finding full agreement with the one-loop results in~\cite{Jenkins:2023bls,Cao:2021cdt}.

\paragraph{\texorpdfstring{$\phi^6 \leftarrow \phi^6$}{phi6fromphi6}:}
This anomalous dimension can be derived from Eq.~\eqref{eq:phi6<-phi6}
\begin{equation}
    \left[\dot C_{\phi^6}\right]_{abcdef} = 
    \frac{1}{2!4!}\sum_{\sigma(\{a,b,c,d,e,f\})}\lambda_{efgh} \left[C_{\phi^6}\right]_{abcdgh} = 6 \lambda  (14+n) C_1 \delta_{(ab}\delta_{cd}\delta_{ef)}\,,
\end{equation}
where we exploited the identity in Eq.~\eqref{eq:identity_permutations}.
Hence, one finds
\begin{equation}
    \dot C_1 = 6\, \lambda\,  (14+n)\, C_1\,.
\end{equation}

\paragraph{\texorpdfstring{$\phi^4 \leftarrow \phi^6$}{phi4fromphi6}:}
From Eq.~\eqref{eq:lambda_CH6} one finds
\begin{equation}
    \dot \lambda_{abcd} = -6!\,m^2 C_1\, \delta_{ef} \delta_{(ab}\delta_{cd}\delta_{ef)} = -144 (4+n)m^2 C_1 \delta_{(ab}\delta_{cd)}\,,
\end{equation}
and therefore
\begin{equation}
    \dot \lambda = -24\, (4+n)\,m^2\, C_1 \,.
\end{equation}

\paragraph{\texorpdfstring{$D^2\phi^4 \leftarrow D^2\phi^4$}{D2phi4fromD2phi4}:}
From the RHS of Eq.~\eqref{eq:D2phi4<-D2phi4_offshell} one finds
\begin{equation}
    -\frac{1}{2}\sum_{\sigma(\{a,b\}\times \{c,d\})}\lambda_{efcd}C_E\left(\delta_{eb}\delta_{af}-\delta_{ef}\delta_{ab}\right) = 4\lambda C_E  [(1+n)\delta_{ab}\delta_{cd}-\delta_{ad}\delta_{bc}-\delta_{ac}\delta_{bd}]\,.
    \label{eq:O(n)_D2phi4}
\end{equation}
This expression allows us to extract the running for both $C_E$ and $[\tildee C_{D^2\phi^4}]_{abcd}$.
For the former, one can apply the substitution $\delta_{ad}\delta_{bc}+\delta_{ac}\delta_{bd} \to -\delta_{ab}\delta_{cd}$, which implies
\begin{equation}
    \dot C_E = 4\,\lambda\, (2+n)\, C_E\,.
\end{equation}
The latter running is obtained by subtracting $\dot C_E \delta_{ab}\delta_{cd}$ from Eq.~\eqref{eq:O(n)_D2phi4}:
\begin{align}
    \left[\dot{\tildee C}_{D^2\phi^4}\right]_{abcd} &= 4\lambda C_E  [(1+n)\delta_{ab}\delta_{cd}-\delta_{ad}\delta_{bc}-\delta_{ac}\delta_{bd}] - \dot C_E \delta_{ab}\delta_{cd}\nonumber \\
    &= -4\lambda C_E (\delta_{ab}\delta_{cd}+\delta_{ad}\delta_{bc}+\delta_{ac}\delta_{bd})\,.
    \label{eq:Ctildedot}
\end{align}
On the other hand $[\dot{\overline C}_{D^2\phi^4}]_{abcd}$ vanishes.

\paragraph{\texorpdfstring{$\phi^6\leftarrow D^2\phi^4$}{phi6fromD2phi4}:}
Using Eqs.~\eqref{eq:phi6<-D2phi4} and \eqref{eq:Ctildedot} to compute the running of $C_1$ induced by $C_E$ one finds:
\begin{align}
    \left[\dot C_{\phi^6}\right]_{abcdef} &= 
    \frac{1}{6!}\sum_{\sigma(\{a,b,c,d,e,f\})}
    \bigg[
    \frac{1}{3!}\lambda_{abcg}\frac{1}{3}\left[\dot{\tildee C}_{D^2\phi^4}\right]_{gdef}
    +
    \frac{1}{2}\lambda_{ghab}\lambda_{ihcd}C_E \delta_{gi}\delta_{ef}\nonumber\\
    &\quad +\frac{1}{6}\lambda_{ghab}\lambda_{icde}C_E\left(\delta_{ih}\delta_{gf}+\delta_{ig}\delta_{hf}-\delta_{hg}\delta_{if}\right)\bigg] \nonumber \\
    &= 20 \lambda^2 C_E \delta_{(ab}\delta_{cd}\delta_{ef)}\,,
\end{align}
and therefore
\begin{equation}
    \dot C_1 = 20\, \lambda^2\, C_E\,.
\end{equation}

\paragraph{\texorpdfstring{$\phi^4 \leftarrow D^2\phi^4$}{phi4fromD2phi4}:}
Also for this anomalous dimension, one can exploit both Eqs.~\eqref{eq:phi4<-D2phi4} and \eqref{eq:Ctildedot}:
\begin{align}
    \dot \lambda_{abcd} &=- \sum_{\sigma(\{a,b,c,d\})} \bigg[ 
    m^2_{ae}\frac{1}{3} \left[\dot{\tildee C}_{D^2\phi^4}\right]_{bcde} +  2 m^2_{ef}\lambda_{gfab}C_E\delta_{eg}\delta_{cd}\nonumber \\
    &\quad +\left(\frac{1}{3}m^2_{ef}\lambda_{gcab}+m^2_{gc}\lambda_{efab}\right)C_E\left(\delta_{gf}\delta_{ed}+\delta_{ge}\delta_{fd}-\delta_{fe}\delta_{gd}\right)\nonumber \\
    &\quad - \frac{4}{3!} \lambda_{ebcd} m^2_{fg}C_E \left(\delta_{fe}\delta_{ag}-\delta_{fg}\delta_{ae}\right)\bigg]\nonumber \\
    &= -96 \lambda m^2 (3+n) C_E \delta_{(ab}\delta_{cd)} \,,
\end{align}
which leads to
\begin{equation}
    \dot \lambda = -16\, \lambda\, m^2\, (3+n)\, C_E\,.
\end{equation}

\paragraph{\texorpdfstring{$\phi^2 \leftarrow D^2\phi^4$}{phi2fromD2phi4}:}
The running of the mass induced by $C_E$ is obtained from Eq.~\eqref{eq:phi2<-D2phi4}:
\begin{align}
    \dot {(m^2)}_{ab} &= -2 C_E \sum_{\sigma(\{a,b\})} \Big[
    m^2_{cd}m^2_{ed} \delta_{ce}\delta_{ab}+m^2_{cd}m^2_{ae}(\delta_{ed}\delta_{bc}+\delta_{ec}\delta_{bd}-\delta_{cd}\delta_{be})
    \nonumber \\
    &\quad -2  m^2_{bc} m^2_{de} \left(\delta_{cd}\delta_{ae}-\delta_{de}\delta_{ac}\right) \Big]\nonumber \\
    &= -8 n m^4 C_E \delta_{ab}\,,
\end{align}
which gives
\begin{equation}
    \dot {(m^2)} = -8\, n\, m^4\, C_E\,.
\end{equation}

\section{Conclusions and outlook}
\label{sec:concl}

In this article we have constructed the physical basis of a general gauge EFT up to mass dimension six, using on-shell methods. For this general theory, we have derived the complete set of one-loop RGEs, where mixing effects between operators of different mass dimensions were taken into account. To demonstrate the utility of these results, we reproduced the complete one-loop bosonic RGEs of the SMEFT. Furthermore, we used our findings to derive results from specific theories such as the $O(n)$ scalar theory, as well as the SMEFT extended by an axion-like particle, finding agreement with the results in the literature.

The derived results for our generic effective gauge theory open up, for instance, the possibility for systematic phenomenological investigations of light-particle extensions of the SM. When such new degrees of freedom are incorporated into the EFT framework, our results allow to compute the corresponding RGE effects. Notably, we show this for the case of axion-like particles with CP-violating interactions, by computing their effects on the SMEFT RGEs which, to the best of our knowledge, have never been derived before. Additionally, the framework can be extended to include new gauge groups, such as additional $U(1)$ symmetries, which could mix with $U(1)_Y$ and lead to non-trivial phenomenological consequences.

Using a generic EFT has highlighted the advantage of performing such general computations, as opposed to working in specific models. One clear benefit is that the RGE for a given class of operators needs to be computed only once. For example, in the case of $\phi^2 F^2 \leftarrow \phi^2 F^2$ the running of four SMEFT Wilson coefficients can be inferred at once, after the term $[C_{\phi^2 F^2}]_{ab}^{A_\alpha B_\beta}$ is decomposed into the corresponding Clebsch-Gordan coefficients. 

Once the complete one-loop anomalous dimension matrix including fermions is known \cite{Aebischer:WIP}, we plan to implement a software containing our results, which allows to perform group theory calculations automatically, like for instance \texttt{GroupMath}~\cite{Fonseca:2020vke}. Such a software will then allow for a complete streamlining of one-loop RGE calculations for gauge theories.

Our findings have primarily been obtained using on-shell, unitarity-based methods. As emphasized in the literature, this method offers the advantage of reusing the same amplitudes to determine multiple anomalous dimensions. Furthermore, the amplitudes we derived will also serve as building blocks for calculating two-loop anomalous dimensions, a goal we plan to pursue in the future. 

\acknowledgments 
JA and NS thank Mikolaj Misiak and Ignacy Nalecz for the collaboration in the early stages of the project. 
LCB and NS thank Paride Paradisi for helpful discussions and encouragement.
JA would like to thank Andreas Helset for useful discussions.
The work of J.A. is supported by the European Union’s Horizon 2020 research and innovation program under the Marie Sk\l{}odowska-Curie grant agreement No.~101145975 - EFT-NLO.
The project received funding from the INFN Iniziativa Specifica APINE.

\section*{Note added}
While finishing this work, the articles \cite{Fonseca:2025zjb,Misiak:2025xzq} which overlap with our study were posted on arXiv. However, our approach differs methodologically as we employ on-shell techniques and a geometric framework, whereas the cited works utilize a diagrammatic approach and functional methods. 

\appendix
\newpage

\section{Conventions and definitions}\label{app:conventions}

In the following, we report a set of conventions and definitions used in this work.

\subsection{Collinear anomalous dimensions}\label{sec:coll_ADM}

The collinear anomalous dimensions for the vector and scalar fields have been computed within the method of form factors, by using as an operator the UV protected stress-energy tensor~\cite{Caron-Huot:2016cwu,EliasMiro:2020tdv,AccettulliHuber:2021uoa,Bresciani:2024shu,Baratella:2022nog}. 
They are respectively given by
\begin{align}
    \gamma_{c,v}^{A_\alpha B_\alpha} = -g_\alpha^2 \left[b_{0,\alpha}\right]^{A_\alpha B_\alpha} \,,  &&
    \gamma_{c,s}^{ab} = -4\sum_{\alpha=1}^{N_G} g_\alpha^2 \left[C_2(R_\alpha)\right]_{a_{\alpha}b_{\alpha}} \,,
    \label{eq:coll_anom_dim}
\end{align}
where
\begin{equation}
    \left[b_{0,\alpha}\right]^{A_\alpha B_\alpha} = \frac{11}{3}\left[C_2(G_\alpha)\right]^{A_\alpha B_\alpha} - \frac{1}{6} \left[S_2(R_\alpha)\right]^{A_\alpha B_\alpha}\,,
\end{equation}
is the one-loop $\beta$-function coefficient of the gauge coupling $g_\alpha$ and
\begin{align}
    \left[C_2(G_\alpha)\right]^{A_\alpha B_\alpha} &= f^{A_\alpha C_\alpha D_\alpha}f^{B_\alpha C_\alpha D_\alpha}\,,\\
    \left[S_2(R_\alpha)\right]^{A_\alpha B_\alpha} &= \Tr(\theta^{A_\alpha}\theta^{B_\alpha})\,, \\ 
    \left[C_2(R_\alpha)\right]_{a_{\alpha}b_{\alpha}} &= \theta^{A_\alpha}_{a_{\alpha}c_{\alpha}} \theta^{A_\alpha}_{c_{\alpha}b_{\alpha}}\,,
\end{align}
respectively define the quadratic Casimir of the adjoint representation of $G_\alpha$, $C_2(G_\alpha)$, the Dynkin index of the representation $R_{\alpha}$, $S_2(R_\alpha)$, and its quadratic Casimir $C_2(R_\alpha)$.

\subsection{Matrix identities}

The matrix identities used for the SMEFT cross-check are:
\begin{gather}
\theta^h \Sigma^I = \theta^I\,, \quad
\theta^h \theta^I = \frac{1}{4} \Sigma^I\,,\quad
\theta^I \Sigma^J = \delta^{IJ}\theta^h+\frac{i}{2}\epsilon^{IJK}\Sigma^K\,,
\label{eq:matrixid1}\\
\theta^I \theta^J = \frac{1}{4}\delta^{IJ}\mathbb 1 + \frac{i}{2}\epsilon^{IJK}\theta^K\,,\quad
    \Sigma^I \Sigma^J = \delta^{IJ}\mathbb 1 + 2i \epsilon^{IJK}\theta^K\,,\label{eq:matrixid2}\\
    \Sigma_{ab}^I\Sigma^I_{cd}+\Sigma_{ac}^I\Sigma^I_{bd}+\Sigma_{ad}^I\Sigma^I_{bc} = \delta_{ab}\delta_{cd}+\delta_{ac}\delta_{bd}+\delta_{ad}\delta_{bc}\,, \label{eq:thetaI_thetaJ}\\
    \theta^h_{ac}\theta^h_{bd}+\theta^h_{ad}\theta^h_{bc}=
    \frac{1}{4} \left(\delta_{ac}\delta_{bd}+\delta_{ad}\delta_{bc}-\delta_{ab}\delta_{cd}-\Sigma_{ab}^I\Sigma^I_{cd}\right)\,, \\
    \theta^I_{ac}\theta^I_{bd}+\theta^I_{ad}\theta^I_{bc}=
    -\frac{1}{4}\left(2 \delta_{ab}\delta_{cd} - \delta_{ac}\delta_{bd}- \delta_{ad}\delta_{bc}\right)\,,\label{eq:thetaI_thetaI_permuted}\\
    \theta^h_{ad}\theta^h_{bc}+2\theta^h_{ac}\theta^h_{bd}+\theta^h_{ab}\theta^h_{cd}=\frac{1}{2}\left(\delta_{ad}\delta_{bc}-\delta_{ab}\delta_{cd}\right)+\frac{1}{4}\left(\Sigma^I_{ad}\Sigma^I_{bc}-\Sigma^I_{ab}\Sigma^I_{cd}\right) \,,\\
    \theta^I_{ad}\theta^I_{bc}+2\theta^I_{ac}\theta^I_{bd}+\theta^I_{ab}\theta^I_{cd} = \frac{3}{4} \left(\delta_{ad}\delta_{bc}-\delta_{ab}\delta_{cd}\right)\,,\\
    \theta^I_{ab}\theta^I_{cd} = -\frac{3}{8}(\delta_{ac}\delta_{bd}-\delta_{ad}\delta_{bc})-\frac{1}{2}(\theta^I_{ac}\theta^I_{bd}-\theta^I_{ad}\theta^I_{bc})\,,\\
    \theta^h_{ab}\theta^h_{cd} = -\frac{1}{2}\delta_{ac}\delta_{bd}+\frac{1}{4}\left(\Sigma^I_{ab}\Sigma^I_{cd}+\Sigma^I_{ad}\Sigma^I_{bc}\right)+\theta^h_{ad}\theta^h_{bc}\,.
\end{gather}
The matrices $\theta^h$, $\theta^I$, and $\Sigma^I$ are defined in Eqs.~\eqref{eq:theta_h_and_I} and~\eqref{eq:Sigma_matrices}.

\subsection{Symmetry factors}
\label{sec:symm_factors}
To simplify the notation of the RGEs we introduce generalized symmetry factors. The following notation is adopted:
\begin{equation}
    \mathcal S_{i_1\dots i_n}=\prod_{\ell=1}^{N_{i_1\dots i_n}} k_\ell !\,,
    \label{eq:symm_fact}
\end{equation}
where $N_{i_1\dots i_n}$ is the number of distinct indices in $I=\{i_1,\dotsc, i_n\}$. Here, $k_\ell$ denotes the cardinality of the $\ell$-th subset of $I$ containing identical indices. Hence, 
\begin{equation}
    \sum_{\ell=1}^{N_{i_1\dots i_n}}k_\ell=n\,.
\end{equation}
Therefore, for $n=2$ and $n=3$ one finds
\begin{align}
    \mathcal S_{ij} = \begin{cases}
        1 & \text{if }i\neq j\,, \\
        2 & \text{if }i=j\,,
    \end{cases} && 
    \mathcal S_{ijk} = \begin{cases}
        1 & \text{if }i\neq j \land j\neq k \land k\neq i\,, \\
        2 & \text{if }i=j\neq k \lor k=i\neq j \lor j=k\neq i \,, \\
        6 & \text{if } i=j=k\,,
    \end{cases}
\end{align}
respectively.

\bibliographystyle{JHEP}
\bibliography{biblio}

\end{document}